\documentclass[11pt,sort&compress]{article}
\usepackage{fullpage}
\usepackage{cite}
\usepackage{graphicx}
\usepackage{amsmath}
\usepackage{amssymb}
\usepackage{hyperref}
\usepackage{color}
\title{Top polarisation studies in $H^-t$ and $Wt$ production}
\date{}
\renewcommand{\vec}[1]{\mbox{\boldmath$ #1 $}}

\begin{document}
\bibliographystyle{utphys}
\newcommand{\msbar}{\ensuremath{\overline{\text{MS}}}}
\newcommand{\DIS}{\ensuremath{\text{DIS}}}
\newcommand{\abar}{\ensuremath{\bar{\alpha}_S}}
\newcommand{\bb}{\ensuremath{\bar{\beta}_0}}
\newcommand{\rc}{\ensuremath{r_{\text{cut}}}}
\newcommand{\Nd}{\ensuremath{N_{\text{d.o.f.}}}}
\setlength{\parindent}{0pt}
\newcommand{\todo}[1]{\textbf{\textcolor{red}{(#1)}}}

\titlepage
\begin{flushright}
Nikhef-2011-029\\
\end{flushright}

\vspace*{0.5cm}

\begin{center}
{\Large \bf Top polarisation studies in $H^-t$ and $Wt$ production}

\vspace*{1cm}
\textsc{R. M. Godbole$^{a}$\footnote{rohini@cts.iisc.ernet.in}, L. Hartgring$^{b}$\footnote{L.Hartgring@nikhef.nl}, I. Niessen$^{c}$\footnote{I.Niessen@science.ru.nl} 
and C.D. White$^{d}$\footnote{Christopher.White@glasgow.ac.uk} } \\

\vspace*{0.5cm} $^a$ Center for High Energy Physics, Indian \\Institute of 
Science, Bangalore 560 012, India\\

\vspace*{0.5cm} $^b$ Nikhef, Science Park 105, 1098 XG \\Amsterdam, The 
Netherlands\\

\vspace*{0.5cm} $^c$ Theoretical High Energy Physics, IMAPP, Faculty of Science, Mailbox 79,\\
P.O. Box 9010, NL-6500 GL Nijmegen, The Netherlands\\

\vspace*{0.5cm} $^d$ School of Physics and Astronomy, Scottish Universities
Physics \\Alliance, University of Glasgow, Glasgow G12 8QQ, Scotland, UK\\

\end{center}

\vspace*{0.5cm}

\begin{abstract}
The polarisation of top quarks produced in high energy processes can 
be a very sensitive probe of physics beyond the Standard Model. 
The kinematical distributions of the decay products of the top quark 
can provide clean information on the polarisation of the produced top and thus can probe new physics effects in the top quark sector. We 
study some of the recently proposed polarisation observables involving the 
decay products of the top quark in the context of $H^-t$ and $Wt$ 
production. 
We show that the effect of the top  polarisation on the decay lepton azimuthal 
angle distribution, studied recently for these processes at leading
order in QCD, is robust with respect to the inclusion of next-to-leading 
order and parton shower corrections.  We also consider the leptonic polar
angle, as well as recently proposed energy-related distributions of the top 
decay products. We construct asymmetry parameters from these observables, which
can be used to distinguish the new physics signal from the $Wt$ background and discriminate between different values of $\tan\beta$
and $m_{H^-}$ in a general type II two-Higgs doublet model. Finally, we
show that similar observables may be useful in separating a Standard Model
$Wt$ signal from the much larger QCD induced top pair production background.
\end{abstract}

\vspace*{0.5cm}

\section{Introduction}

The top quark $t$ is the heaviest known fundamental particle. Its mass is similar to the energy scale of electroweak symmetry breaking. Given that 
physics beyond the Standard Model (BSM)  may describe the origin of this symmetry
breaking, it is widely hoped that new physics will show itself by leaving 
an imprint in the behaviour of the top quark. In most BSM scenarios, top
quarks play a special role and arise prominently in the decays of new 
particles, e.g. new gauge bosons, gluinos, top-partners or heavy resonances 
involving the $t$. The Large Hadron Collider (LHC) offers top quark production rates 
far in excess of  those at the Tevatron, allowing detailed scrutiny of the top 
quark and its interactions. Usually, the biggest background to such new physics searches are top quarks produced by QCD processes within Standard Model (SM). It then becomes imperative to
look for criteria that can discriminate efficiently between the two sources
of the produced top quarks.\\

Polarisation of the top quark can be one very important handle to identify new physics signals for two reasons. Firstly, it is well known that the polarisation of produced particles can
provide more information about the dynamics of the production process than
total cross-sections, since it can probe the chiral structure of the 
interaction responsible. Even more importantly, for the QCD induced $t \bar t$
production, which forms the bulk of the top production at the LHC, the top 
quark is unpolarised on average. In contrast, if a top is produced in association with the $W$, the $V$--$A$ nature of the
weak interaction implies that the produced top quark is
always left-handed, so the top quark is completely polarised. 
Top quarks coming from BSM processes often can have a different polarisation as well. Hence, the polarisation of the produced top can help
to distinguish the SM top quarks from the BSM top quarks.\\

Fortunately, the top polarisation is also a quantity which is amenable to an
experimental measurement. Due to its large mass, the top quark decays before it hadronises. Therefore the top polarisation state can leave an imprint in the kinematic distributions of its decay products. The correlation between the top spin direction
and these kinematic distributions can be used 
effectively to get information about the former and hence about the 
dynamics responsible for producing the top in a specific state of polarisation.
In fact, many studies have explored the use of the top polarisation as 
a probe and discriminator of new  physics~\cite{Harlander:1994ac,Hikasa:1999wy,Rindani:1999gd,Boos:2003vf,Gajdosik:2004ed,Allanach:2006fy,Godbole:2006eb,Godbole:2006tq,Li:2006he,Najafabadi:2006um,BhupalDev:2007is,Eriksson:2007fx,Perelstein:2008zt,Nojiri:2008ir,Shelton:2008nq,Godbole:2009dp,Arai:2010ci,Djouadi:2009nb,Krohn:2009wm,Godbole:2010kr,AguilarSaavedra:2010nx,Degrande:2010kt,Huitu:2010ad,Cao:2010nw,Jung:2010yn,Choudhury:2010cd,Gopalakrishna:2010xm,Godbole:2011hw,Krohn:2011tw,Baumgart:2011wk,Rindani:2011pk,Baglio:2011ap,Rindani:2011gt,Ananthanarayan:2010xs}. 
Uses of top polarisation as a means to obtain information on the mechanism of  
$t \bar t$ pair production~\cite{Harlander:1994ac,Hikasa:1999wy,Allanach:2006fy,Godbole:2006eb,BhupalDev:2007is,Godbole:2009dp,Godbole:2010kr,Degrande:2010kt,Cao:2010nw,Jung:2010yn,Choudhury:2010cd,Godbole:2011hw,Krohn:2011tw,Ananthanarayan:2010xs} 
and that of single top production~\cite{Arai:2010ci,Huitu:2010ad,Rindani:2011pk,Rindani:2011gt} or to sharpen up the signal of new physics~\cite{Perelstein:2008zt,Nojiri:2008ir,Agashe:2006hk} by reducing the 
background from unpolarised tops, exist in the literature. Of particular 
interest for the purposes of this note, 
are the investigations of Refs.~\cite{Huitu:2010ad,Baglio:2011ap}, 
which showed that top polarisation can be used to extract information on the
model parameters of a two Higgs Doublet model via a study of associated 
production of a charged Higgs  and the  $t$ quark.  Different probes of the top 
polarisation, using the above mentioned correlation between the top spin
direction and decay product kinematic distributions have been 
constructed~\cite{Allanach:2006fy,Godbole:2006tq,Shelton:2008nq,Godbole:2009dp,Godbole:2010kr,Krohn:2009wm,Falkowski:2011zr}.
The angular distributions of the decay leptons provide a particularly 
robust probe due to their insensitivity to higher order  
corrections~\cite{Jezabek:1988ja,Czarnecki:1990pe,Brandenburg:2002xr} 
and to possible new physics in the $tbW$ 
vertex~\cite{Grzadkowski:1999iq,Grzadkowski:2002gt,Grzadkowski:2001tq,Hioki:2002vg,Ohkuma:2002iv,Rindani:2000jg,Godbole:2002qu}.\\

As will be discussed later, the traditional probe of polarisation requires a
measurement of the angular distribution of the decay products in the rest frame
of the decaying top and thus reconstruction of the top quark rest frame is needed. It helps if the top polarisation observables one considers can 
be constructed in the lab frame, thereby avoiding the uncertainties
which might arise from having to  reconstruct the top 
quark rest frame. One such observable for a top quark that decays
leptonically  was presented 
in~\cite{Allanach:2006fy,Godbole:2006tq,Godbole:2009dp,Godbole:2010kr}. 
In this case the authors considered the azimuthal angle of the decay lepton
in the lab frame, and showed that this can be a sensitive probe of top quark
polarisation and, consequently, new physics effects. \\

As mentioned above, the angular observables are independent of corrections to the {\it decay} of the top quark to a good approximation, so they depend only on nonzero polarisation contributions to  the {\it production} of the top\footnote{Throughout
the paper, we will adopt the framework of the narrow width approximation,
in which production and decay are explicitly disentangled.}.
However, for the case of heavily boosted tops, the decay products of the top
quark get collimated. While in principle, it may be possible to construct the
angular observables in this case as well \cite{atlasnote}, additional 
polarisation observables constructed using energies of the top decay products 
as measured in the laboratory  can be of interest and use in this case. 
Such observables were recently proposed and studied in~\cite{Shelton:2008nq,Krohn:2009wm} and take the form of energy ratios of various top decay products. These
observables are sensitive to corrections to both the production and decay 
of the top quark~\cite{Godbole:2006tq,AguilarSaavedra:2010nx} and thus can potentially offer a 
complementary window on new physics in the top quark sector.\\

The observable based on the azimuthal angle of the decay 
lepton~\cite{Godbole:2006tq,Godbole:2010kr} was 
further exploited in~\cite{Huitu:2010ad} for the specific case of top quark 
production in association with a charged Higgs boson. It was shown that 
azimuthal observables are potentially efficient in discriminating between 
different regions of the charged Higgs parameter space and in
separating the $Ht$ production process from SM single top production in association with a $W$. However, this analysis was carried out at leading order (LO) in perturbation 
theory only. The decay product kinematic distributions in the {\it lab} receive 
both polarisation dependent and independent contributions. The latter depend
on the kinematics of the decaying top, such as its transverse momentum and 
the boost parameter.  While the higher order corrections coming from the
chirality and parity conserving QCD  interactions will not affect the top 
polarisation, they can change the kinematics of the produced top quark and 
hence it is  important to verify that the conclusions of the LO analysis are 
robust against next-to-leading order (NLO) corrections. \\

The aim of this paper is to study all the observables mentioned above in two
different contexts. Firstly, we reconsider $H^-t$ production, in the setup
of a general type II two Higgs doublet model. We confirm the results 
of~\cite{Huitu:2010ad} and, importantly, demonstrate explicitly that polarisation effects are still prevalent when NLO corrections are included,
together with a parton shower for estimating the effect of higher order
quark and gluon radiation. To this end, we use the recently developed MC@NLO
software of~\cite{Weydert:2009vr}. We furthermore extend the analysis 
of~\cite{Huitu:2010ad} by including polar angle distributions, and examining
the energy-related observables of~\cite{Shelton:2008nq}. We use our results
to motivate the definition of certain asymmetry parameters, all of which are
shown to give markedly different values for different regions of the charged
Higgs model parameter space, as well as for the main background of Standard 
Model $Wt$ production.\\

The second context we consider is that of $Wt$ production itself. This is an
important background for a number of new physics searches, but is also an 
interesting production channel in its own right~\cite{Tait:1999cf,
Campbell:2005bb,Zhu:2002uj,Frixione:2008yi,White:2009yt}, and one of three 
different single top production modes in the Standard Model, such that it
represents approximately 20\% of the total rate. 
Whilst the other two, $s-$ and $t-$channel production, are sensitive to the existence of 
both four fermion operators and corrections to the $Wtb$ vertex, $Wt$ 
production only depends on the latter. Thus it offers a useful comparison 
with the other production modes from a new physics point of view. It is
also important to verify the Standard Model, and $Wt$ production has
yet to be observed. A significant background to this process comes from the top pair
production. It is of interest to examine observables which may enhance the signal to background ratio of the $Wt$ 
mode. Polarisation-dependent observables are potentially useful because
a top quark that is produced in association with a $W$ boson is completely polarised, while in top pair production the top quarks are
unpolarised on average. We will indeed see that the same observables that we study in the
context of $H^-t$ production are also useful in the $Wt$ case.\\

The structure of the paper is as follows. In section~\ref{sec:define}, we
define the various observables which we consider throughout the rest of the 
paper and briefly discuss the general effects one expects when including NLO corrections. In section~\ref{sec:Ht}, we present results for these observables from
$H^-t$ production, and use the distributions we obtain in order to construct 
asymmetry parameters, which distil the difference between different charged
Higgs parameters, or between $H^-t$ and $Wt$ production. In 
section~\ref{sec:Wt} we examine the use of similar observables in trying to
separate $Wt$ from top pair production. Finally, in section~\ref{sec:conclude}
we discuss our results and conclude. 

\section{Polarisation dependent observables in top quark production}
\label{sec:define}

In this section, we briefly review the observables we will consider throughout 
the paper. We will study both angular and energy observables. 
The starting point of construction of all the polarisation observables is
the angular distribution of the decay products in the rest frame 
of the $t$ quark:  
\begin{displaymath}
t\rightarrow Wb\rightarrow i\, i'\, b,
\end{displaymath}
where $i$ and $i'$ denote the decay products of the $W$. 
Throughout the paper we will neglect off-diagonal elements of the CKM matrix,
considering only the decay to $b$ quarks. Furthermore, we will 
explicitly  talk about single top quark production for the time being, given 
that single antitop quark production can be distinguished from this by 
considering the sign of the lepton from the top quark decay. 
The polarisation of the produced quark is given by,
\begin{equation}
P_t=\frac{\sigma(+,+)-\sigma(-,-)}{\sigma(+,+)+\sigma(-,-)},
\label{Ptdef}
\end{equation}
where $\sigma(\pm,\pm)$ is the cross-section for a positive or 
negative helicity top quark respectively. In general, the transverse polarisation is negligible.\\

The effect that the polarisation of the top quark ensemble has on its decay 
products is most easily studied in the top quark rest frame, where the angular distribution of the decay product $f$ 
is given by:
\begin{equation}
\frac{1}{\Gamma_l}\frac{\mathrm{d}\Gamma_l}{\mathrm{d}\cos\theta_{f,\rm rest}}=\frac{1}{2}\left(1+ \kappa_f P_t\cos\theta_{f,\rm rest}\right).\label{eq:topdecay}
\end{equation}
Here $\Gamma_l$ is the partial decay width, $P_t$ is the degree of 
polarisation in the top quark ensemble and the polar angle 
$\theta_{f,\rm rest}$ is the angle between the decay product $f$ and the top spin vector. $\kappa_f$ is the analysing power of the decay product $f$. It is $1$ for 
a positive lepton and a $d$ quark. For the $u$ quark and $\nu_l$ its value 
is -0.31 and for the $b$ and $W$ the values are $-0.4$ and $0.4$ 
respectively~\cite{Bernreuther:2008ju}.
Thus we see that a positively charged lepton is the most efficient
polarisation analyser. Corrections to
these values of $\kappa$ can originate from any nonstandard  $tbW$ couplings and/or
from higher order QCD and QED corrections. The leading 
QCD corrections to 
$\kappa_b$, $\kappa_d$ and $\kappa_u$ are of the order of a few percent,
decreasing its magnitude somewhat~\cite{Brandenburg:2002xr}. As shown 
explicitly in~\cite{Godbole:2010kr} the value of $\kappa_l$ does not
receive any corrections from the anomalous $tbW$ coupling at leading order. 
Thus the angular distribution of the decay lepton in the rest frame reflects 
the polarisation of the decaying quark faithfully even in the presence of 
such corrections, and hence is a good measure of polarisation effects
in the top production process.\\

However, we want to use polarisation-dependent observables in the lab frame. 
The correlation between the polarisation of the decaying top and the different 
kinematic variables of the decay product are then obtained by using 
eq.~(\ref{eq:topdecay}) and appropriate Lorentz transformations. As already
mentioned in the introduction, a series of investigations indicate that
analagously to the situation in the top rest frame 
the energy integrated decay lepton angular distributions in the lab frame are unaltered to  linear order in the anomalous $tbW$ coupling.
Thus the correlation between the top polarisation and angular distributions 
of the decay lepton is unchanged to the same order. It is important to note that
the decay lepton distributions in the lab frame are influenced not only by the top quark 
polarisation, but also by the boost $B$ from the top quark rest frame 
to the laboratory frame and by the transverse momentum
of the top quark $p_t^T$. Here we will use a boost parameter based on the total momentum of the top $|\vec{p}_{\rm top}|$ and the top energy $E_t$
\begin{equation}
B=\frac{|\vec{p}_{\rm top}|}{E_t}.
\label{Bdef}
\end{equation}

As an example we consider the lab frame polar angle $\theta_l$ of the lepton w.r.t. the top quark direction. Due to the top boost, $\theta_l$ is smaller than its counterpart in the rest frame $\theta_{l,\rm rest}$. Thus, the distribution of 
$\theta_l$ in the lab frame is more strongly peaked towards $0$ for a stronger  top boost as well as for a more positively 
polarized top quark. 
\\

In addition to the polar angle, one can study the azimuthal angle. To this end, the $z$ axis is chosen to be the beam axis. Together with the top quark 
direction this defines the top quark production plane, containing the $z$ 
and $x$ axes, the $x$-axis chosen such that the top quark momentum has a 
positive $x$ component. We then construct a right-handed coordinate system 
and define the azimuthal angle $\phi_l$ as the angle of the decay lepton in 
the ($x$,$y$) plane. In the rest frame this variable does not depend on the 
longitudinal polarisation, but in the lab frame it picks up a dependence on 
$\theta_{l,\rm rest}$ through the top boost. Consequently it can be used as 
a probe for the top quark polarisation. An example shape of the $\phi_l$ 
distribution may be seen in figure 4 of~\cite{Godbole:2010kr}, or 
in figure~\ref{thetal1} of this paper. For positively polarized tops it is peaked at $\phi_l=0$ and 
$\phi_l=2\pi$, with a minimum at $\phi_l=\pi$. It should be noted that 
nonzero $p_t^T$ also causes the $\phi_l$ distributions to peak near $\phi_l=0$
and $\phi_l = 2\pi$, {\it independent} of the polarisation state of the
$t$ quark. In other words, the peaking at $\phi_l=0$ and $2 \pi$ is caused by kinematic effects, even for an unpolarised top. It is enhanced even further for a positively polarised top. For a completely negatively polarised top, the pure polarisation dependent effects can sometimes even overcome the peaking caused by kinematical effects. The peaks
of the distribution then shift a little away from $\phi=0$ and $2 \pi$. More importantly they lie below those expected for the positively polarised and unpolarised top. The relative number of leptons near $\phi=0$ and $2 \pi$ is thus reduced progressively as we go from a positively polarised to unpolarised to a negatively polarised top. For normalised distributions the ordering is exactly the
opposite at $\phi =\pi$ where the relative number of leptons increases as we go from a positively polarised top to a negatively polarised top.\\

This shape then motivates the 
definition of the asymmetry parameter~\cite{Godbole:2010kr}:
\begin{equation}
A_\phi=\frac{\sigma(\cos\phi_l>0)-\sigma(\cos\phi_l<0)}{\sigma(\cos\phi_l>0)
+\sigma(\cos\phi_l<0)},
\label{Aldef}
\end{equation}
where $\sigma$ is the fully integrated cross-section. A higher top quark polarisation or a stronger top boost will result in a more sharply peaked $\phi_l$ distribution and thus yield a higher value of $A_\phi$. This parameter has been considered for the specific case of $H^-t$ production in~\cite{Huitu:2010ad}, in a LO analysis at parton level (i.e. without a parton shower). There it was found that typical values of $A_\phi$ are very different to those obtained for $Wt$ production. Furthermore, there is pronounced variation of $A_\phi$ as both 
$\tan\beta$ (the ratio of Higgs VEVs) and the charged Higgs mass $m_{H}$ are 
varied. We reconsider these results in section~\ref{sec:Ht}.\\

Although energy observables are not independent of the top quark decay, 
they can provide additional information about the production process and may 
be of particular use when the top quarks are highly boosted. 
It was shown in~\cite{Shelton:2008nq} that in a kinematic regime where the 
tops are heavily boosted the following ratios are sensitive to the polarisation state of the top quark:
\begin{equation}
z=\frac{E_b}{E_t},\quad u=\frac{E_l}{E_l+E_b},
\label{zudef}
\end{equation}
where $E_t$, $E_b$ and $E_l$ are respectively the (lab frame) energies of the 
top quark, and the $b$ quark and lepton coming from its decay. The analysis
of~\cite{Shelton:2008nq} was at the LO parton level, but in practical 
applications one may also consider $E_b$ to be the energy of e.g. a $b$ jet. 
Note that the ranges of $z$ and $u$ are given in principle by
\begin{equation}
0\leq z,u\leq 1,
\label{zurange}
\end{equation}
although there will be a cut-off at high and low values due to the finite $b$
quark and $W$ boson masses. One may define these observables for
any value of a cut on the top quark boost parameter, but at low values of the boost, both $z$ and $u$ are
increasingly contaminated with contributions that are insensitive to the
top quark polarisation, thus reducing their effectiveness as discriminators
of new physics parameters etc. We will see this explicitly in 
section~\ref{sec:Ht}.\\

\subsection{Differences between leading order and next-to-leading order}
\label{sec:LOvsNLO}

So far these polarisation-dependent observables have been studied only at leading order (LO) accuracy. For a given polarisation-dependent observable, such a calculation
represents a best case scenario in which polarisation effects in the
production of the top quark are the least diluted by kinematic effects. Beyond this order in 
perturbation theory, additional radiation may carry away energy and/or 
angular momentum. The goal of this paper is to extend the study to next-to-leading order (NLO) accuracy, including also the effects of a parton shower. Studying the observables at NLO + shower level and comparing them to the LO result provides a handle on their robustness.\\

The NLO calculation includes QCD interactions, which conserve parity and chirality. Therefore, the NLO corrections cannot change the 
polarisation of the top quark. Kinematic 
effects on the other hand do change when going to NLO + shower accuracy. In particular, as will be shown explicitly in figure~\ref{BplotHt},
the boost of the top quark, as measured by the $B$ parameter of eq.~(\ref{Bdef}), increases a few percent due to the higher order corrections. \\

For the LO computation of the $H^-t$ production process, we use MadGraph 5~\cite{Alwall:2011uj,Alwall:2007st}, where we extended the Standard Model to include the charged Higgs coupling. The NLO calculation matched to a parton shower was performed using the
MC@NLO software package described in~\cite{Weydert:2009vr,Frixione:2008yi,
Frixione:2002ik,Frixione:2008ym,Frixione:2010wd,Frixione:2003ei}, with spin
correlations implemented according to the algorithm 
of~\cite{Frixione:2007zp}\footnote{Alternative methods for matching NLO 
computations with a parton shower have been presented 
in~\cite{Frixione:2007vw,Giele:2007di}. See also~\cite{Frixione:2007nw,
Re:2010bp,Weydert:2010cx} for implementations of the processes discussed in 
this paper.}.\\

The $Wt$ production process poses a conceptual problem at NLO, due to the fact that some of the real emission diagrams beyond LO involve an intermediate top quark pair.
The contribution from such diagrams is large when the $\bar{t}$
becomes resonant, reflecting an interference between the $Wt$ and top-pair 
production processes. How to most accurately model the sum of $Wt$ and top-pair production then becomes a somewhat controversial matter of opinion, and
there are two main points of view. The first is that all singly and doubly 
resonant diagrams must be combined, thus including all interference (and
off-shell) effects (see, for example,~\cite{Kauer:2001sp,Kersevan:2006fq}). 
A major deficiency of such calculations, however, is that
they typically do not include NLO corrections, which for top pair production
are known to be large. Recently, NLO corrections for the $WWb\bar{b}$ final
state have been presented~\cite{Denner:2010jp}, also including decay of the 
$W$ bosons~\cite{Bevilacqua:2010qb}, in the so-called four flavour scheme in 
which all initial state $b$ quarks are explicitly generated via gluon 
splitting, although these results have yet to be interfaced with a parton
shower.\\

The second point of view is that singly and doubly
resonant contributions may be safely regarded as separate production processes,
which may be meaningfully combined subject to suitable analysis cuts, an 
approach followed by e.g.~\cite{Campbell:2005bb,Zhu:2002uj,Frixione:2008yi,
Re:2010bp}. This amounts to defining a subtraction term, which removes
doubly resonant contributions from the $Wt$ cross-section. A potential 
deficiency of such an approach is that gauge invariance is violated by
terms $\sim{\cal O}(\Gamma_t/m_t)$, where $\Gamma_t$ is the top quark width, 
although it is usually argued that this is more a problem of principle than
one of practice. Another way to think about this procedure is that the 
subtraction term avoids the double counting that would result upon na\"{i}vely
adding the $Wt$ and top pair cross-sections at NLO. Such on-shell subtraction
schemes are in fact a common feature in many NLO calculations involving 
extensions to the Standard Model, in which intermediate heavy particles
abound (see e.g.~\cite{Beenakker:1996ch,Plehn:1998nh,Plehn:2010gp,
Binoth:2011xi}). Indeed, in this context, the interference problem is usually 
referred to in terms of being a double counting issue.\\

It is not our intention to reignite the debate on the validity of on-shell
subtraction schemes. But, in order to discuss $Wt$ production at all, we must
necessarily take the view that it makes sense to separate singly and doubly
resonant production modes. For a detailed recent discussion of this viewpoint,
see~\cite{White:2009yt}. In that paper, it was argued that $Wt$ is unambiguous
for suitable analysis cuts, and we will assume the validity of this approach in
what follows.\\

The MC@NLO code for $Wt$ production includes two definitions of $Wt$ 
production, labelled Diagram Removal (DR) and Diagram Subtraction (DS), where 
the difference between these is intended to represent the systematic 
uncertainty due to interference with top pair production. Roughly speaking, 
DS subtracts doubly resonant (i.e. top pair) contributions at the 
cross-section level (thus is gauge invariant up to 
terms $\sim{\cal O}(\Gamma_t/m_t)$), and DR subtracts such contributions at 
the amplitude level. The difference between these then mostly measures the 
interference between $Wt$ and $t\bar{t}$ production, up to ambiguities in the 
subtraction term. However, one only formally trusts each calculation if the 
DR and DS results agree closely, which relies upon the imposition of suitable 
analysis cuts for reducing the interference. We will not implement such cuts in the calculation of the observables for $H^-t$ production. Despite this, we will show the results obtained from both the DR and DS calculations.\\

\section{Results for $H^-t$ production}
\label{sec:Ht}
In the previous section, we briefly reviewed the observables which have been
presented in~\cite{Shelton:2008nq,Godbole:2010kr}, and which are designed to
be sensitive to the polarisation state of produced top quarks. In this section,
we study these observables for single top production in association with a 
charged Higgs boson. The latter does not occur in the Standard Model of 
particle physics, but exhibits a somewhat generic presence in possible
extensions, including supersymmetry.\\

We will consider a type II two Higgs doublet model, where the
coupling of the charged Higgs to the top and bottom quarks is given by
\begin{equation}
G_{H^- t\bar{b}}=
-\frac{i}{v\sqrt 2}V_{tb}\Big[m_b\tan\beta(1-\gamma_5)+m_t\cot
\beta(1+\gamma_5)\Big].
\label{vertex}
\end{equation}
Here the vacuum expectation values of the two Higgs doublets are $v\cos\beta$
and $v\sin\beta$, such that $\tan\beta$ is their ratio~\footnote{For a 
pedagogical review of Higgs physics within and beyond the Standard Model,
see~\cite{Djouadi:2005gi,Djouadi:2005gj}.}.\\

The top quark polarisation in the $H^-t$ production process does not follow directly from eq.~(\ref{vertex}). As explained in detail in Ref.~\cite{Huitu:2010ad}, the polarisation vanishes if $m_H=6m_t$ and if $\tan\beta=\sqrt{m_t/m_b}$. In addition, it was shown in figure 4 of that paper that the $\tan\beta$ dependence of the polarisation is different for different Higgs masses. For Higgs masses below $6m_t$ it is negative if $\tan\beta<\sqrt{m_t/m_b}$ and positive for higher values of $\tan\beta$. The polarisation for higher Higgs masses has the opposite behaviour. Following Ref.~\cite{Huitu:2010ad} we will plot observables for extremal charged Higgs mass values of 200 GeV and 1500 GeV \footnote{However, see Ref.~\cite{Bona:2009cj} for current constraints on charged Higgs models from $B$ physics.}. In the rest of this section, we will often show distributions for $m_H=200$~GeV and $m_H=1500$~GeV as representative examples. For a given value of $\tan\beta$, the former is more strongly polarised than the latter.\\

One may study how the observables
of section~\ref{sec:define} vary throughout the two dimensional parameter 
space $(m_H,\tan\beta)$. In what follows, we will do this at LO and NLO, as specified in section \ref{sec:LOvsNLO}. Note that the aim of this section is not to undertake a fully 
comprehensive phenomenological analysis, including all relevant backgrounds 
together with realistic experimental cuts. Rather, we wish to study the efficacy of the different observables that reflect the polarisation of the parent top, and in particular 
their robustness when one includes higher order effects.\\

In order to present results, we consider the LHC with a centre of mass energy
of 14 TeV, and define parameters as follows: the top mass and width are 
$m_t=172.5$ GeV and $\Gamma_t=1.4$ GeV respectively. The $W$ mass and width are respectively $m_W=80.42$ GeV and $\Gamma_W=2.124$ GeV. Factorization and
renormalization scales are set to $\mu_r=\mu_f=m_t$. We calculate LO and 
MC@NLO results using MSTW 2008 LO and NLO parton sets~\cite{Martin:2009iq,
Martin:2009bu,Martin:2010db}. Note that the $b$ mass entering the Yukawa
coupling is run as in~\cite{Djouadi:1997yw}, from a pole mass of $m_b=4.95$ 
GeV~\footnote{Strictly speaking, one should run the $b$ mass at one-loop order
for the LO results, and two-loop order for the NLO results. We do not do this 
here in order to facilitate a more direct comparison between the LO and MC@NLO
results, given that the relative proportion of right- and left-handed $H^-t$
couplings is governed by the value of $m_b(\mu_r)/m_t(\mu_r)$. 
We have checked that the difference in running is a small effect.}. \\

\begin{figure}[!h]
\begin{center}
\scalebox{0.4}{\includegraphics{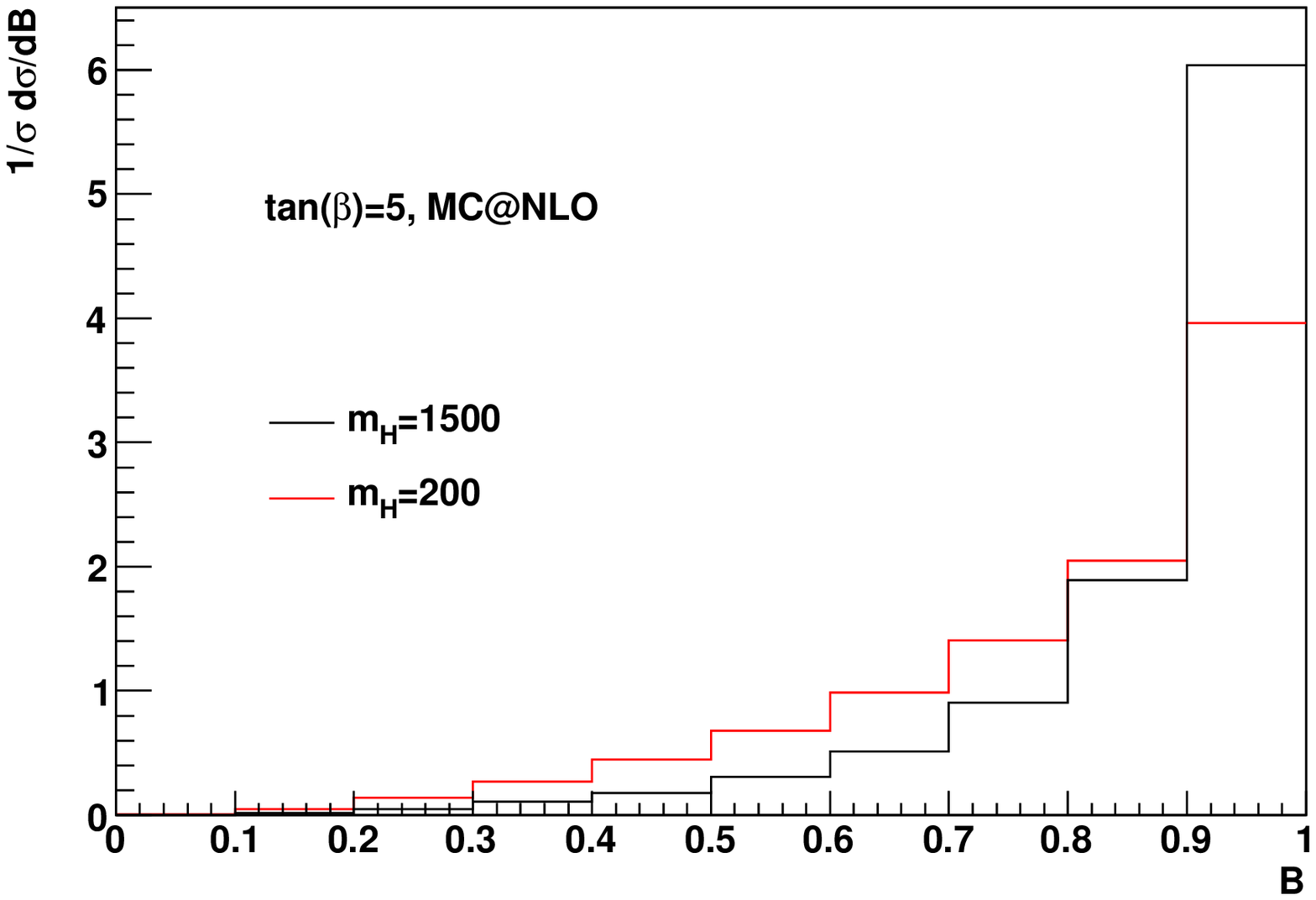}}
\scalebox{0.4}{\includegraphics{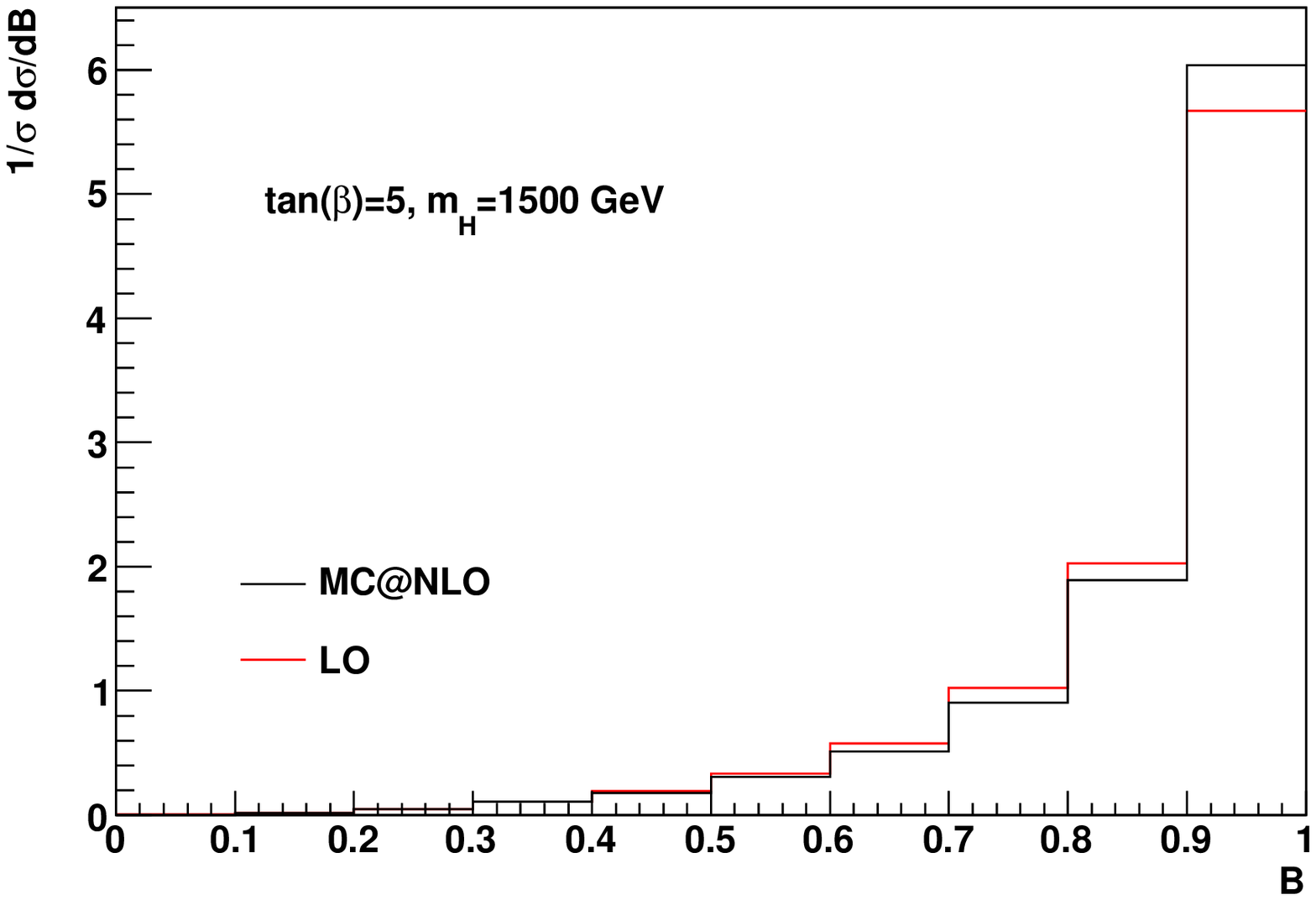}}
\caption{The distribution of the boost parameter of in $H^-t$ production for $\tan\beta=5$ and two different Higgs masses is shown on the left-hand side. On the right-hand side the boost parameter is shown at LO and NLO plus parton shower level.}
\label{BplotHt}
\end{center}
\end{figure}

As explained in section~\ref{sec:define}, the polarisation-dependent observables are affected considerably by the kinematics of the top. Therefore we first briefly discuss the boost parameter $B$ and the top transverse momentum $p_t^T$. On the left-hand side of figure~\ref{BplotHt}, the distribution of the boost parameter is shown for two different values of the charged Higgs mass. On the right-hand side, the LO and NLO + parton shower distributions are compared. The distribution is much more strongly peaked for the high
Higgs mass, as expected from the fact that the top quark must recoil against
the heavy particle. In addition we see that the NLO+parton shower effects increase the boost parameter slightly. This can be traced back to the definition of
eq.~(\ref{Bdef}), coupled with the fact that the energy of the top quark softens more on average than its momentum when higher order effects are included. 
\\

\subsection{Azimuthal angle $\phi_l$}

\begin{figure}[!h]
\begin{center}
\scalebox{0.4}{\includegraphics{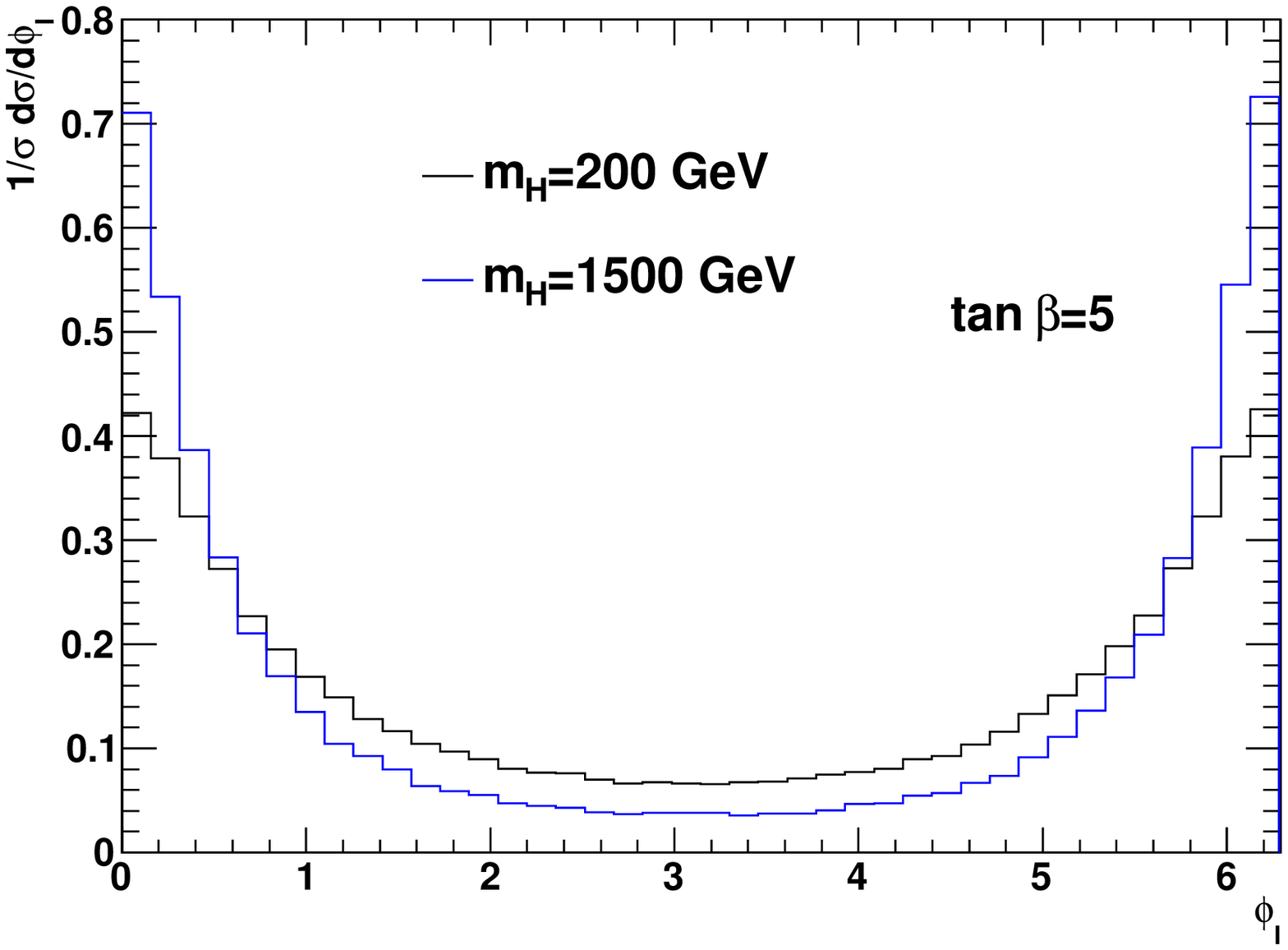}}
\scalebox{0.4}{\includegraphics{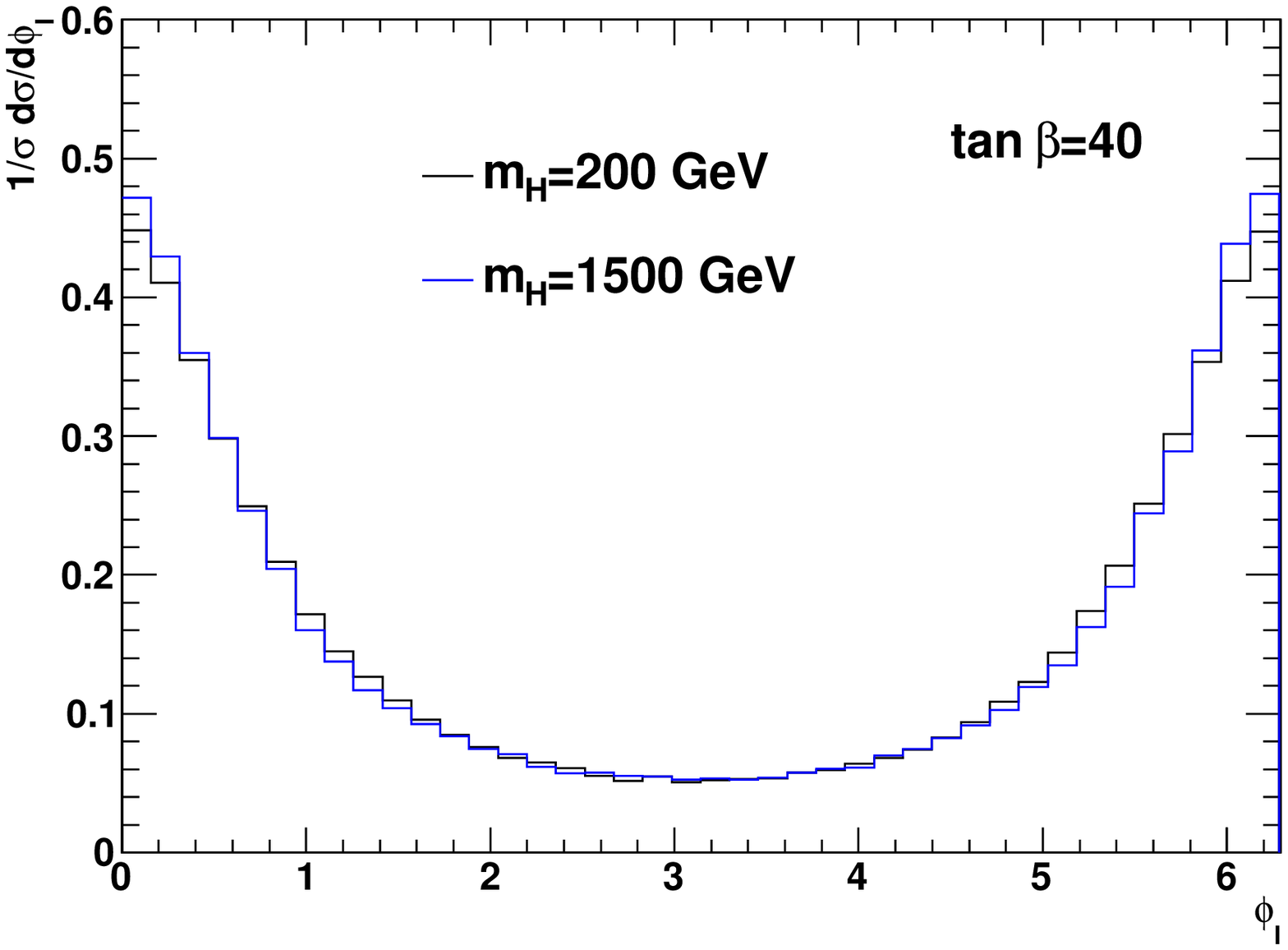}}
\caption{Azimuthal angle ($\phi_l$) of the decay lepton from the top quark, as
defined in the text, at NLO plus parton shower level.}
\label{phil1}
\end{center}
\end{figure}

Figure~\ref{phil1} shows the $\phi_l$ distribution for two different 
values of $\tan\beta$, and two different charged Higgs masses at NLO + parton shower. For $\tan\beta=5$, there is a pronounced difference 
between the two $\phi_l$ distributions at different mass values, with the 
higher mass value showing more asymmetry. 
At high $\tan\beta$, there is very little 
difference between the two Higgs mass values. The reason for this behaviour can be traced back to the polarisation of the top. At low $\tan\beta$ a light Higgs yields a negatively polarised top, so in the rest frame the lepton tends to be emitted in the backward direction (cf. eq.~(\ref{eq:topdecay})). For a heavy Higgs the top is positively polarised for low values of $\tan\beta$, so the lepton is emitted in the forward direction. Since the top is boosted more for higher Higgs masses, the kinematics enhance this polarisation effect. For large $\tan\beta$, the top polarisation has the opposite sign, so in that case the kinematics cancel the effect of the polarisation. 
\\

\begin{figure}[!h]
\begin{center}
\scalebox{0.4}{\includegraphics{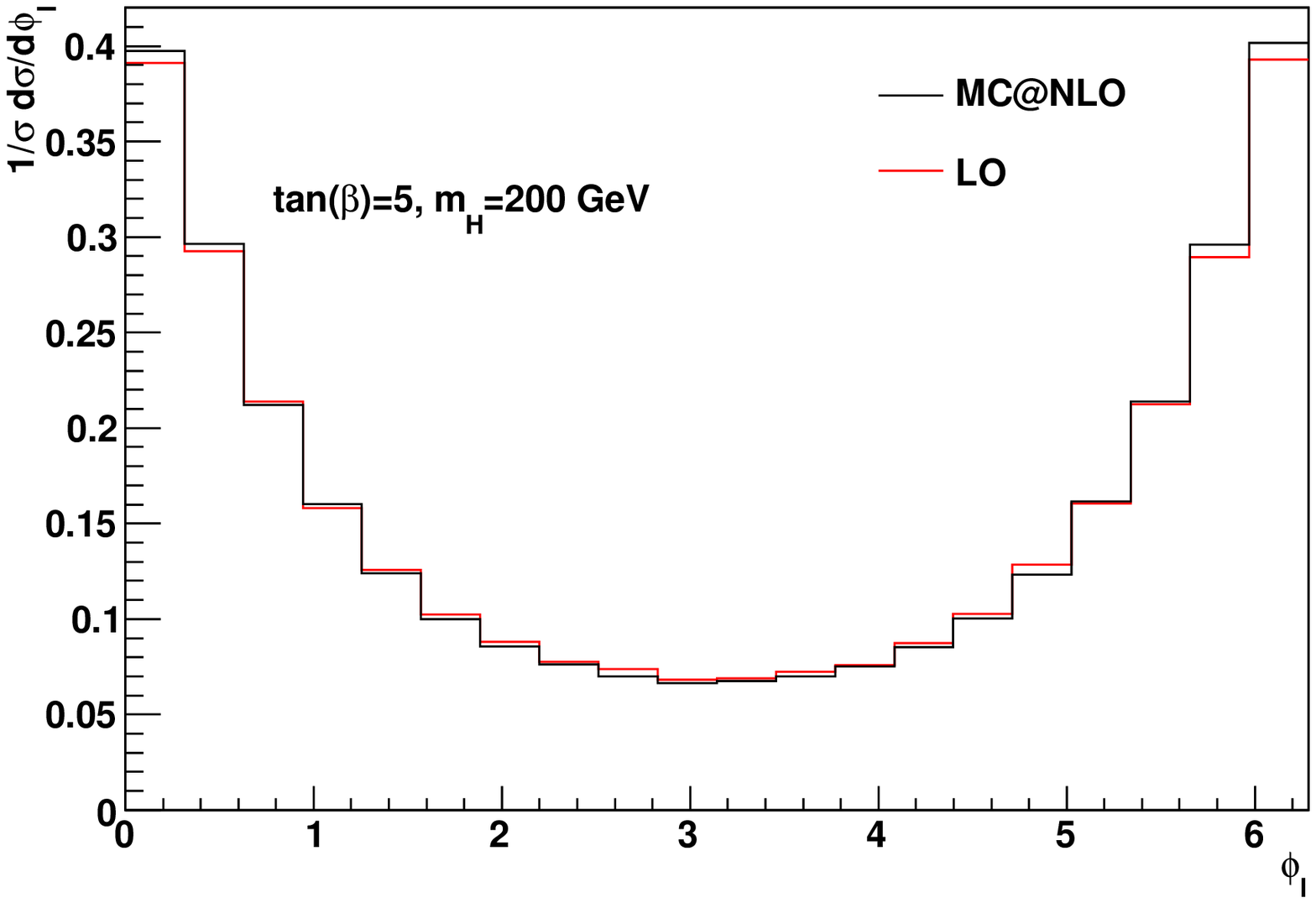}}
\scalebox{0.4}{\includegraphics{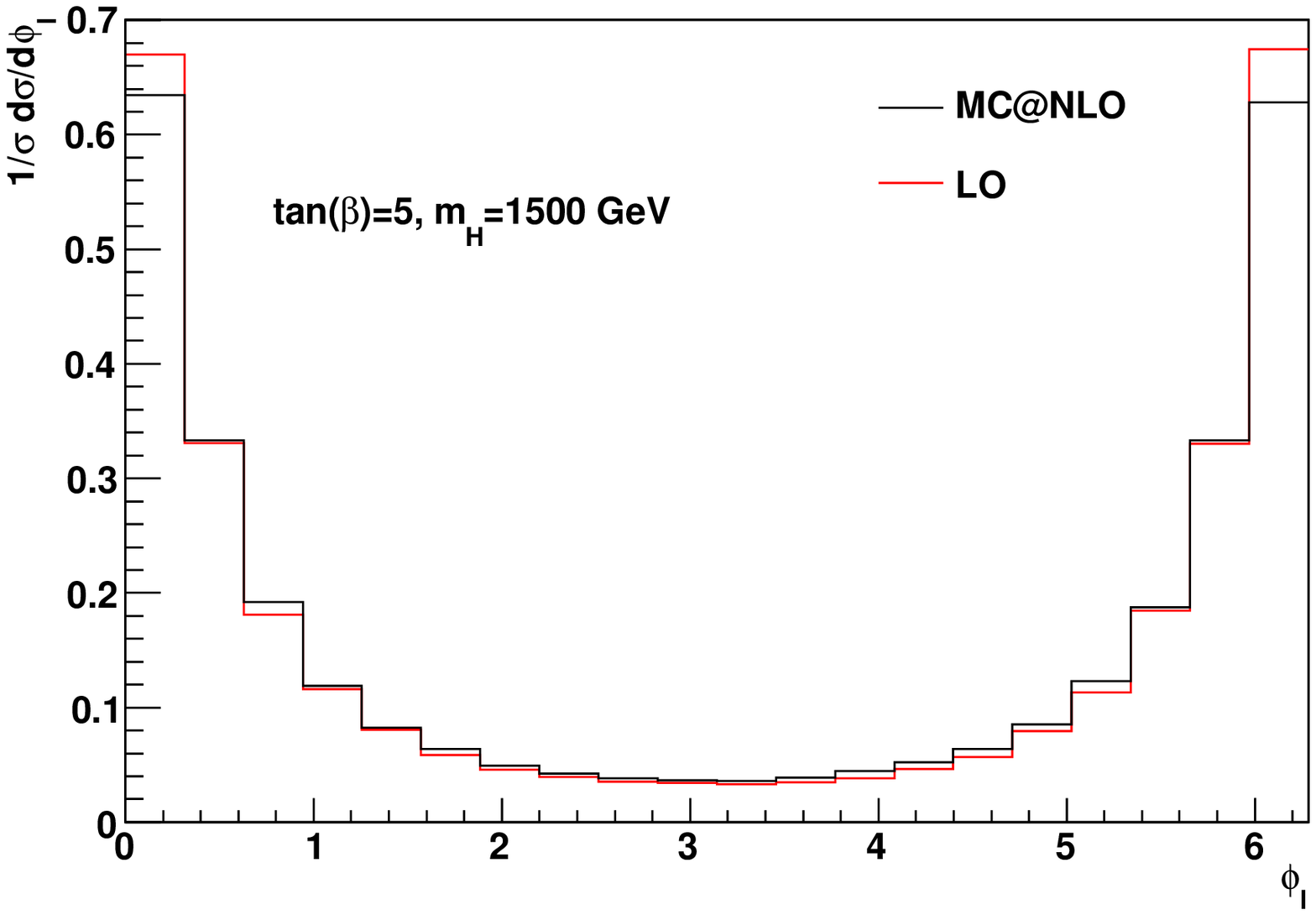}}
\caption{Azimuthal angle ($\phi_l$) of the decay lepton from the top quark, as defined in the text, comparing LO and NLO + parton shower.}
\label{phiLOvsNLO}
\end{center}
\end{figure}

In figure~\ref{phiLOvsNLO} the $\phi_l$ distribution is shown at LO and MC@NLO level for tan$(\beta)=5$ and two different charged Higgs masses. The results 
can be compared to figure 6 of~\cite{Huitu:2010ad}, and indeed the 
qualititative trend of the curves is the same as 
in~\cite{Huitu:2010ad}. In the case of a high Higgs mass the distribution becomes slightly flatter due to the NLO corrections and parton shower. This is caused by competing kinematic effects. As shown in figure~\ref{BplotHt}, the top boost increases slightly due to the higher order corrections, but the $p_t^T$ distribution is typically softer compared to LO, and progressively more so for higher Higgs masses as the top then showers more on average.
The higher top boost leads to a sharper $\phi_l$ distribution, but for high Higgs masses the effect of the softer $p_t^T$ distribution is stronger, resulting in a flatter distribution in the end.
\\

\begin{figure}[!h]
\begin{center}
\scalebox{0.5}{\includegraphics{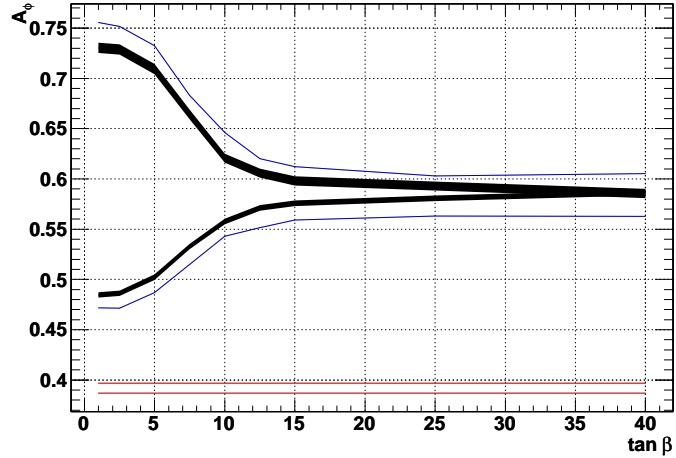}}
\caption{Azimuthal asymmetry parameter for $H^-t$ production, as defined in 
eq.~(\ref{Aldef}). LO (MC@NLO) results are shown in blue (black), for 
$m_H=200$ GeV (lower curves) and $m_H=1500$ GeV (upper curves). The error
band is statistical. Results for $Wt$ production, using both the DR and DS 
approaches in~\cite{Frixione:2008yi}, are shown in red.}
\label{Alplot}
\end{center}
\end{figure}

We can quantify this further by calculating the asymmetry parameter of 
eq.~(\ref{Aldef}). We show this in figure~\ref{Alplot}, for the two Higgs 
mass values used above and a range of $\tan\beta$ values.
Both LO and MC@NLO results are shown for comparison, where for the MC@NLO 
results we include an error band stemming from statistical uncertainty. 
The shape of figure~\ref{Alplot} is very similar to the results 
of~\cite{Huitu:2010ad}:
for the large charged Higgs mass value, a high 
asymmetry is observed for low $\tan\beta$, which decreases at large 
$\tan\beta$. For the low charged Higgs mass value, the opposite trend is
seen.\\

The MC@NLO results show less of a difference between the two Higgs mass
values than the LO results. This is caused by the competing kinematic effects we already saw in figure \ref{phiLOvsNLO}. The higher top boost leads to a larger value of the asymmetry $A_\phi$, but for high Higgs masses the effect of the softer $p_t^T$ distribution is stronger, yielding a net reduction of $A_\phi$. At NLO, the difference between the two Higgs mass values is smaller than at LO, even at low $\tan\beta$. However, a pronounced asymmetry is 
still visible, with a strong dependence on the charged Higgs parameters, so 
the azimuthal asymmetry appears to be quite robust with respect to higher 
order corrections.\\

We see that the difference between the DR and DS results is much less
than the difference between $Wt$ and $H^-t$ production, which gives us 
confidence that the interference issue does not get in the way of getting an 
estimate of the asymmetry parameter for $Wt$. Thus, the fact that $Wt$ and 
$H^-t$ production lead to rather different $A_\phi$ values 
(for essentially any choice of $m_H$ or $\tan\beta$), as has already been 
observed at LO~\cite{Huitu:2010ad}, remains true at NLO and after a parton 
shower has been applied.

\subsection{Polar angle $\theta_l$}

\begin{figure}[!h]
\begin{center}
\scalebox{0.4}{\includegraphics{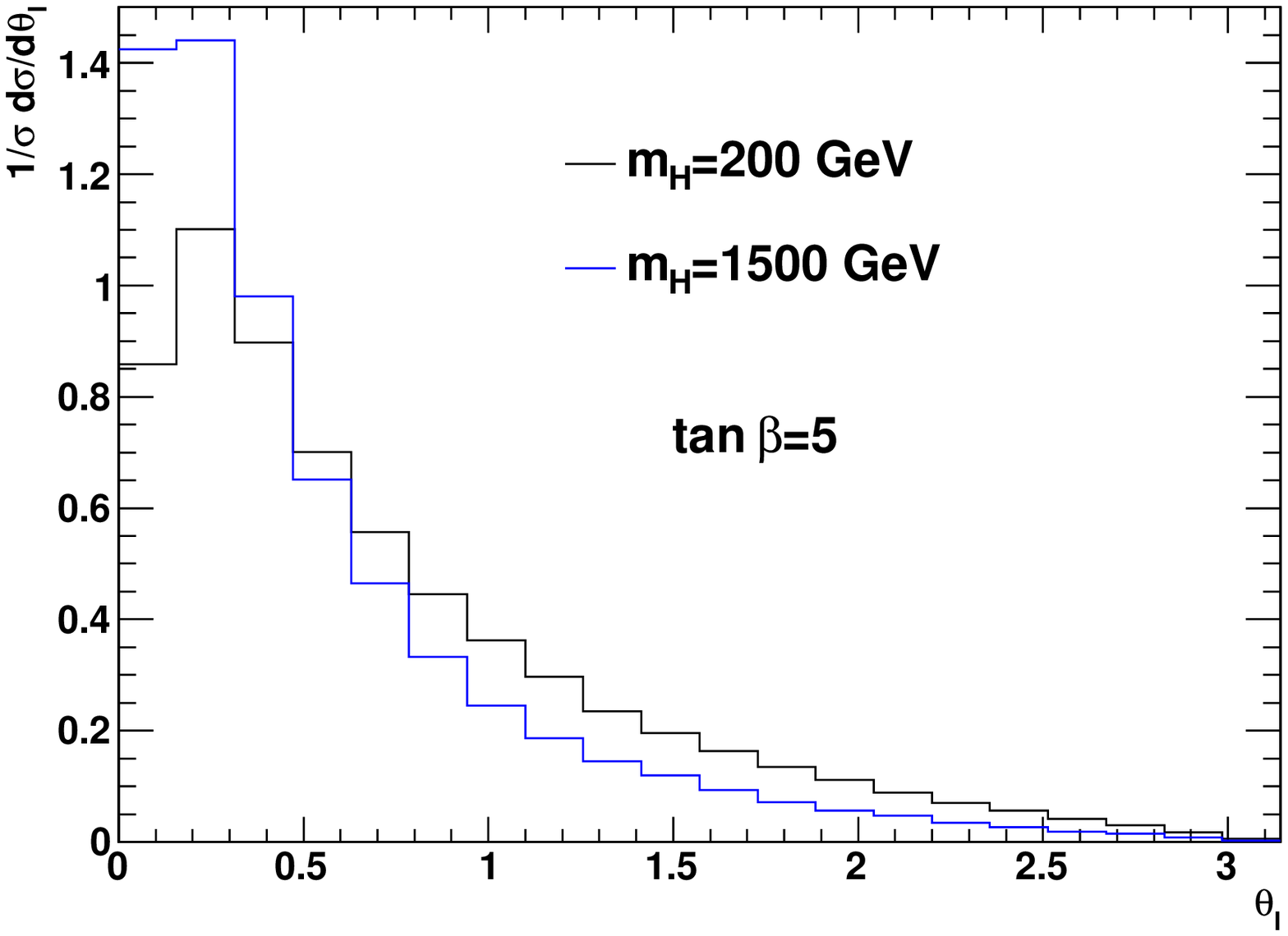}}
\scalebox{0.4}{\includegraphics{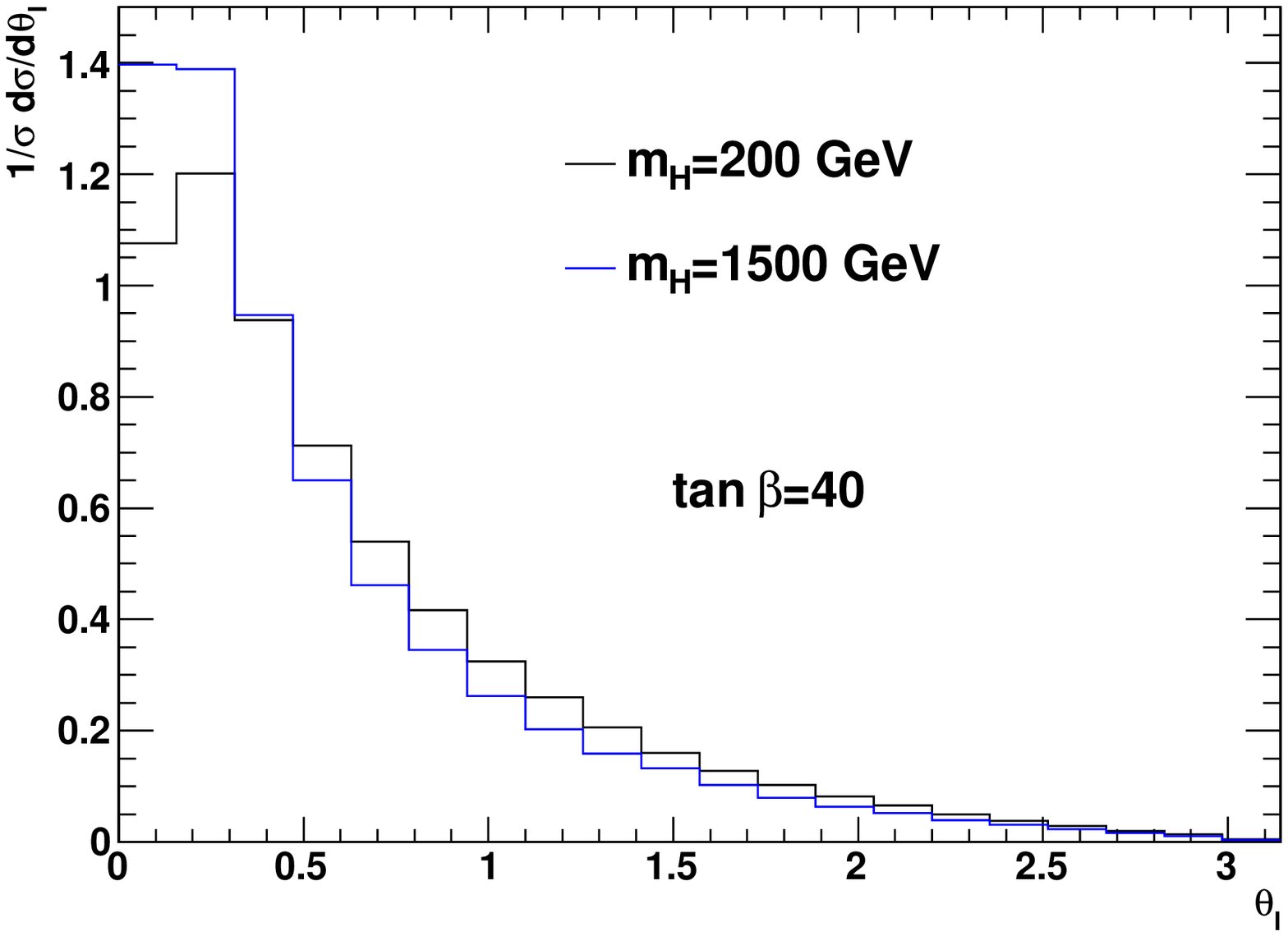}}
\caption{Polar angle ($\theta_l$) of the decay lepton from the top quark, 
measured with respect to the top quark direction, at NLO plus parton shower
level.}
\label{thetal1}
\end{center}
\end{figure}

One may also consider the polar angle between the decay lepton and the top
quark direction. Figure~\ref{thetal1} shows the NLO+parton shower results for the same extremal values of $\tan\beta$ and $m_H$ as
in figure~\ref{phil1}. We see that the distribution is more sensitive to 
the Higgs mass at small $\tan\beta$ than at large $\tan\beta$, which is again due to the enhancement (cancellation) of the polarisation effects by the kinematics at low (high) $\tan\beta$ .
\\

\begin{figure}[!h]
\begin{center}
\scalebox{0.4}{\includegraphics{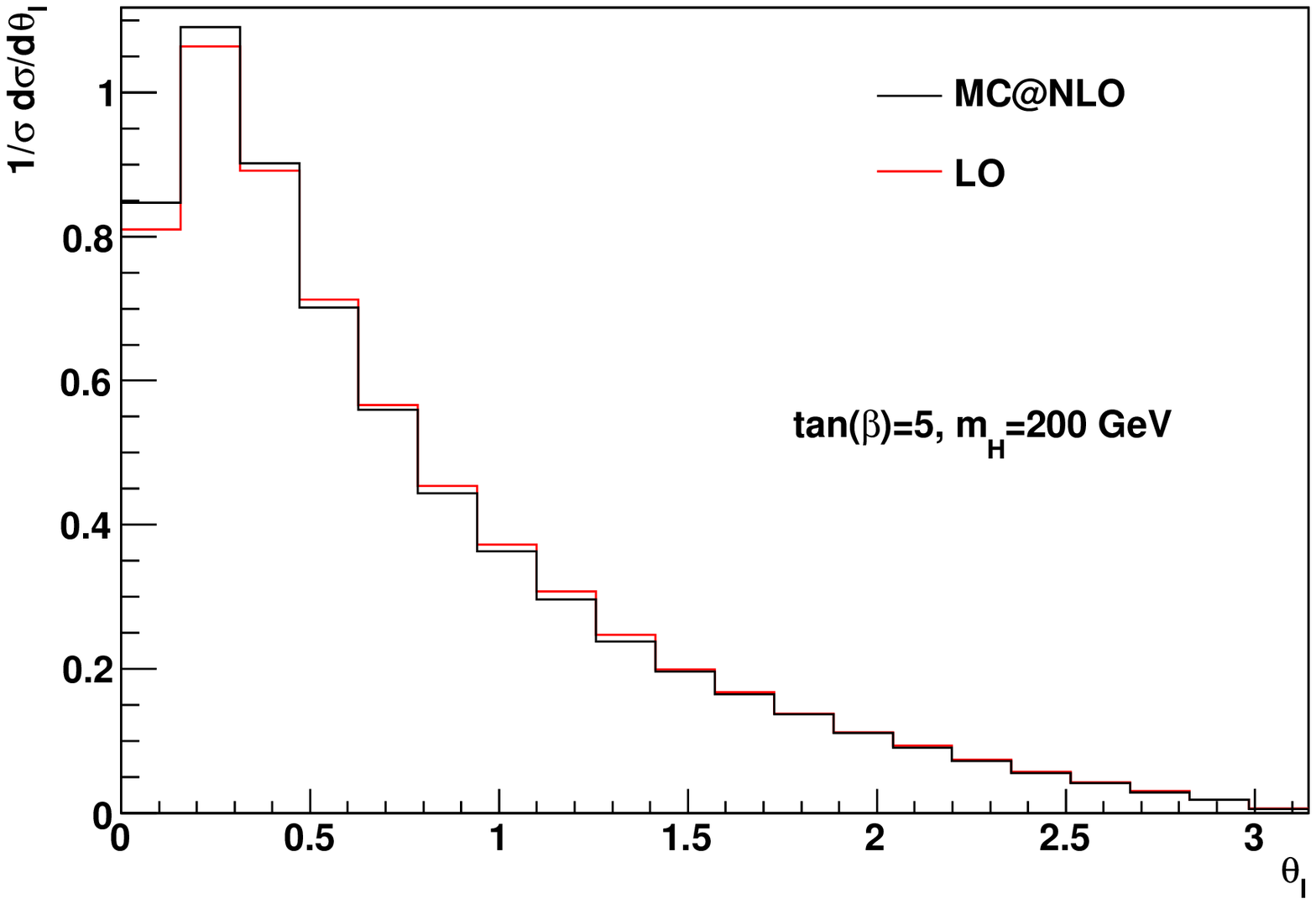}}
\scalebox{0.4}{\includegraphics{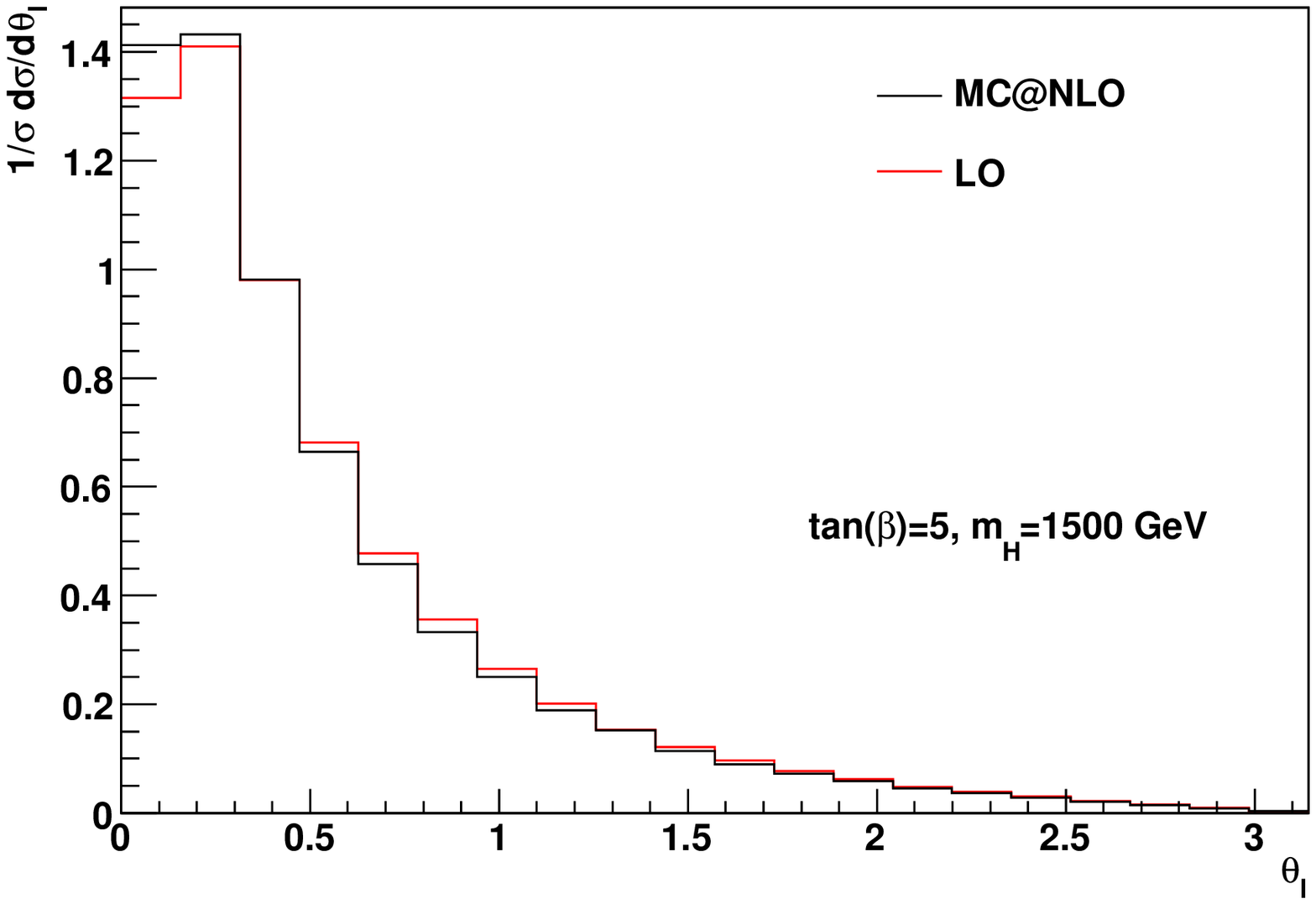}}
\caption{Polar angle ($\theta_l$) of the decay lepton from the top quark, 
measured with respect to the top quark direction, at LO and NLO plus parton shower
level.}
\label{thetaLOvsNLO}
\end{center}
\end{figure}

The distribution of $\theta_l$ at LO and MC@NLO level is shown in figure~\ref{thetaLOvsNLO}. As with the $\phi_l$ distribution, the NLO distribution strongly resembles the LO results. The NLO distribution is peaked towards $\theta_l=0$ somewhat more due to the slight increase in the top boost parameter.
\\

In all cases,
the distribution shows a strong peak at low values of $\theta_l$, with a 
fall-off at higher values. Given that the distribution must be normalised,
a distribution which has a slower fall-off must correspondingly have a lesser
peak, and vice versa. This motivates the definition of the following asymmetry
parameter:

\begin{equation}
A_{\theta}=\frac{\sigma(\theta_l<\pi/4)-\sigma(\theta_l>\pi/4)}
{\sigma(\theta_l>\pi/4)+\sigma(\theta_l<\pi/4)}.
\label{Athetadef}
\end{equation}
We have here used $\pi/4$ as representative of the point at which distributions
corresponding to different points in parameter space cross each other. 
However, we have found no obvious analytic justification for this result, 
so this number can in principle be varied in order to enhance the 
asymmetry. \\

\begin{figure}[!h]
\begin{center}
\scalebox{0.5}{\includegraphics{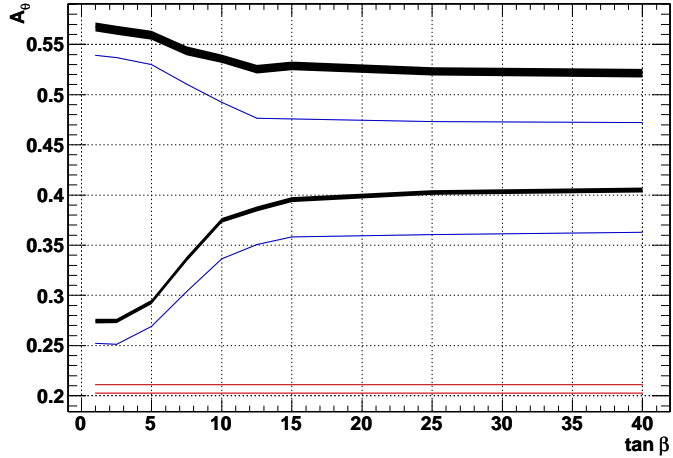}}
\caption{Polar asymmetry parameter for $H^-t$ production, as defined in 
eq.~(\ref{Athetadef}). LO (MC@NLO) results are shown in blue (black), for 
$m_H=200$ GeV (lower curves) and $m_H=1500$ GeV (upper curves). The error
band is statistical. Results for $Wt$ production, using both the DR and DS 
approaches in~\cite{Frixione:2008yi}, are shown in red.}
\label{Athetaplot}
\end{center}
\end{figure}

Results for the polar asymmetry parameter are shown in figure~\ref{Athetaplot}.
Again we show both LO and MC@NLO results, where a statistical uncertainty band
is included for the latter. 
One sees that the MC@NLO values of $A_\theta$ are higher than the LO results, as expected from the higher value of the top boost at MC@NLO level compared to 
LO. In contrast to the azimuthal 
asymmetry, there is a significant difference between the extremal charged 
Higgs mass values at large $\tan\beta$. This makes the polar angle extremely 
useful as a complementary observable to the azimuthal angle, as the latter is 
relatively insensitive to the charged Higgs mass at large $\tan\beta$. \\

Similarly to the azimuthal case, one sees from figure~\ref{Athetaplot} that 
typical values for the polar asymmetry are markedly different to the result
obtained for $Wt$ production, as estimated by the DR and DS results. Again this
is presumably a reliable conclusion, given that the difference between the two
$Wt$ results is much less than the difference between the $H^-t$ and $Wt$ 
results. This information is a potentially valuable tool in being able to 
distinguish charged Higgs boson production from the $Wt$ background.

\subsection{Energy ratio observables}
In the previous sections, we presented results for angular distributions of the
decay lepton in $H^-t$ and $Wt$ production, finding these to be robust 
discriminators of the charged Higgs parameter space, as well as of use in 
distinguishing a charged Higgs signal from the Standard Model background. 
In this section, we consider the energy ratios
of eq.~(\ref{zudef}), which were first defined in~\cite{Shelton:2008nq}.\\

Note that both the $z$ and $u$ observables depend on the energy of the $b$ 
quark emanating from the top quark decay. In a leading order calculation,
this can be straightforwardly identified. In an experimental environment,
one must use event selection cuts which require the presence of a tagged  
$b$ jet, and use the energy of this jet in constructing eq.~(\ref{zudef}).
A full phenomenological analysis is beyond the scope 
of this paper: we here wish to present a first analysis of the $z$ and $u$
parameters in the context of $H^-t$ production, unshrouded by the full 
complications of an experimental analysis. There is then a choice to be made
regarding which energy to use in presenting results from MC@NLO. One option 
is to use the energy of the $b$-flavoured hadron that contains the $b$ quark
from the top decay, requiring this to be stable. However, to facilitate a more 
direct comparison with the LO results, we instead define $E_b$ via the energy 
conservation relation
\begin{equation}
E_b=E_t-E_l-E_\nu,
\label{Ebdef}
\end{equation}
where $E_t$, $E_l$ and $E_\nu$ are the energies of the top quark, decay lepton
and decay neutrino respectively. The latter is, of course, unmeasurable in a 
real experiment but can be identified in a Monte Carlo event generator. Our
definition of $E_b$ then means that our comparisons between LO and MC@NLO 
results measure the collective effect of a single hard additional emission 
(from the NLO matrix element), together with the parton shower, but with no 
non-perturbative contributions from e.g. hadronization or the underlying event.
We deem such an approach to be valid in assessing the robustness of energy 
ratio observables against perturbative higher order corrections, which is our
present aim. \\

\begin{figure}[!h]
\begin{center}
\scalebox{0.4}{\includegraphics{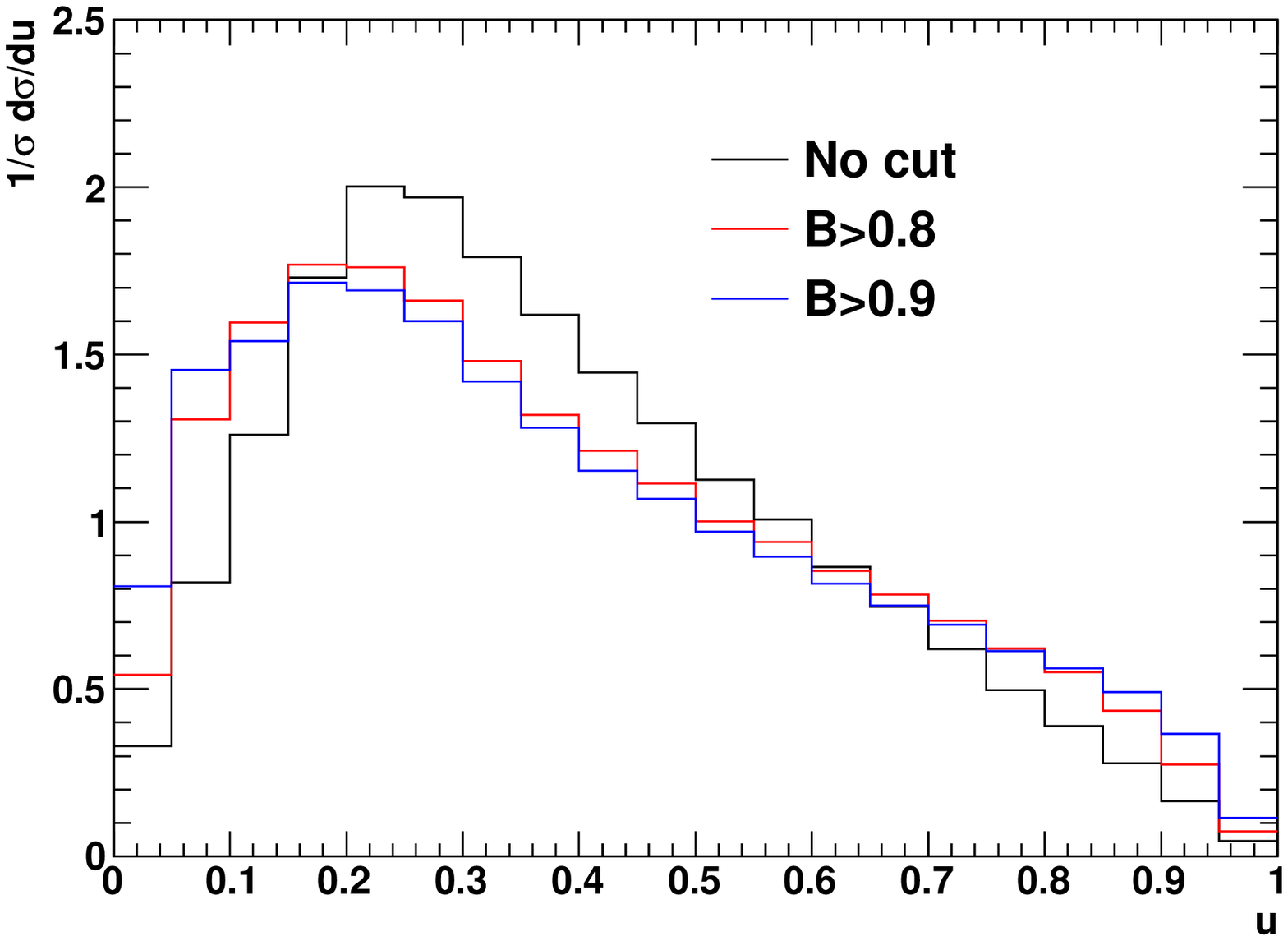}}
\scalebox{0.4}{\includegraphics{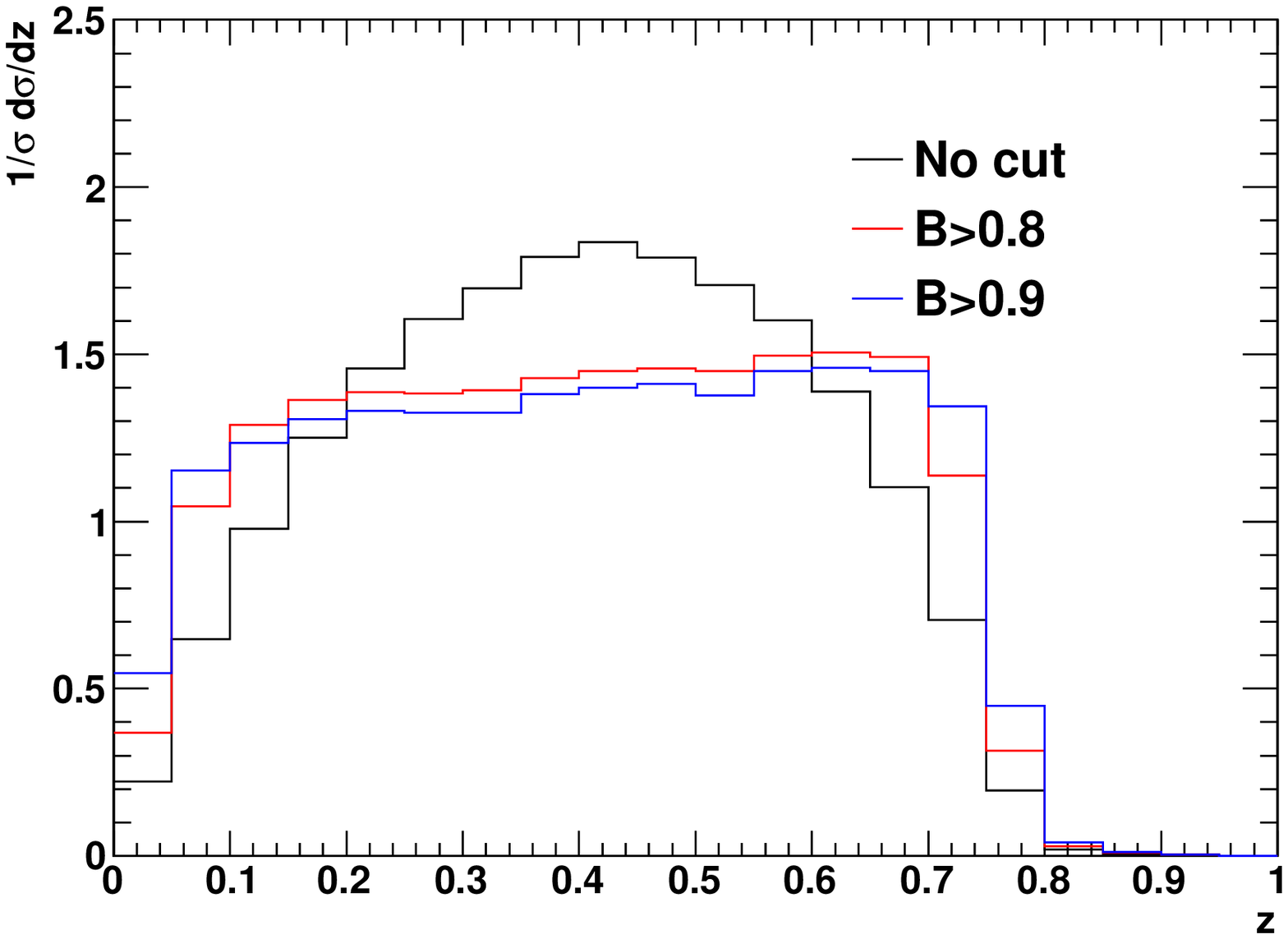}}
\caption{Distribution of $u$ (left-hand plot) and $z$ (right-hand plot) 
for $\tan\beta=1$ and $m_H=200$ GeV, at NLO plus parton shower level. 
Results are shown for different cut 
values on the boost parameter $B$ of eq.~(\ref{Bdef}).}
\label{EratboostplotsHt}
\end{center}
\end{figure}

The energy ratios of eq.~(\ref{zudef}) are more sensitive to the top quark
polarisation in the kinematic region in which the decaying top quark is highly
boosted. It is important to check which values of a cut on the boost parameter are
sufficient in order to isolate the desired sensitivity to the top quark
polarisation. To this end, we plot the energy ratios $z$ and $u$ of
eq.~(\ref{zudef}) for different values of this cut in 
figure~\ref{EratboostplotsHt}. 
One sees that the results with a cut are markedly different to those with no 
cut (as expected). However, the difference between results with $B>0.9$ and 
$B>0.8$ is much less, suggesting that a cut of $B>0.8$ is sufficient. \\

\begin{figure}[!h]
\begin{center}
\scalebox{0.4}{\includegraphics{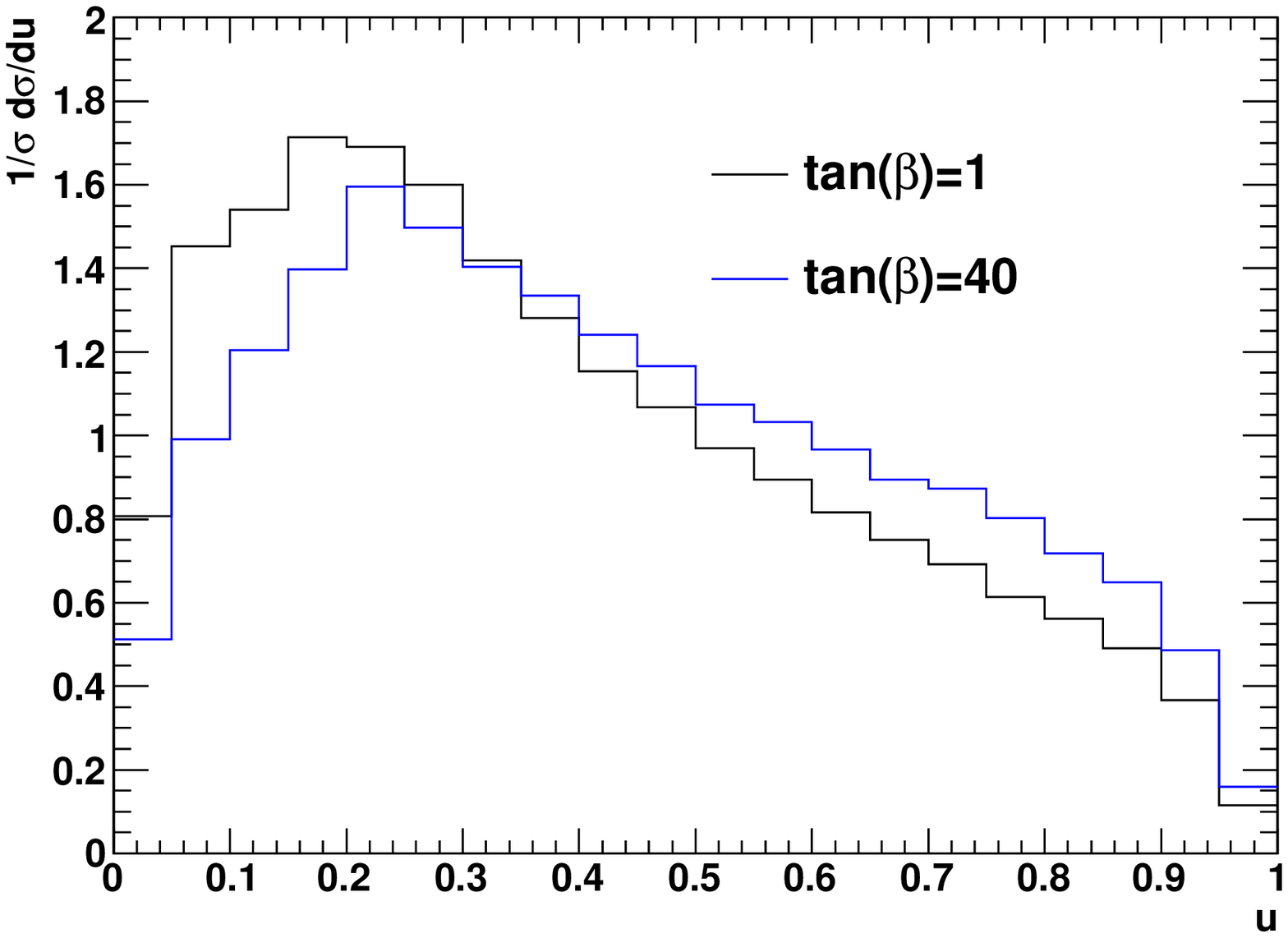}}
\scalebox{0.4}{\includegraphics{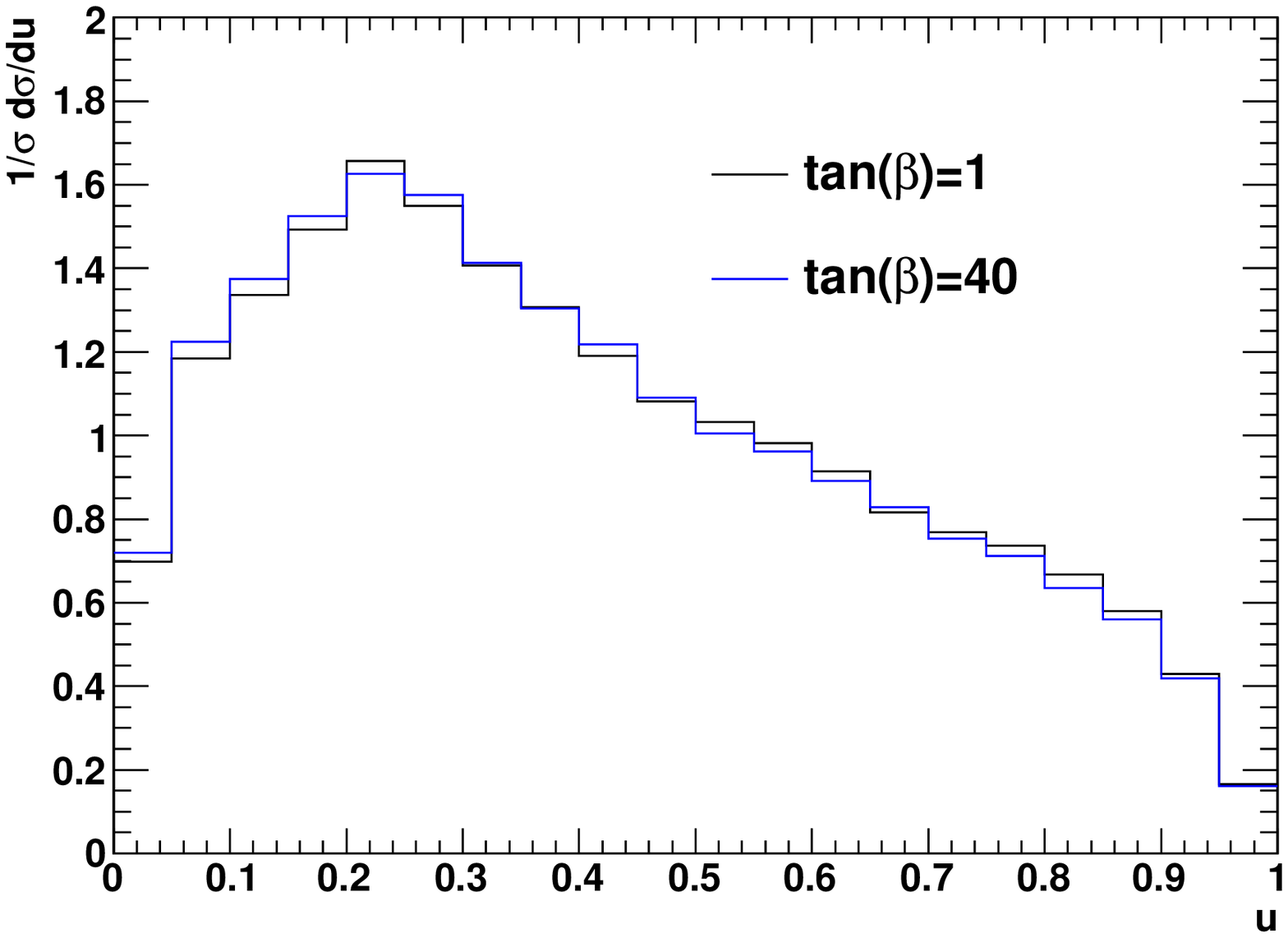}}
\caption{Distribution of $u$, as defined in eq.~(\ref{zudef}), where a cut on 
the boost parameter $B>0.8$ has been applied, at NLO plus parton shower level.
Results are shown for $m_H=200$ 
GeV (left-hand plot) and $m_H=1500$ GeV (right-hand plot).}
\label{EratplotsHt2}
\end{center}
\end{figure}

The distribution of $u$ at MC@NLO level after the cut $B>0.8$ is applied is shown in figure~\ref{EratplotsHt2} for two values of $m_H$. 
The shape of the plots can be compared to the corresponding figures 
in~\cite{Shelton:2008nq}, which are presented for the ideal case in which the 
top quark is completely polarized and infinitely boosted, i.e. $P_t=\pm1$ and $B\to1$. The latter seem to show a much more pronounced difference between the curves for positive and 
negative helicity top quarks. This is mostly due to the fact that in our case the top quarks are not completely polarized. The high Higgs mass in particular does not yield a strong top quark polarization. For the lower Higgs mass, the shapes are broadly consistent with the results of~\cite{Shelton:2008nq}: for the negatively polarised top quarks
($\tan\beta=1$), the distribution falls off more sharply for higher values of 
$u$. Also, the curvature of the distributions is different for lower values of
$u$ for the two different $\tan\beta$ values. \\

\begin{figure}[!h]
\begin{center}
\scalebox{0.4}{\includegraphics{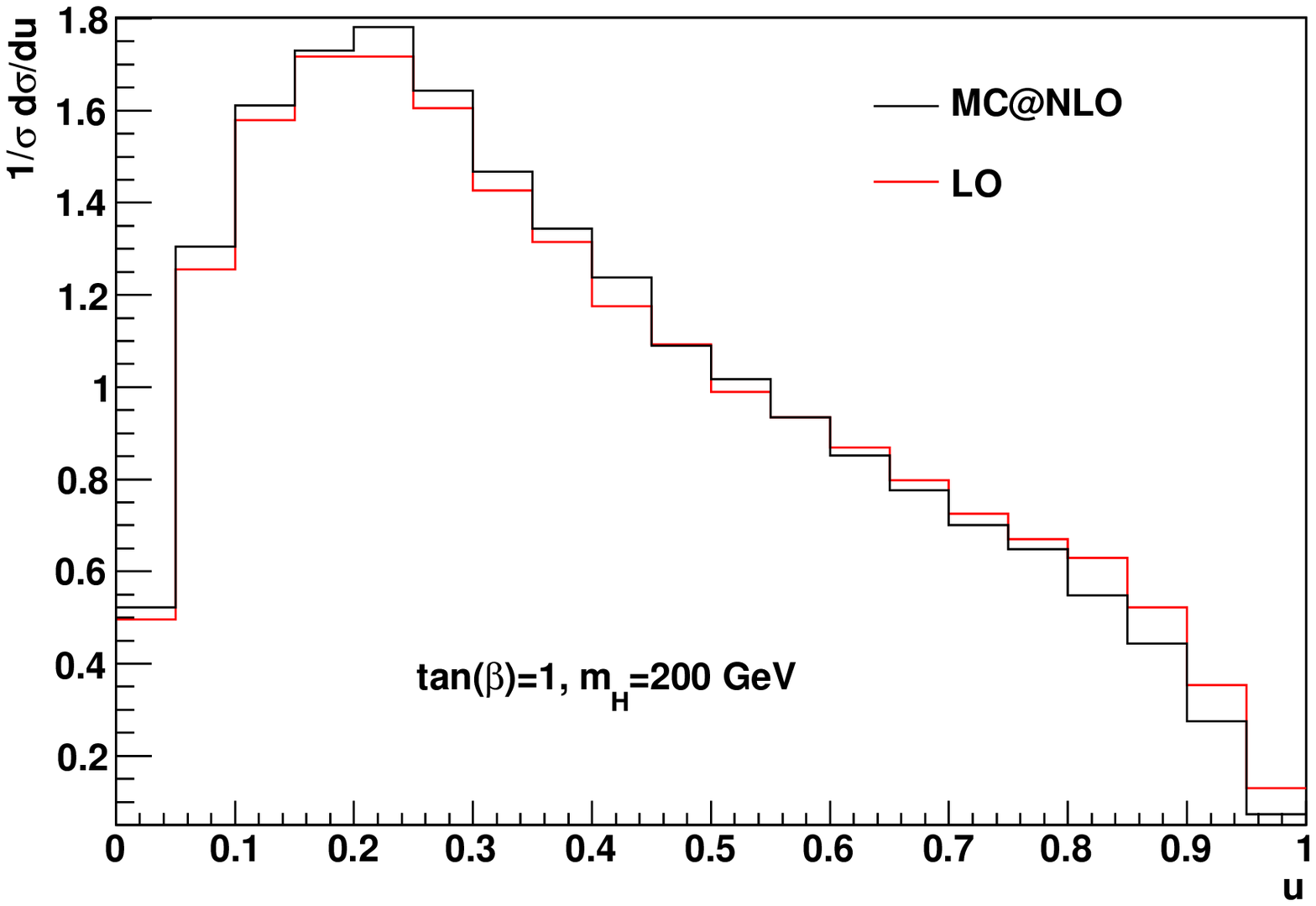}}
\scalebox{0.4}{\includegraphics{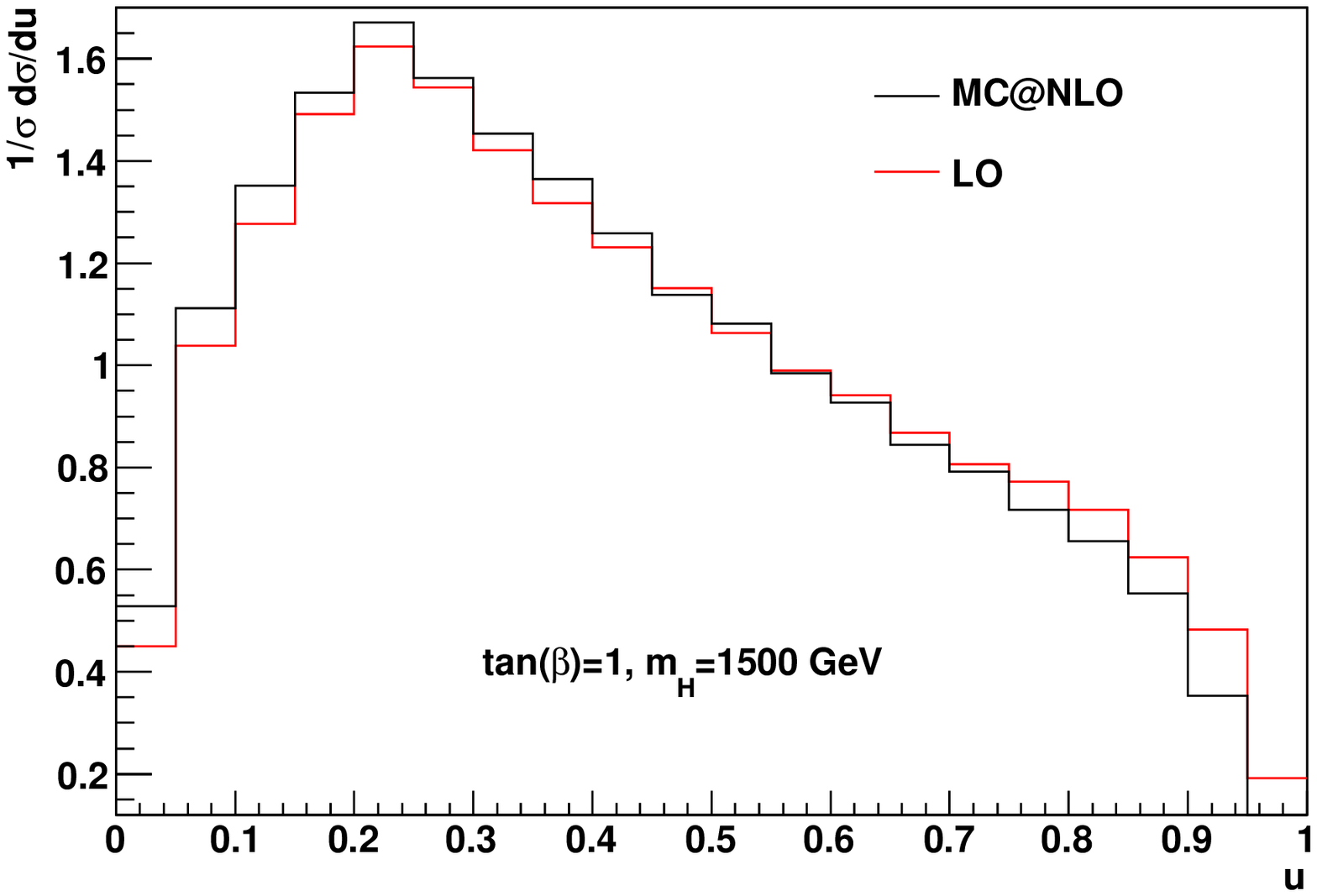}}
\caption{Distribution of $u$ with a boostcut of $B>0.8$.}
\label{uLOvsNLO}
\end{center}
\end{figure}

The $u$ variable at LO and MC@NLO level with a boostcut of $B>0.8$ is shown in figure~\ref{uLOvsNLO}. We see that the general shape does not change when including NLO+parton shower corrections. However, the difference between the LO and MC@NLO distributions is more pronounced than for the angular variables, indicating that this distribution might be slightly less robust w.r.t. higher order corrections.
\\

\begin{figure}[!h]
\begin{center}
\scalebox{0.4}{\includegraphics{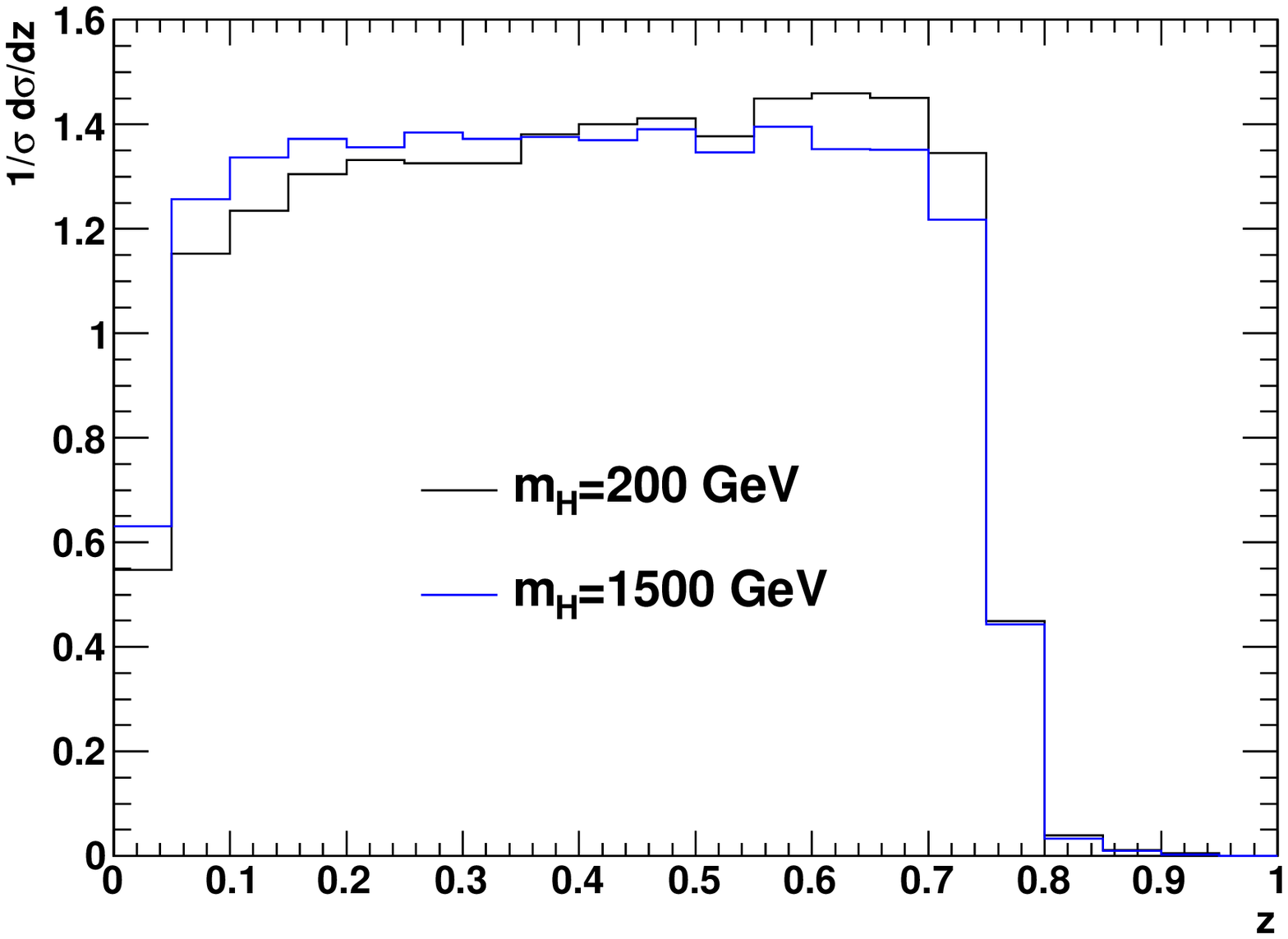}}
\scalebox{0.4}{\includegraphics{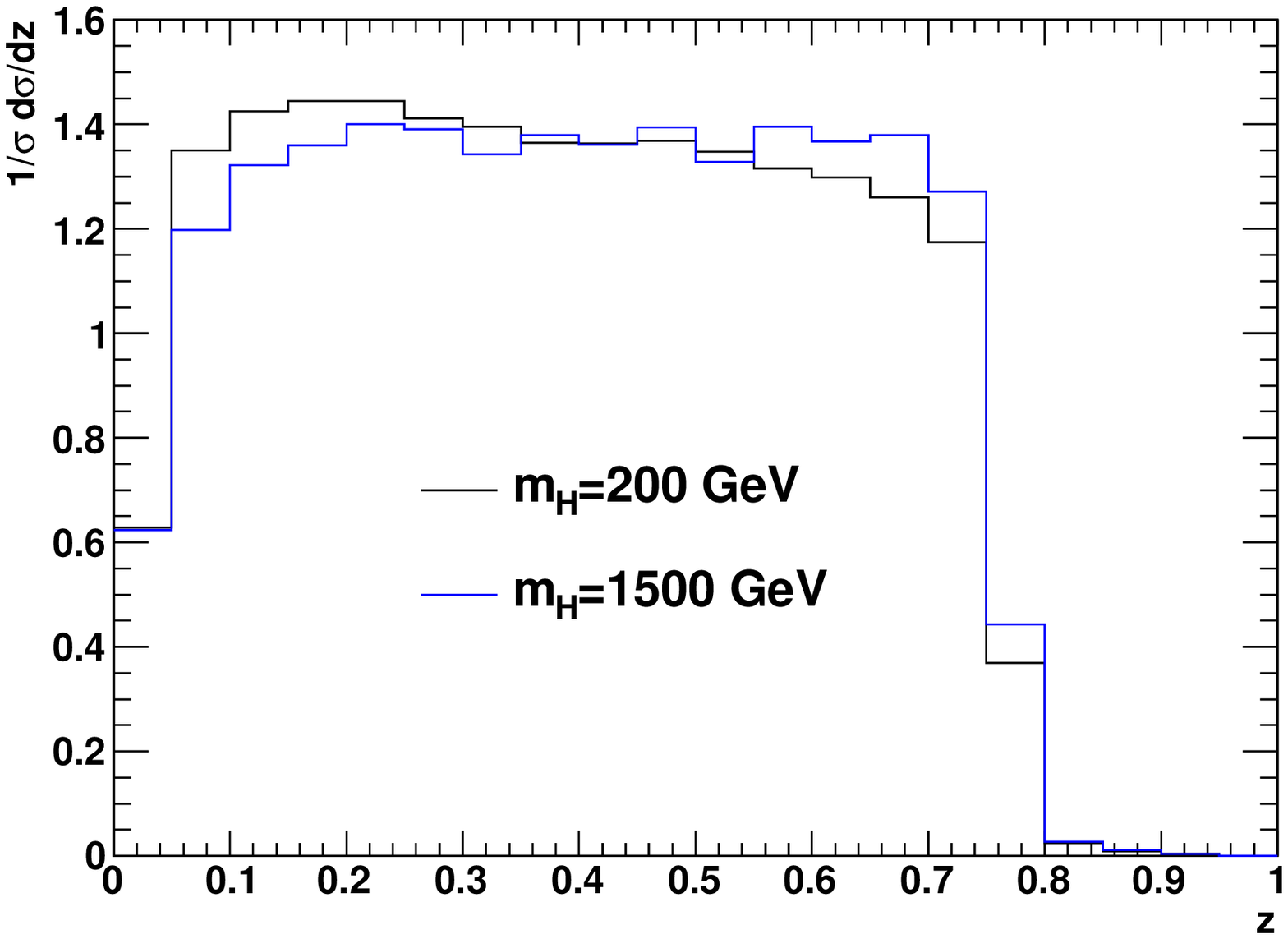}}
\caption{Distribution of $z$, as defined in eq.~(\ref{zudef}), where a cut on 
the boost parameter $B>0.8$ has been applied, at NLO plus parton shower 
level. Results are shown for 
$\tan\beta=1$ (left-hand plot) and $\tan\beta=40$ (right-hand plot).}
\label{E2ratplotsHt}
\end{center}
\end{figure}

We may also consider the $z$ distribution, which is shown for our two 
extremal $\tan\beta$ values in figure~\ref{E2ratplotsHt}. 
The plots have three distinct regimes. Firstly, there is a sharp fall-off as 
$z\rightarrow0$, due to the finite mass of the $b$ quark. Then, there is an 
intermediate regime $0.1\lesssim z\lesssim0.7$, over which the $z$ 
distribution is approximately linear, with the sign of the slope correlated 
with the polarisation of the top quark (i.e. positive and 
negative for negatively and positively polarised top quarks respectively). 
Finally, there is another fall-off as $z\rightarrow 1$, due to the finite $W$ 
boson mass. Again one sees very little correlation for the charged Higgs mass 
of 1500 GeV due to the small value of the polarisation.\\

\begin{figure}[!h]
\begin{center}
\scalebox{0.4}{\includegraphics{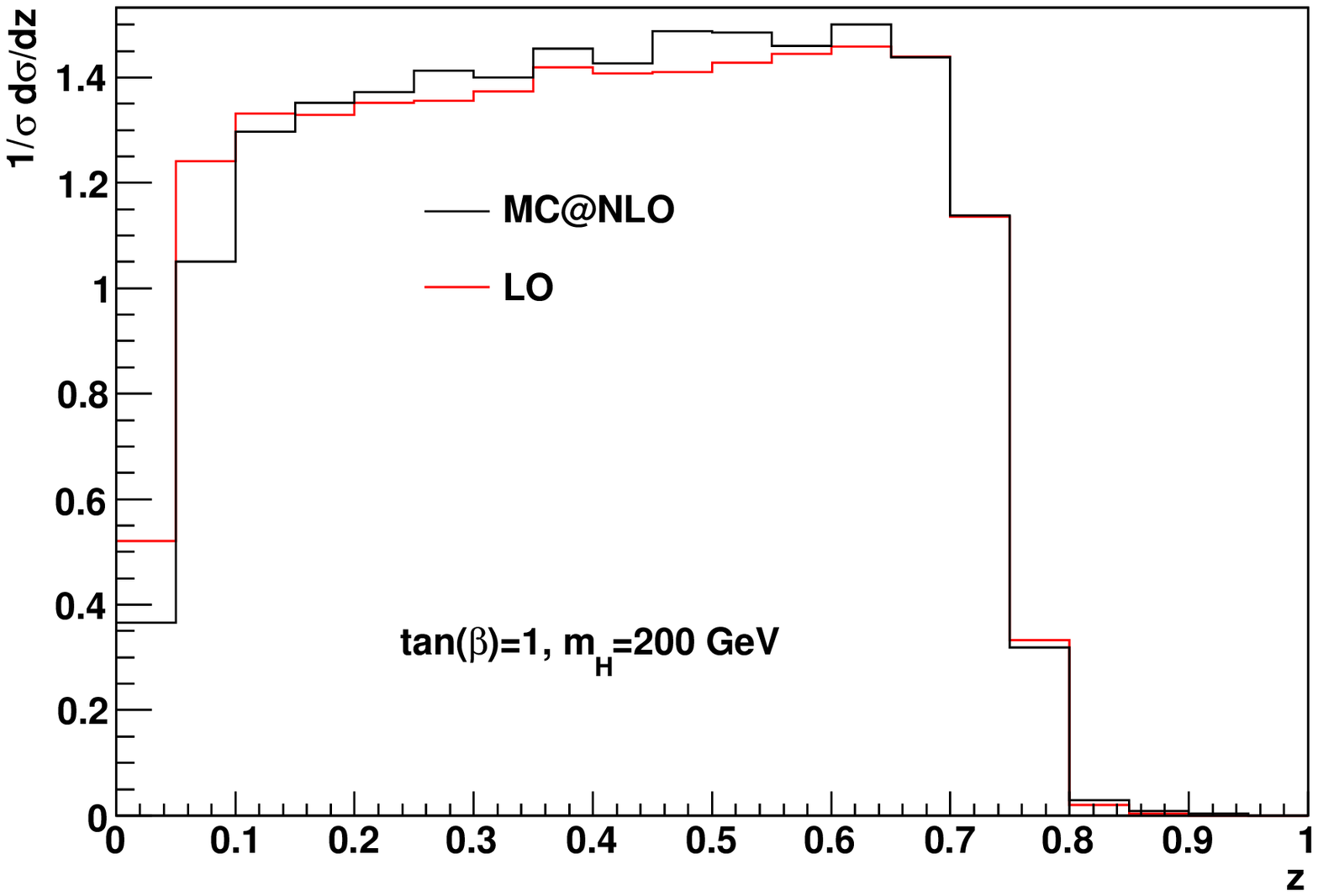}}
\scalebox{0.4}{\includegraphics{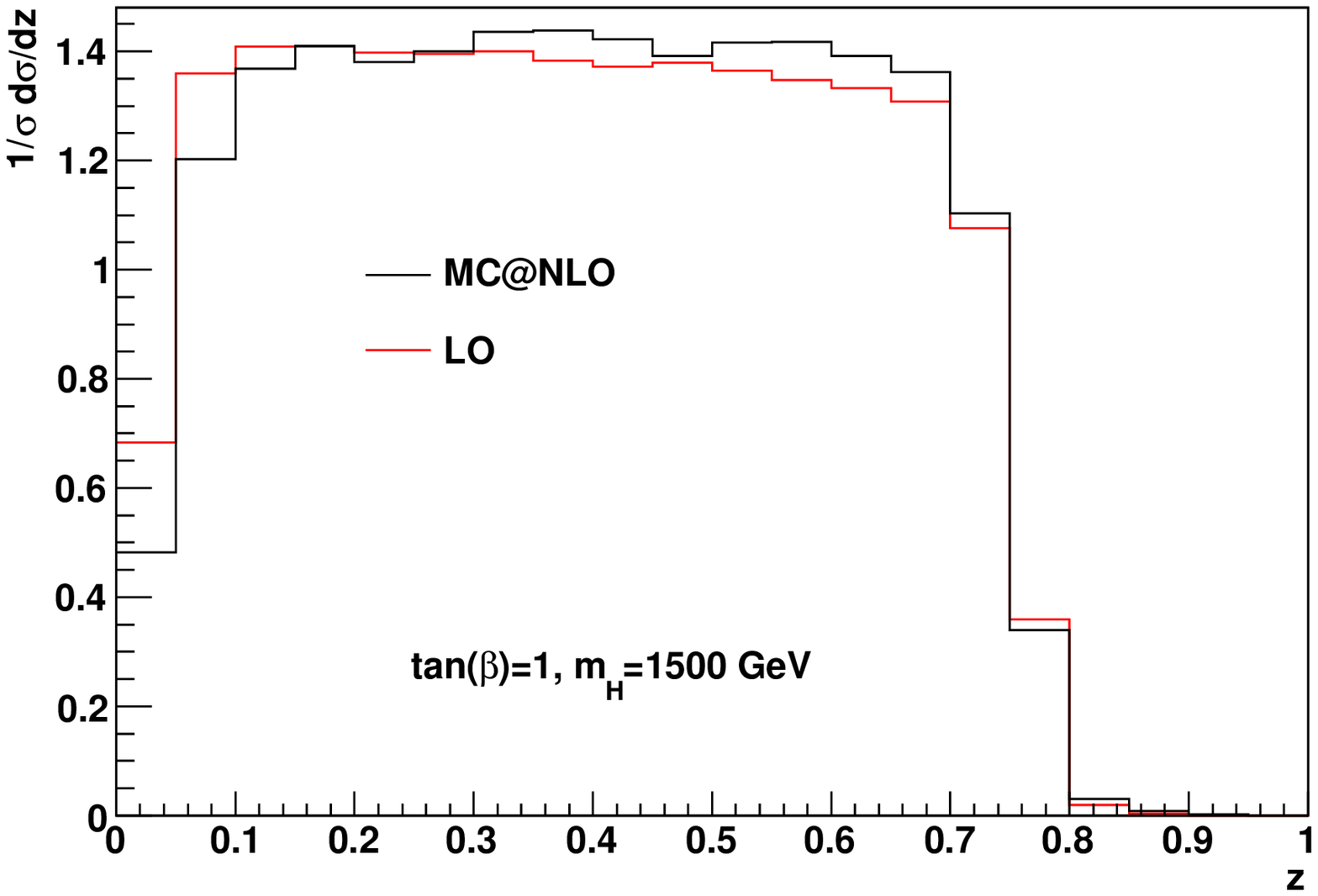}}
\caption{Distribution of $z$ at LO and MC@NLO level, with a boostcut of $B>0.8$.}
\label{zLOvsNLO}
\end{center}
\end{figure}

In figure~\ref{zLOvsNLO} we see that this is not due to the NLO and parton shower effects. The distribution is changed by these effects, but the correlation is not very strong even at LO. For the lower Higgs mass we also see that the NLO+parton shower corrections change the distribution more than for the angular distributions.
\\

For the angular observables of the previous section, we defined asymmetry
parameters which efficiently distil the difference between different regions
of the charged Higgs parameter space into single numbers. It is perhaps useful
to also adopt this strategy for the energy ratios $u$ and $z$. Regarding the 
former, one may first note that the normalisation of the distribution means
that a slower fall-off above the peak region entails less events below the
peak region. One may exacerbate this effect by defining the corresponding
asymmetry parameter
\begin{equation}
A_u=\frac{\sigma(u>0.215)-\sigma(u<0.215)}{\sigma(u>0.215)+\sigma(u<0.215)}.
\label{Audef}
\end{equation}
Here $u\simeq 0.215$ is chosen as the approximate position of the peak, 
motivated by the analysis of~\cite{Shelton:2008nq}. As in the case of the
polar angle asymmetry of eq.~(\ref{Athetadef}), however, this choice can in
principle be varied in order to enhance the result. \\

\begin{figure}[!h]
\begin{center}
\scalebox{0.4}{\includegraphics{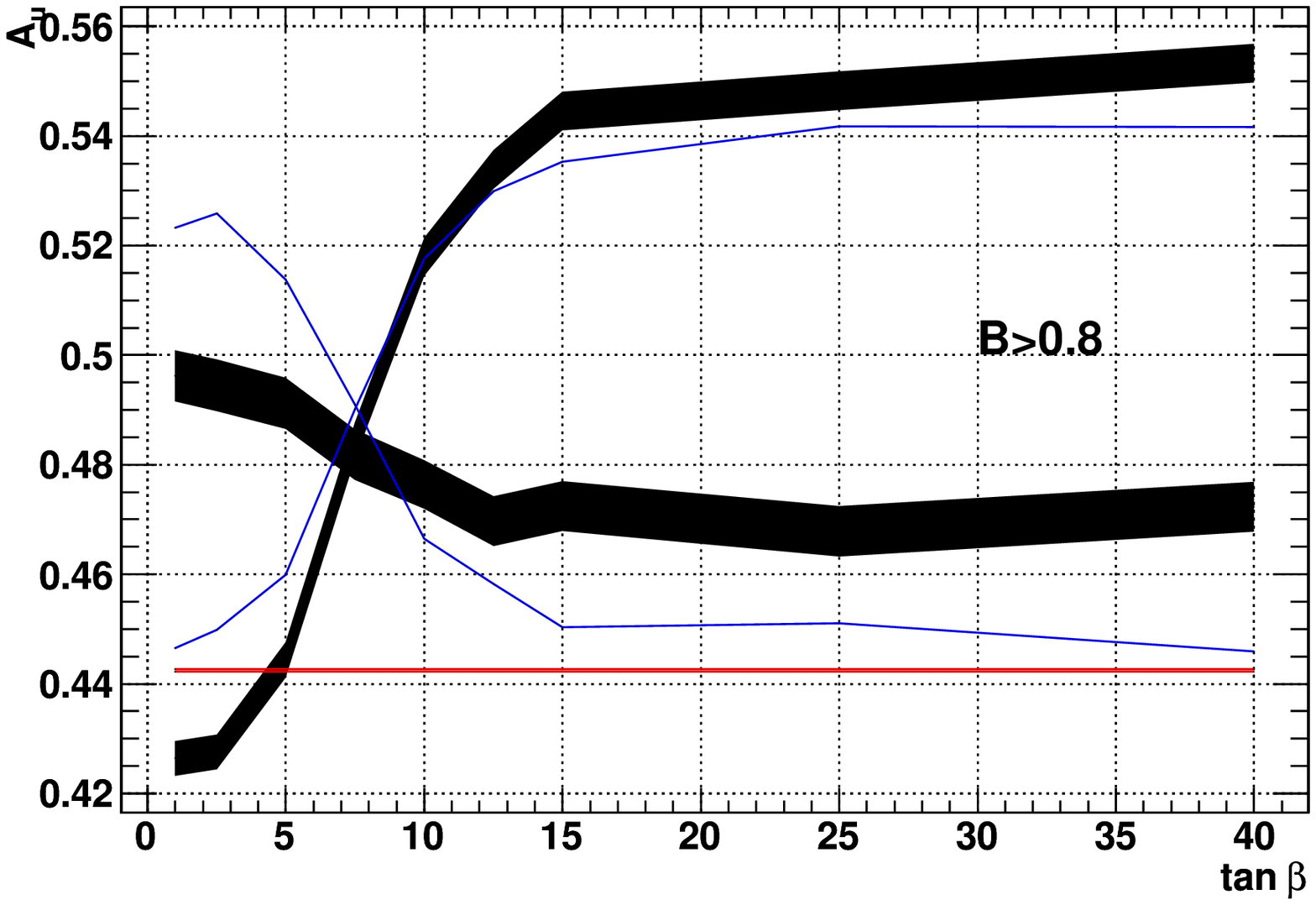}}
\scalebox{0.4}{\includegraphics{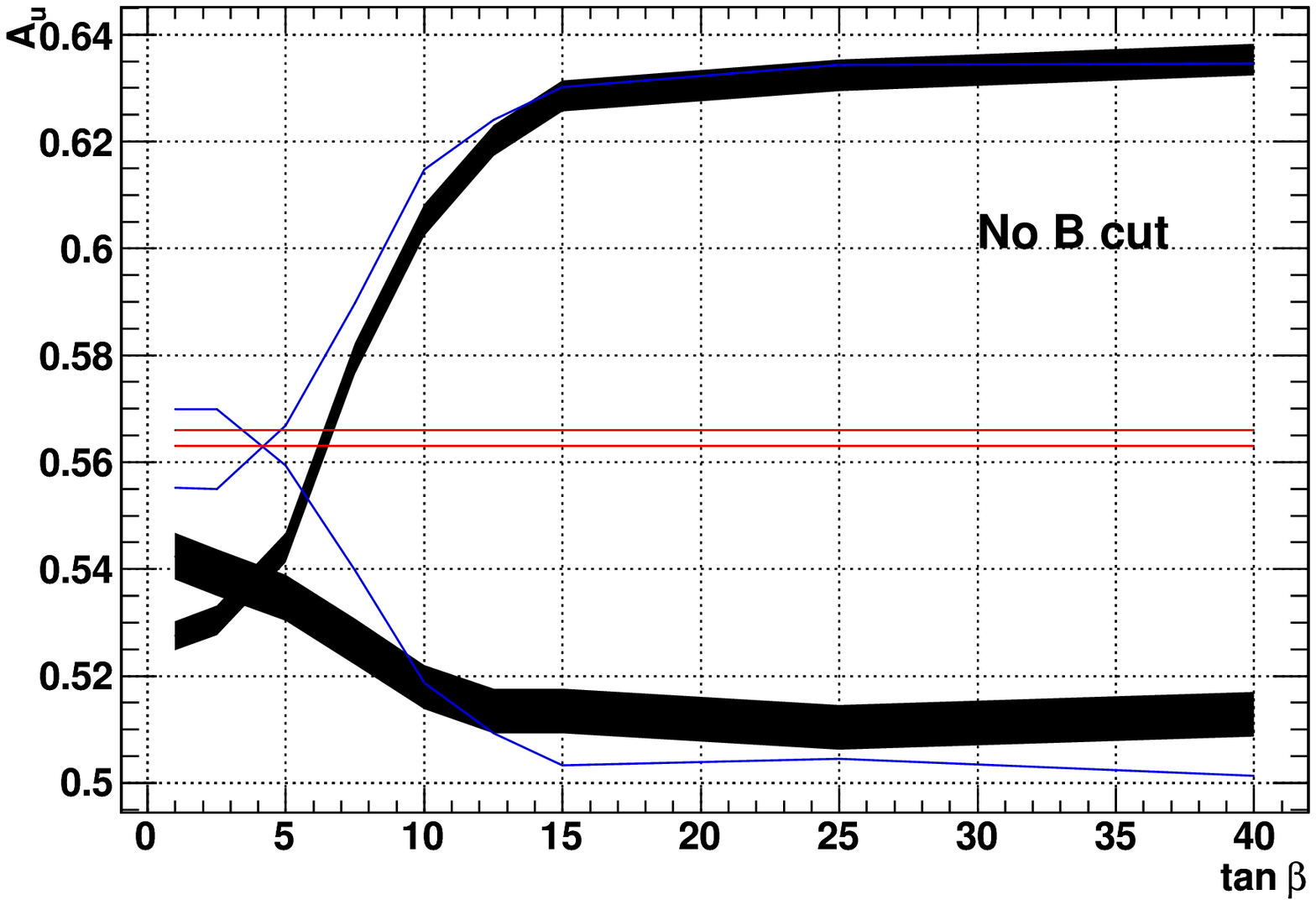}}
\caption{The asymmetry parameter $A_u$ for $H^-t$ production, as defined in 
eq.~(\ref{Audef}). LO (MC@NLO) results are shown in blue (black), for 
$m_H=200$ GeV (upper curves at large $\tan\beta$) and $m_H=1500$ GeV 
(lower curves at large $\tan\beta$). The error
band is statistical. Results for $Wt$ production, using both the DR and DS 
approaches in~\cite{Frixione:2008yi}, are shown in red (in the left-hand plot the DS and DR results are on top of each other).}
\label{Auplot}
\end{center}
\end{figure}

The behaviour of $A_u$ is shown in figure~\ref{Auplot}, for a cut on the boost
parameter of $B>0.8$. For comparison purposes, we also show the result one 
would obtain with no cut on the boost parameter, where the $u$ observable 
suffers significant contamination from contributions which are insensitive
to the top quark polarisation. 
As expected, the $A_u$ variable has more discriminating power for the lower Higgs mass, since the top is more strongly polarised in that case. In addition one sees that the cut on the boost parameter has a larger effect for the lower Higgs mass than for the higher one, although this effect is somewhat weaker at MC@NLO level, where the top is more boosted on average. Generally, there is more of a 
pronounced difference between the LO and MC@NLO values than in the case of the
angular asymmetries considered in the previous section. Furthermore, decorrelation is more pronounced for heavier Higgs masses, due presumably to the fact that the top quark showers more on average. \\

As for the angular asymmetry, we also show results for $Wt$ production
in figure~\ref{Auplot}. Before a cut on the boost parameter is applied,
the $Wt$ result sits more or less in the middle of the $H^-t$ results over
most of the range in $\tan\beta$. This is not the case once a cut is applied,
and indeed a significant difference is observed between the $Wt$ and $H^-t$
results. Admittedly, this difference appears larger (and thus more useful)
for smaller charged Higgs masses, and is only 3\% or so for the largest
Higgs mass we consider.\\

We may also define an asymmetry parameter for the energy ratio $z$ of 
eq.~(\ref{zudef}). This is perhaps most conveniently done by considering only
the linear regime in figure~\ref{E2ratplotsHt}, occuring at intermediate
values of $z$, as it is the sign of the slope in this kinematic region that
distinguishes the cases of positive and negatively polarised tops. We therefore
define
\begin{equation}
A_z=\frac{\sigma(0.1\leq z\leq0.4)-\sigma(0.4<z\leq 0.7)}
{\sigma(0.1\leq z\leq0.4)+\sigma(0.4<z\leq 0.7)}.
\label{Azdef}
\end{equation}
We have chosen the values at which to define the intermediate region by eye
from figure~\ref{E2ratplotsHt}. Again, these could be varied in order to 
maximise the resulting asymmetry. \\

\begin{figure}[!h]
\begin{center}
\scalebox{0.4}{\includegraphics{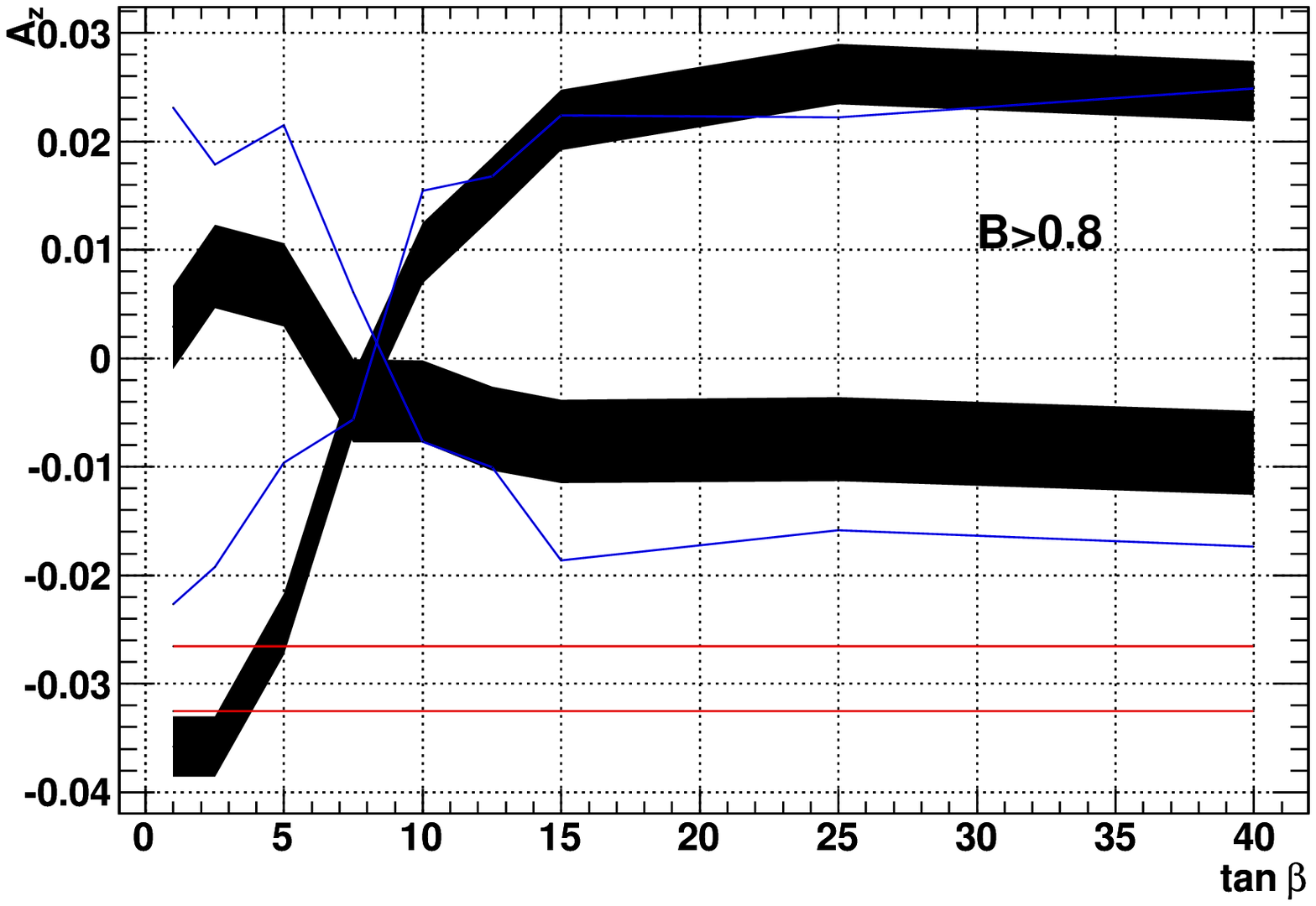}}
\scalebox{0.4}{\includegraphics{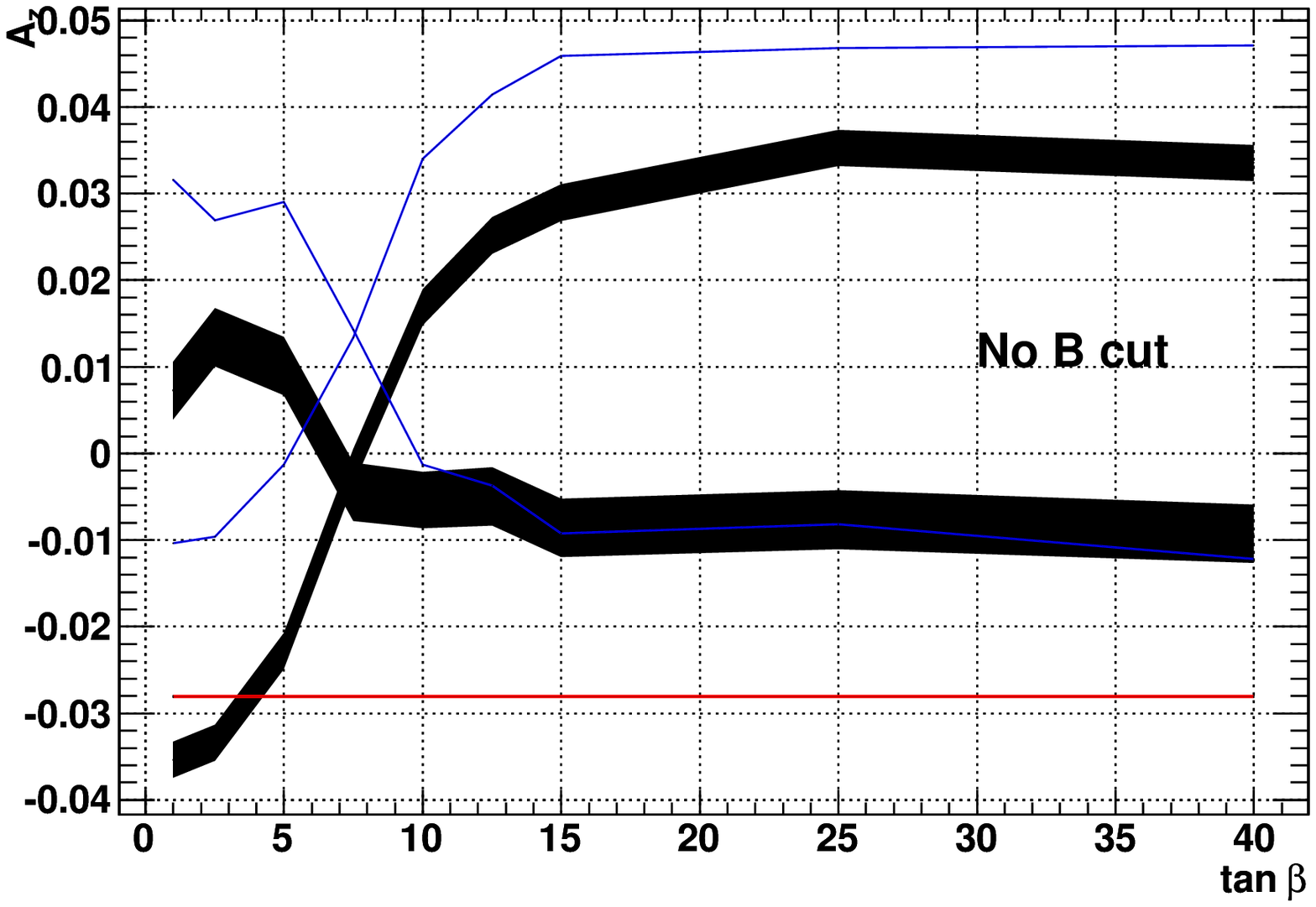}}
\caption{The asymmetry parameter $A_z$ for $H^-t$ production, as defined in 
eq.~(\ref{Audef}). LO (MC@NLO) results are shown in blue (black), for 
$m_H=200$ GeV (upper curves at large $\tan\beta$) and $m_H=1500$ GeV 
(lower curves at large $\tan\beta$). The error
band is statistical. Results for $Wt$ production, using both the DR and DS 
approaches in~\cite{Frixione:2008yi}, are shown in red (in the right-hand plot the DR and DS results are on top of each other).}
\label{Azplot}
\end{center}
\end{figure}

The behaviour of $A_z$ is shown in figure~\ref{Azplot}. A first notable feature
is the lack of smoothness, even in the LO results. This is due to the fact that
the boundaries of the intermediate regime will themselves depend on the value 
of $\tan\beta$, leading to fluctuations such as those observed in the figure.
It may be that such fluctuations can be ameliorated by tuning of these 
boundaries, with a corresponding trade-off in the size of the asymmetry
observed. The sign of the asymmetry flips for each charged Higgs mass 
as the full range in $\tan\beta$ is scanned, which is expected since the sign of the polarisation changes.
Note that there is again a marked difference between the LO
and NLO results, particularly for the higher Higgs mass, and that the boost cut has a larger effect for the lower Higgs mass.
\\

As before, one may compare the $H^-t$ and $Wt$ results. Here, though, a note of
caution is necessary, because the difference between the DR and
DS results for $Wt$ appears more pronounced for this parameter. In 
particular, it varies considerably before and after the boost cut is applied. 
This greater variation is perhaps exacerbated by the smallness of the
asymmetry (which is at best only a few percent), but also suggests that
interference with top pair production may be an issue in interpreting the
$Wt$ results. It is nevertheless the case that the difference with $Wt$ is
most pronounced at either low Higgs mass and high $\tan\beta$, or high Higgs
mass and low $\tan\beta$. In both these cases, the sign of the top polarisation in $H^-t$ production is opposite to the one in $Wt$ production. This results in a small asymmetry
of opposite sign to the $Wt$ case, but roughly comparable in size.\\

To summarise, we have here presented results for a number of angular and
energy-related distributions and, building upon the analysis 
of~\cite{Godbole:2010kr,Huitu:2010ad},  defined a corresponding asymmetry 
parameter for each that efficiently encodes the difference in these 
distributions for different regions in the charged Higgs parameter space, as 
well as the differences between $Wt$ and $H^-t$ production. All of these 
asymmetries seem to be fairly robust against NLO and parton shower 
corrections. In addition, they complement each other, since different observables are sensitive to different parts of the parameter space. This suggests that they may indeed be very useful in isolating 
a charged Higgs boson, with subsequent identification of its properties. In 
the following section, we consider a second context in which such observables 
may be useful, namely that of isolating $Wt$ production itself as a signal.\\

\section{Results for $Wt$ production}
\label{sec:Wt}
In the previous section, we examined the angular and energy distributions
introduced in section~\ref{sec:define} in $H^-t$ production, and defined
asymmetry parameters which are potentially highly useful in elucidating the
properties of a charged Higgs boson. In this section, we investigate whether
these same observables have anything useful to say about Standard Model $Wt$ 
production. \\

There are three production modes for a single top quark in the Standard Model.
Two of these, the so-called $s-$ and $t-$ channel modes, have been observed
in combination at both the Tevatron~\cite{Aaltonen:2009jj,Abazov:2009ii,
Abazov:2011rz} and LHC~\cite{Chatrchyan:2011vp,Schwienhorst:2011ng}. 
The theoretical state of
the art is also highly advanced, and includes fixed order 
computations~\cite{Harris:2002md,Campbell:2004ch,Cao:2004ky,
Cao:2004ap,Cao:2005pq}, NLO plus parton shower 
implementations~\cite{Frixione:2005vw,Alioli:2009je}, resummed 
results~\cite{Kidonakis:2011wy}, and finite top width 
corrections~\cite{Falgari:2011qa,Falgari:2010sf}. For related phenomenological
studies, see~\cite{Cao:2007ea,Sullivan:2005ar,Sullivan:2004ie,
Motylinski:2009kt}. As already stated in the introduction, $Wt$ production 
offers a complementary window through which to look at top quark interactions,
being sensitive to corrections to the $Wtb$ vertex, but not to four fermion 
operators which may affect the $s-$ and $t-$ channel modes. The investigation
of $Wt$ production as a signal in its own right was first explored 
in~\cite{Tait:1999cf}. Since then, computations have been carried out at
NLO~\cite{Zhu:2002uj,Campbell:2005bb}, and also matched to a parton shower at 
this accuracy~\cite{Frixione:2008yi,Re:2010bp}. \\

The aim of this section is to examine angular observables and energy ratios
for both $Wt$ and top pair production, for semi-realistic analysis cuts, and
to reflect upon whether these results may be useful in enhancing the signal to 
background ratio of the former process. To this end, we adopt the following
$Wt$ signal cuts, similar to those used in~\cite{White:2009yt}:

\begin{center}
\textbf{$Wt$ signal cuts}
\end{center}
\begin{enumerate}
\item The presence of exactly 1 $b$ jet with $p_t^T>50$ GeV and $|\eta|<2.5$. 
No other $b$ jets with $p_t^T>25$ GeV and $|\eta|<2.5$.
\item The presence of exactly 2 light flavor jets with $p_t^T>25$ GeV and 
$|\eta|<2.5$. In addition, their invariant mass should satisfy $55$ 
GeV$<m_{j_1j_2}<85$ GeV.
\item Events are vetoed if the invariant mass of the $b$ jet and light jet 
pair satisfies
\begin{displaymath}
150 \quad{\rm GeV}\quad<\sqrt{(p_{j_1}+p_{j_2}+p_{b})^2}\quad<\quad190\quad 
{\rm GeV}.
\end{displaymath}
\item The presence of exactly 1 isolated lepton with $p_t^T>25$ GeV and 
$|\eta|<2.5$. The lepton should satisfy $\Delta R>0.4$ with respect to the 
two light jets and the $b$ jet, where $R$ is the distance in the $(\eta,\phi)$
plane.
\item The missing transverse energy should satisfy $E^{miss}_{T}>$25 GeV.
\end{enumerate}
Here the first cut is the most useful in getting rid of top pair production, 
as one expects two $b$ jets on average in $t\bar{t}$ production, but only 
one $b$ jet in $Wt$. The other cuts pick out semi-leptonic 
decays\footnote{Note that to increase the statistics in our analysis, we will
explicitly generate semi-leptonic decays using MC@NLO. The above analysis cuts,
however, will still affect the shapes of distributions.}. That is, 
one $W$ boson decays to leptons (we would want this to be the $W$ boson from 
the top quark decay), and the other decays to quarks. We thus expect
two light jets whose invariant mass reconstructs the $W$ mass, as well as a 
lepton and missing energy from the neutrino. The only difference with respect
to the cuts used in~\cite{White:2009yt} is the presence of an additional cut
involving the invariant mass of the $b$ jet and light jet pair, restricting
this to lie away from the top mass. This ensures that the selected 
semi-leptonic events are such that the top quark in $Wt$ decays leptonically, 
and the $W$ hadronically, as is required in order to use the decay lepton
as a marker of top quark polarisation effects. \\

It was shown in~\cite{White:2009yt} that, for these signal cuts (minus the 
invariant mass requirement for the three jets, which was unnecessary in that
analysis), $Wt$ is a well-defined scattering process in that interference 
with pair production can be neglected. This was found by comparing the DR and 
DS results from MC@NLO. The results in this section were obtained using the DR subtraction method. Furthermore, the $Wt$ cross-section was found to be 
larger than the scale-variation uncertainty associated with the top pair 
cross-section. If this had not been true, then $Wt$ production would be 
swallowed up in the uncertainty of the top pair prediction, and much more 
care would be needed in order to be able to claim that it can be observed 
independently. We thus use the above cuts as an example of a fairly minimal
analysis which guarantees that $Wt$ is a well-defined signal. We will see that
even for this analysis, the angular and energy-related observables defined
in section~\ref{sec:define} display pronounced differences between $Wt$ and
top pair production. \\

Note that in this section, in order to be more realistic, we consider 
distributions constructed from the isolated lepton entering the cuts.
This is not guaranteed to be the decay lepton from the top quark, although
the likelihood of this is increased by the event selection cuts. 
Also, we assume that the top quark direction is reconstructed with perfect
resolution. In practice this would be done by considering the four-momenta
of the $b$ jet and isolated lepton passing the cuts, together with missing 
energy. A full determination of the uncertainty induced in the reconstruction
of the top quark (also including detector effects) is beyond the scope of the 
present study. Note that in $Wt$ and $W\bar{t}$ production, we assume that
the top and antitop quark is reconstructed respectively. In top pair 
production, one constructs either the top or antitop quark which decays to
give the isolated lepton passing the selection cuts.
In contrast to the $H^-t$ results of the previous section,
we present results for a centre of mass energy of 7 TeV. Jets are clustered using the $k_T$ algorithm~\cite{Catani:1993hr} with D=0.7.\\

\begin{figure}
\begin{center}
\scalebox{0.5}{\includegraphics{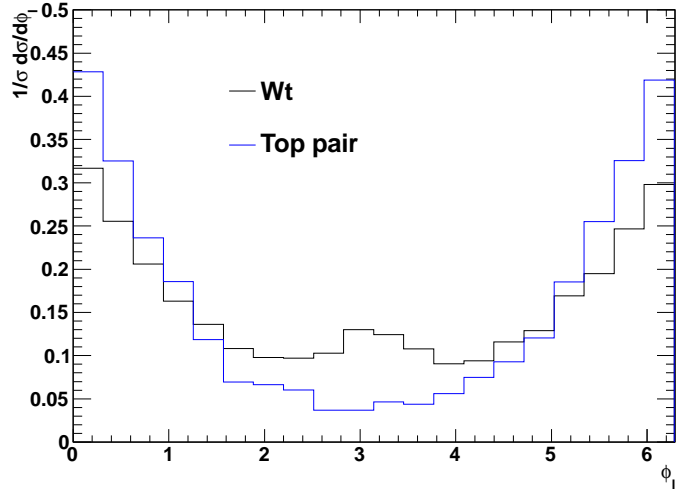}}
\caption{Azimuthal angle distribution of the isolated lepton which enters the
$Wt$ signal cuts, for both $Wt$ and top pair production, at NLO plus parton 
shower level.}
\label{philWtfig}
\end{center}
\end{figure}

We first consider the azimuthal angle $\phi_l$, whose distribution is shown in 
figure~\ref{philWtfig} for both $Wt$ and top pair production.
The first thing to notice is that there is a distinct shape difference between
the $Wt$ and top pair curves. The $Wt$ results include a slight peak structure
at $\theta=\pi$, due to the contribution from events in which the $W$ boson
decays leptonically, rather than the top quark. This structure is missing in
the case of top pair production, due to the symmetrical nature of the 
final state. For the choice of analysis cuts 
given above, one may evaluate the asymmetry parameter $A_\phi$, which is shown in
table~\ref{Alresults}.
\begin{table}
\begin{center}
\begin{tabular}{c|c|c}
$B_{cut}$ & $Wt$ & Top pair\\
\hline
0     & 0.33 $\pm$ 0.01   & 0.63 $\pm$ 0.02 \\
0.8   & 0.41 $\pm$ 0.02   & 0.70 $\pm$ 0.05 \\
0.9   & 0.42 $\pm$ 0.03   & 0.70 $\pm$ 0.07 \\
0.95  & 0.44 $\pm$ 0.04  & 0.68 $\pm$ 0.08 \\
\end{tabular}
\caption{Results for the azimuthal asymmetry parameter $A_\phi$ 
of eq.~(\ref{Aldef}), evaluated using the isolated lepton entering the 
$Wt$ selection cuts, and for different values of a cut $B>B_{cut}$ on the 
boost parameter of the top quark.}
\label{Alresults}
\end{center}
\end{table} 
The values for $Wt$ and top pair production are significantly different. This 
is potentially a useful distinguishing feature between the two production 
processes. \\

\begin{figure}
\begin{center}
\scalebox{0.5}{\includegraphics{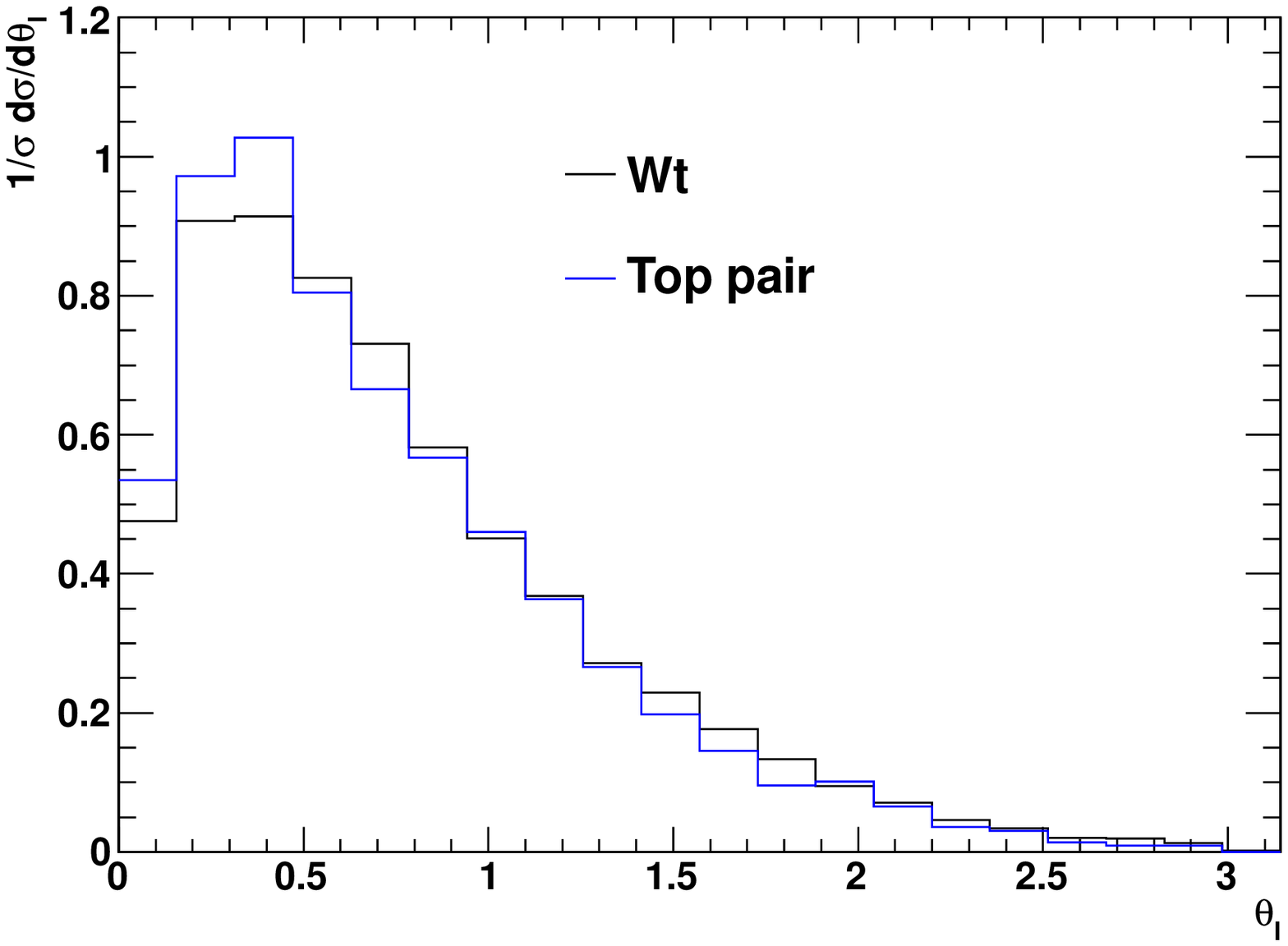}}
\caption{Polar angle distribution of the isolated lepton which enters the
$Wt$ signal cuts, for both $Wt$ and top pair production, at NLO plus parton
shower level.}
\label{thetaplotWt}
\end{center}
\end{figure}

Next, we consider the polar angle $\theta_l$, again defined in terms of the
isolated lepton entering the $Wt$ signal cuts. The distribution of this angle
is shown in figure~\ref{thetaplotWt}.
There is a notable difference between the $Wt$ and top pair production, due to the negative polarisation of the top in the former case.
The corresponding asymmetry parameters $A_\theta$ are shown in 
table~\ref{Athetaresults}.
\begin{table}
\begin{center}
\begin{tabular}{c|c|c}
$B_{cut}$ & $Wt$ & Top pair\\
\hline
0     & 0.02 $\pm$ 0.01   & 0.26 $\pm$ 0.02 \\
0.8   & 0.18 $\pm$ 0.02   & 0.38 $\pm$ 0.04 \\
0.9   & 0.49 $\pm$ 0.03   & 0.75 $\pm$ 0.07 \\
0.95  & 0.70 $\pm$ 0.05   & 0.97 $\pm$ 0.10 \\
\end{tabular}
\caption{Results for the polar asymmetry parameter $A_\theta$ 
of eq.~(\ref{Athetadef}), evaluated using the isolated lepton entering the 
$Wt$ selection cuts, and for different values of a cut $B>B_{cut}$ on the 
boost parameter of the top quark.}
\label{Athetaresults}
\end{center}
\end{table} 
Again the results are different between the two production processes which,
as in the azimuthal case, is a potentially useful discriminator between the
two processes. \\

In the case of $H^-t$ production considered in section~\ref{sec:Ht}, we also
considered various observables which depended upon the boost of the top quark.
This is clearly of practical importance for heavy charged Higgs masses, which
do indeed lead to heavily boosted top quarks in a sizeable fraction of events,
as is clear from figure~\ref{BplotHt}. One expects
boosted top observables to be less useful in $Wt$ production, due to the
fact that the $W$ boson is much lighter. Nevertheless, it is perhaps worth
examining the dependence of various observables on the boost parameter of the
top quark. If sizeable differences between $Wt$ and top pair production were
to be observed, the impact on the signal to background ratio would then 
outweigh the loss in signal cross-section. \\

\begin{figure}
\begin{center}
\scalebox{0.5}{\includegraphics{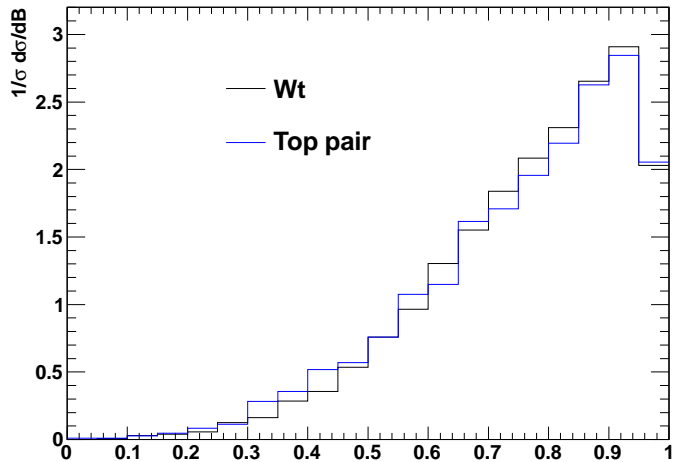}}
\caption{Distribution of the boost parameter $B$ of eq.~(\ref{Bdef}), at NLO
plus parton shower level.}
\label{BfigWt}
\end{center}
\end{figure}

\begin{figure}
\begin{center}
\scalebox{0.4}{\includegraphics{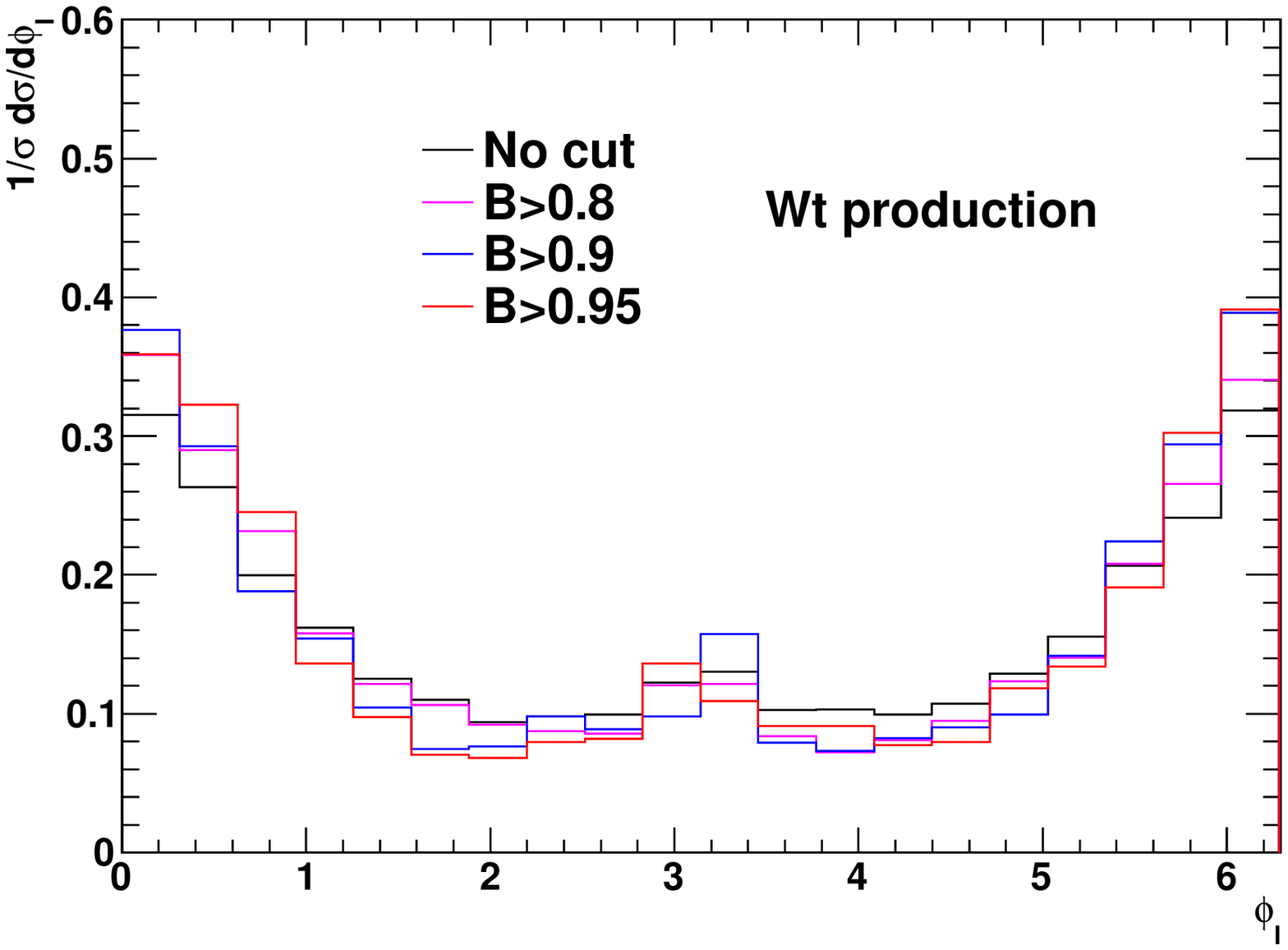}}
\scalebox{0.4}{\includegraphics{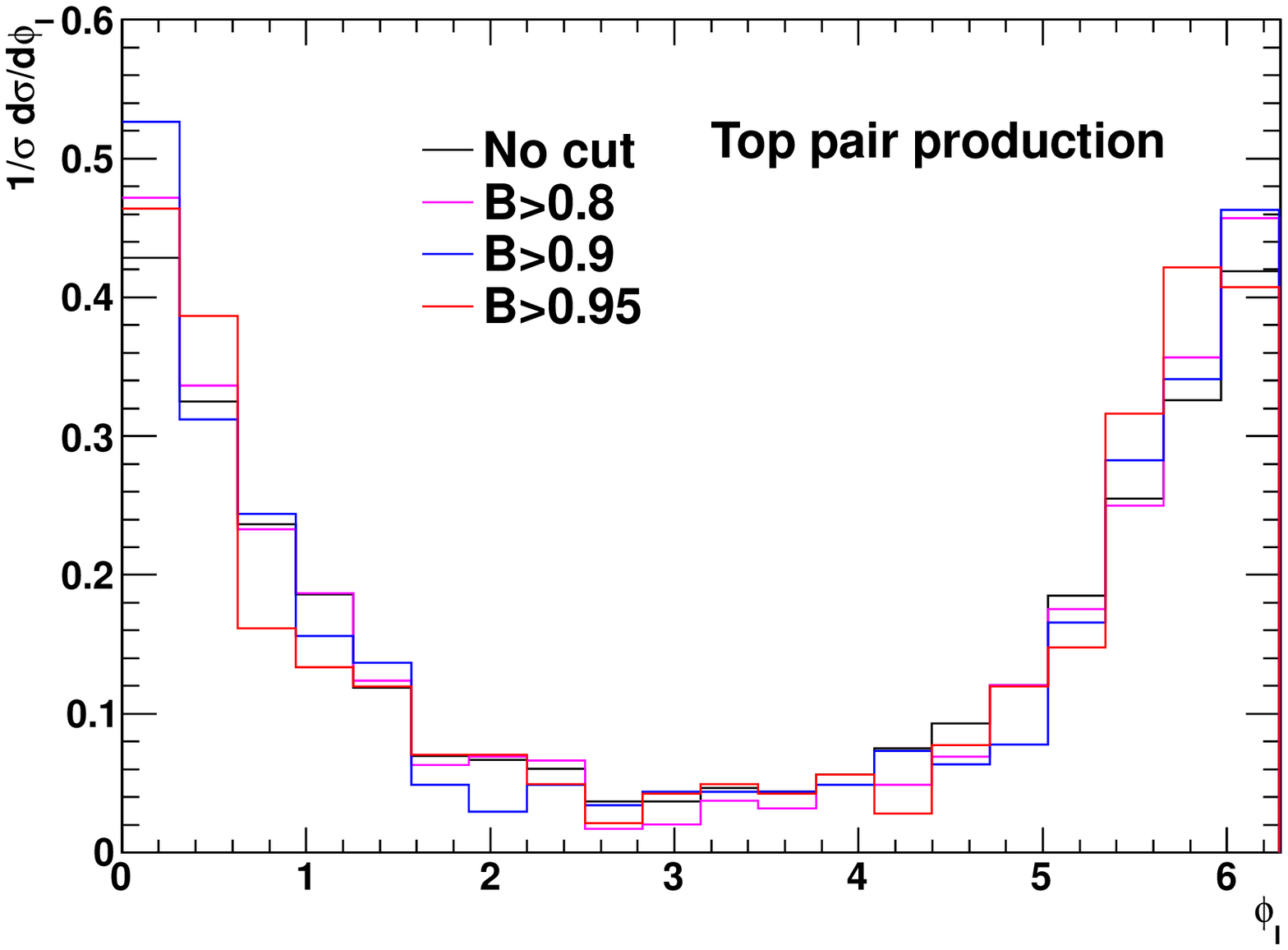}}
\caption{Azimuthal angle distribution of the isolated lepton which enters the
$Wt$ signal cuts, for $Wt$ and top pair production, for different values
of a cut $B>B_{cut}$ on the boost parameter of eq.~(\ref{Bdef}), at NLO
plus parton shower level.}
\label{AlboostplotWt}
\end{center}
\end{figure}

The distribution of the boost parameter $B$ of eq.~(\ref{Bdef}) is shown for 
both $Wt$ and top pair production in figure~\ref{BfigWt}, and one sees that 
there is a reasonable fraction of events in both cases which have $B>0.8$, 
albeit not as many as in the $H^-t$ case of the previous section. This is not 
surprising, given that charged Higgs masses of at least 200 GeV were 
considered there, so that the top recoiled against a much more massive 
particle than a $W$ boson. Here we also have a lower centre of 
mass energy. The $\phi_l$ distributions for the two 
processes are shown in figure~\ref{AlboostplotWt} for different values of a 
cut $B>B_{cut}$. 
One sees that, whilst there is some dependency on the boost parameter, the
qualitative features remain identical. The corresponding asymmetries $A_\phi$ are
given in table~\ref{Alresults}. One sees that the absolute value of the 
difference between the asymmetries for the two processes is roughly independent
of the boost cut. However, the relative difference decreases.
\\

\begin{figure}
\begin{center}
\scalebox{0.4}{\includegraphics{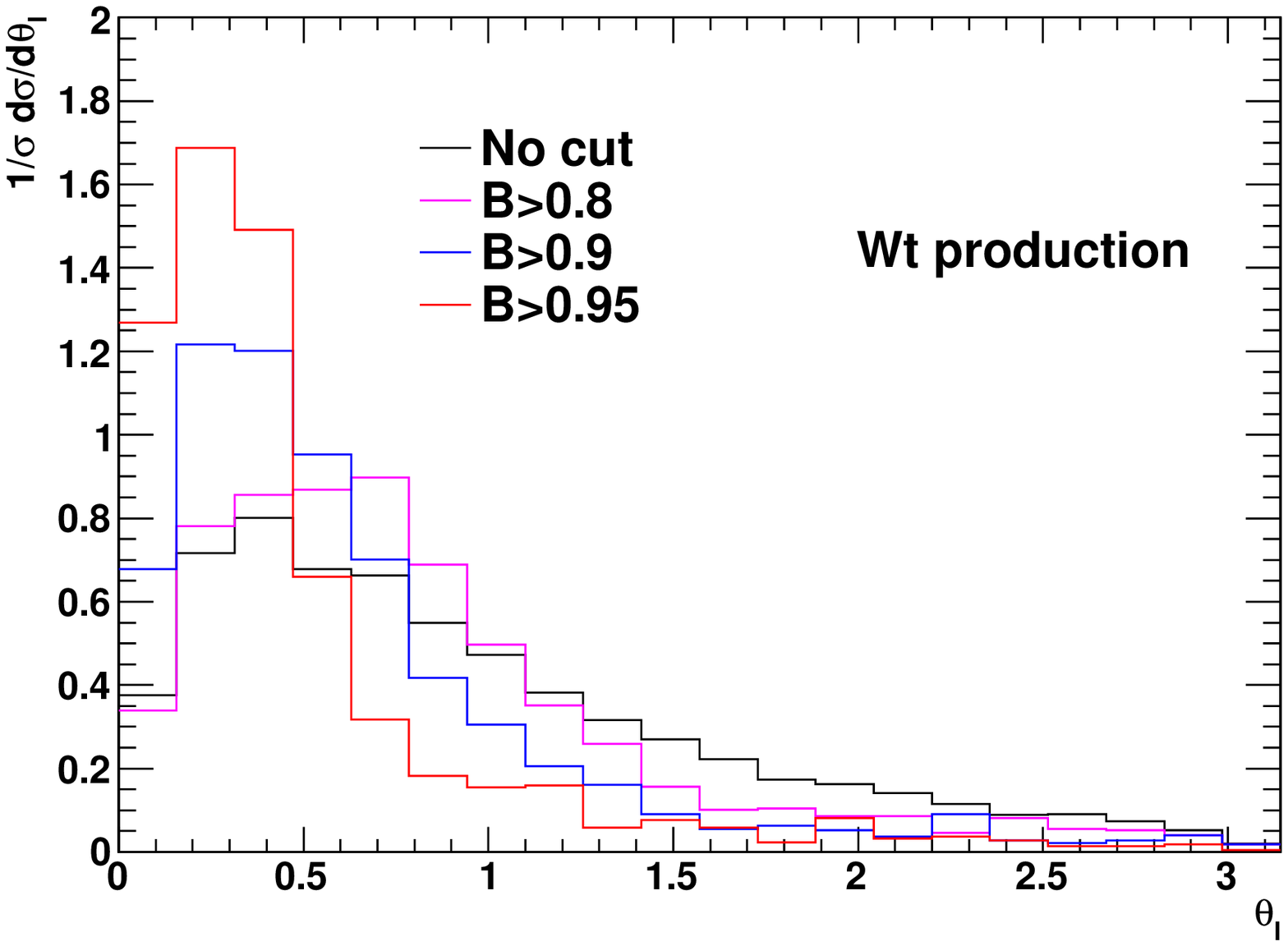}}
\scalebox{0.4}{\includegraphics{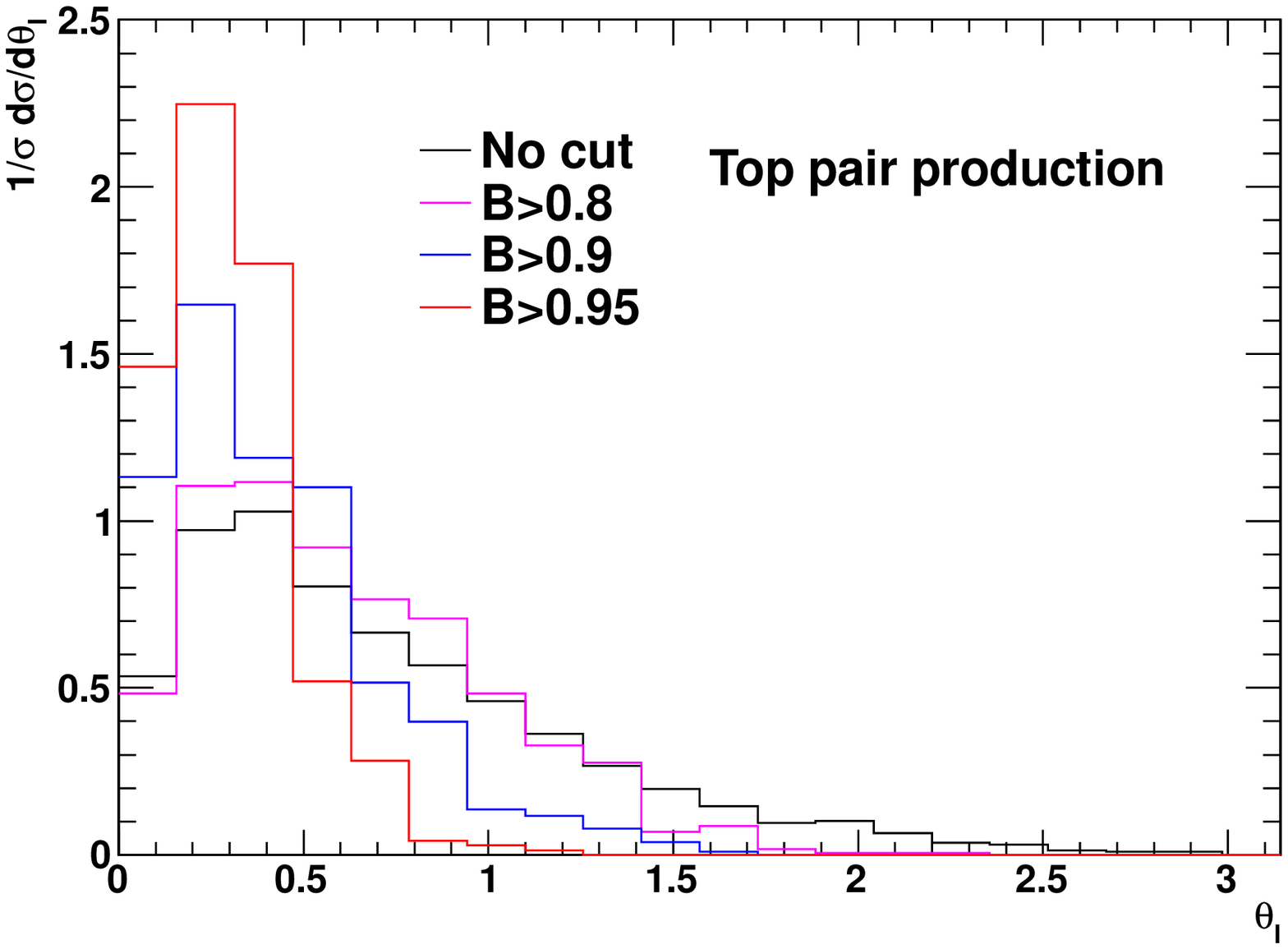}}
\caption{Polar angle distribution of the isolated lepton which enters the
$Wt$ signal cuts, for $Wt$ and top pair production, for different values
of a cut $B>B_{cut}$ on the boost parameter of eq.~(\ref{Bdef}), at NLO
plus parton shower level.}
\label{AthetaboostplotWt}
\end{center}
\end{figure}

One expects a much greater effect from the boost on the polar angle 
distribution, as the requirement of a boosted top will concentrate the decay 
products in polar angle. The $\theta_l$ distributions as a function of 
$B_{cut}$ are shown in figure~\ref{AthetaboostplotWt}.
The effect of the higher boost cut is to increase the peak region of the 
distribution at the expense of the tail, as expected. The corresponding 
$A_\theta$ values are collected in table~\ref{Athetaresults}. Unsurprisingly,
both sets of results display an increase in $A_\theta$ as the boost cut is
increased. This implies that a boost cut is actually detrimental in this case,
as the relative difference between the asymmetry parameters in the two 
processes decreases.\\

\begin{figure}
\begin{center}
\scalebox{0.4}{\includegraphics{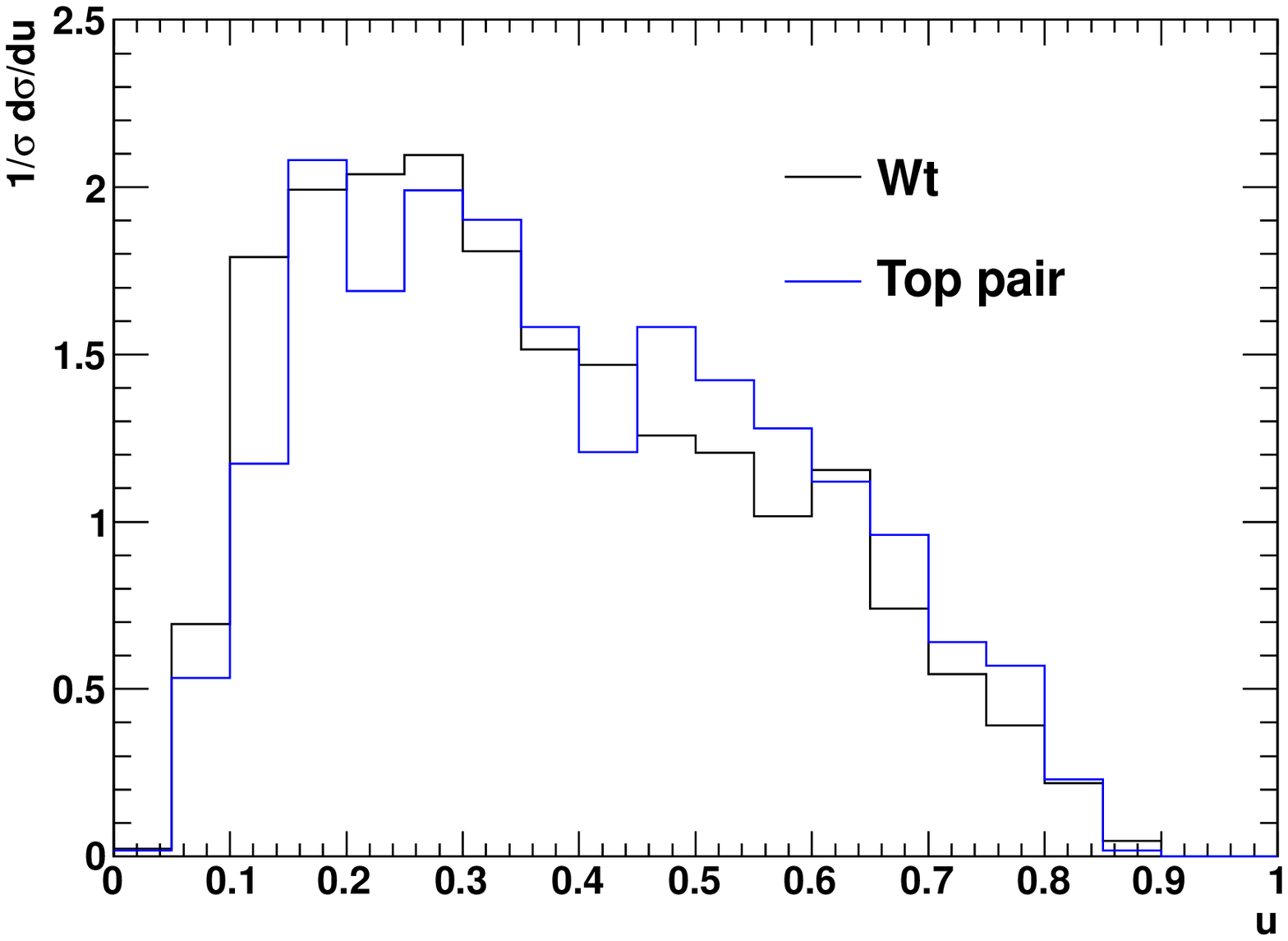}}
\scalebox{0.4}{\includegraphics{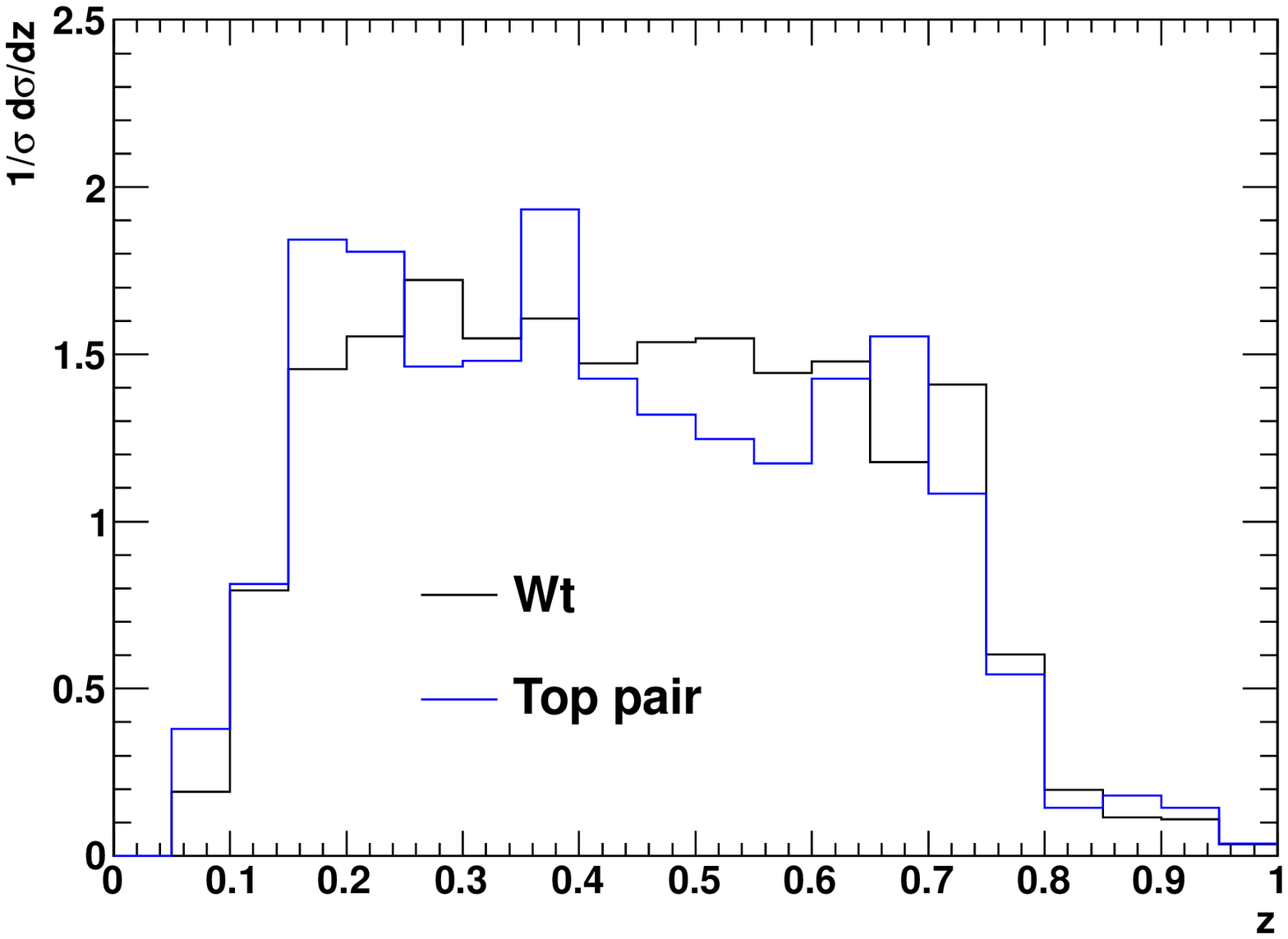}}
\caption{Distributions of $u$ and $z$, as defined in eq.~(\ref{zudef}), 
where a cut on the boost parameter $B>0.8$ has been applied, at NLO plus
parton shower level.}
\label{EratplotsWt}
\end{center}
\end{figure}

Finally, we present results for the energy ratios of eqs.~(\ref{zudef}), which
were shown to be useful for $H^-t$ production in section~\ref{sec:Ht}. In that 
case, we defined the energy of the $b$ quark via eq.~(\ref{Ebdef}), which is
possible in a Monte Carlo study but not in a real experiment. Here, given that
we have explicitly implemented analysis cuts in terms of jets, we define 
$E_b$ to be the energy of the $b$ jet which enters the cuts. Then the  
distributions of $z$ and $u$, with a cut on the boost parameter of $B>0.8$, are
shown in figure~\ref{EratplotsWt}.
The first thing to note is that the results for the $u$ distribution do not
show a significant difference between $Wt$ and top pair production. This is
perhaps not so surprising given that we have already seen in 
section~\ref{sec:Ht} that oppositely polarised top quarks tend to exhibit
smaller differences in energy-related distributions than in angular 
distributions. Here we are essentially probing the difference between a 
polarised top quark and one which is unpolarised on average, and thus one 
expects an even smaller difference in behaviour. \\

The $z$ distribution in figure~\ref{EratplotsWt} shows some 
difference between the $Wt$ and top pair distributions. However, the top
pair result does not closely resemble the flat profile one would expect
for unpolarised top quarks, due presumably to that fact that the shape has
been sculpted somewhat by the event selection cuts, in particular those which
implement restrictions on jet invariant masses.\\

Given the above results,  it does not seem particularly useful to examine the 
asymmetry parameters of eqs.~(\ref{Audef}, \ref{Azdef}) in the present context.
Nevertheless, the fact that a shape difference persists in the 
$z$ distribution between $Wt$ and top pair production still makes this a 
potentially useful observable in discriminating the two processes. One must
also bear in mind the result for the polar asymmetry from above, namely that
a boost cut will decrease the relative difference between the angular
asymmetries in $Wt$ and top pair production. Thus, and perhaps unsurprisingly,
the utility of boost cuts in $Wt$ production is somewhat limited.

\section{Conclusion}
\label{sec:conclude}
In this paper, we have examined the role that observables which are sensitive
to top quark polarisation can play in exploring the parameter space of charged
Higgs models, and also in distinguishing $H^-t$ production from (Standard 
Model) $Wt$ production. In particular, we examined the azimuthal and polar 
angles $\phi_l$ and $\theta_l$ of~\cite{Godbole:2010kr,Huitu:2010ad}, and the 
energy ratios $z$ and $u$ of~\cite{Shelton:2008nq}, defining corresponding 
asymmetry parameters analagous to that already defined for the azimuthal angle 
in~\cite{Huitu:2010ad}. Importantly, we found that polarisation effects are 
robust up to NLO and including parton shower corrections\footnote{A similar
robustness has already been observed in (Standard Model) $s-$ and $t-$ 
channel single top production~\cite{Motylinski:2009kt}.}. At this level, each
of the asymmetry parameters showed significant difference between different
regions in the charged Higgs parameter space $(m_H,\tan\beta)$, and also 
between $H^-t$ and $Wt$ production. The full set of asymmetries taken together
thus provides a potentially highly useful probe of charged Higgs properties.
Angular observables are sensitive only to corrections to the production of a 
top quark, and the polar angle is able to discriminate between charged Higgs 
masses at high $\tan\beta$ values, where the azimuthal angle cannot. Energy
observables are sensitive to corrections to both the production and decay of
top quarks. Although more difficult to construct (owing to the need for a 
cut on the boost parameter of the top quark), they give useful complementary
information, particularly on the value of the charged Higgs mass at 
intermediate and high $\tan\beta$ values.\\

As a second application of these observables, we considered the problem of
distinguishing Standard Model $Wt$ production from top pair production, which
is a significant background. Under the assumption that it is meaningful to
separate $Wt$ and top pair production, we observed significant differences,
for semi-realistic $Wt$ analysis cuts, between angular distributions relating
to the isolated lepton entering the cuts. It is worth pointing out that the 
cuts we used are fairly minimal in terms of signal to background 
ratio~\cite{White:2009yt}. Nevertheless, large differences are obtained between
the two production processes, which suggests that our findings would persist
in a more realistic study, including detector effects etc. \\

One may also consider boosted top quark observables in Standard Model $Wt$
production, and we gave a couple of examples in section~\ref{sec:Wt}. These
seem less useful than in $H^-t$ production, however. In the angular 
observables, a cut on the boost parameter does not increase the absolute 
difference between the asymmetry parameters for $Wt$ and top pair production,
and decreases the relative difference. For energy observables, 
one sees only a small difference between the $u$ distributions even when a 
boost cut is applied. This is due mainly to the fact that one is comparing a 
polarised top quark in $Wt$ with an (on average) unpolarised top quark in top 
pair production, rather than an oppositely polarised top quark. A larger 
difference is observed in the $z$ distribution, which may yet be a useful 
observable in distinguishing $Wt$ and top pair production.\\

To summarise, the observables studied in this paper are useful probes of both 
$H^-t$ and $Wt$ production, and seem to be robust against higher order 
perturbative corrections. They therefore deserve further investigation.

\section*{Acknowledgments}
We thank Wim Beenakker, Craig Buttar, James Ferrando and Eric Laenen for many useful 
discussions. CDW is supported by the STFC Postdoctoral Fellowship ``Collider 
Physics at the LHC'', and is very grateful to the theory group at Nikhef for 
warm hospitality. IN and LH are supported by the Foundation for Fundamental
Research of Matter (FOM), program 104 ``Theoretical Particle Physics in the Era of the LHC''. IN would like to thank the Centre of High Energy Physics at the Indian Institute of Science for their hospitality.
R.G. wishes to thank University of Utrecht (UU) for the
award of a Utrecht-Asia  visiting professorship and for
hospitality during her stay at UU. Further, she wishes to acknowledge 
support from the Department of  Science and Technology, India under Grant 
No. SR/S2/JCB-64/2007, under the J.C. Bose Fellowship scheme.

\bibliography{refs.bib}

\providecommand{\href}[2]{#2}\begingroup\raggedright\begin{thebibliography}{10%
0}

\bibitem{Harlander:1994ac}
R.~Harlander, M.~Jezabek, J.~H. Kuhn, and T.~Teubner, ``{Polarization in top
  quark pair production near threshold},'' {\em Phys.Lett.} {\bf B346} (1995)
  137--142, \href{http://www.arXiv.org/abs/hep-ph/9411395}{{\tt
  hep-ph/9411395}}.

\bibitem{Hikasa:1999wy}
K.-i. Hikasa, J.~M. Yang, and B.-L. Young, ``{R-parity violation and top quark
  polarization at the Fermilab Tevatron collider},'' {\em Phys.Rev.} {\bf D60}
  (1999) 114041, \href{http://www.arXiv.org/abs/hep-ph/9908231}{{\tt
  hep-ph/9908231}}.

\bibitem{Rindani:1999gd}
S.~D. Rindani and M.~M. Tung, ``{Longitudinal quark polarization in e+ e- $\to$
  t anti-t and chromoelectric and chromomagnetic dipole couplings of the top
  quark},'' {\em Eur.Phys.J.} {\bf C11} (1999) 485--493,
  \href{http://www.arXiv.org/abs/hep-ph/9904319}{{\tt hep-ph/9904319}}.

\bibitem{Boos:2003vf}
E.~Boos, H.~Martyn, G.~A. Moortgat-Pick, M.~Sachwitz, A.~Sherstnev, {\em et
  al.}, ``{Polarization in sfermion decays: Determining tan beta and trilinear
  couplings},'' {\em Eur.Phys.J.} {\bf C30} (2003) 395--407,
  \href{http://www.arXiv.org/abs/hep-ph/0303110}{{\tt hep-ph/0303110}}.

\bibitem{Gajdosik:2004ed}
T.~Gajdosik, R.~M. Godbole, and S.~Kraml, ``{Fermion polarization in sfermion
  decays as a probe of CP phases in the MSSM},'' {\em JHEP} {\bf 0409} (2004)
  051, \href{http://www.arXiv.org/abs/hep-ph/0405167}{{\tt hep-ph/0405167}}.

\bibitem{Allanach:2006fy}
B.~C. Allanach, C.~Grojean, P.~Z. Skands, E.~Accomando, G.~Azuelos, {\em et
  al.}, ``{Les Houches physics at TeV colliders 2005 beyond the standard model
  working group: Summary report},''
  \href{http://www.arXiv.org/abs/hep-ph/0602198}{{\tt hep-ph/0602198}}.

\bibitem{Godbole:2006eb}
R.~M. Godbole, S.~Kraml, S.~D. Rindani, and R.~K. Singh, ``{Probing
  CP-violating Higgs contributions in $\gamma\gamma\to f\bar f$ through fermion
  polarization},'' {\em Phys.Rev.} {\bf D74} (2006) 095006,
  \href{http://www.arXiv.org/abs/hep-ph/0609113}{{\tt hep-ph/0609113}}.

\bibitem{Godbole:2006tq}
R.~M. Godbole, S.~D. Rindani, and R.~K. Singh, ``{Lepton distribution as a
  probe of new physics in production and decay of the t quark and its
  polarization},'' {\em JHEP} {\bf 12} (2006) 021,
\href{http://www.arXiv.org/abs/hep-ph/0605100}{{\tt hep-ph/0605100}}.

\bibitem{Li:2006he}
P.-Y. Li, G.-R. Lu, J.~M. Yang, and H.~Zhang, ``{Probing R-parity Violating
  Interactions from Top Quark Polarization at LHC},'' {\em Eur.Phys.J.} {\bf
  C51} (2007) 163--168, \href{http://www.arXiv.org/abs/hep-ph/0608223}{{\tt
  hep-ph/0608223}}.

\bibitem{Najafabadi:2006um}
M.~Mohammadi~Najafabadi, ``{Secondary particles spectra in decay of polarized
  top quark with anomalous tWb coupling},'' {\em J.Phys.G} {\bf G34} (2007)
  39--46, \href{http://www.arXiv.org/abs/hep-ph/0601155}{{\tt hep-ph/0601155}}.

\bibitem{BhupalDev:2007is}
P.~Bhupal~Dev, A.~Djouadi, R.~Godbole, M.~Muhlleitner, and S.~Rindani,
  ``{Determining the CP properties of the Higgs boson},'' {\em Phys.Rev.Lett.}
  {\bf 100} (2008) 051801, \href{http://www.arXiv.org/abs/0707.2878}{{\tt
  0707.2878}}.

\bibitem{Eriksson:2007fx}
D.~Eriksson, G.~Ingelman, J.~Rathsman, and O.~Stal, ``{New angles on top quark
  decay to a charged Higgs},'' {\em JHEP} {\bf 01} (2008) 024,
\href{http://www.arXiv.org/abs/0710.5906}{{\tt 0710.5906}}.

\bibitem{Perelstein:2008zt}
M.~Perelstein and A.~Weiler, ``{Polarized Tops from Stop Decays at the LHC},''
  {\em JHEP} {\bf 0903} (2009) 141,
  \href{http://www.arXiv.org/abs/0811.1024}{{\tt 0811.1024}}.

\bibitem{Nojiri:2008ir}
M.~M. Nojiri and M.~Takeuchi, ``{Study of the top reconstruction in top-partner
  events at the LHC},'' {\em JHEP} {\bf 0810} (2008) 025,
  \href{http://www.arXiv.org/abs/0802.4142}{{\tt 0802.4142}}.

\bibitem{Shelton:2008nq}
J.~Shelton, ``{Polarized tops from new physics: signals and observables},''
  {\em Phys. Rev.} {\bf D79} (2009) 014032,
\href{http://www.arXiv.org/abs/0811.0569}{{\tt 0811.0569}}.

\bibitem{Godbole:2009dp}
R.~M. Godbole, S.~D. Rindani, K.~Rao, and R.~K. Singh, ``{Top polarization as a
  probe of new physics},'' {\em AIP Conf.Proc.} {\bf 1200} (2010) 682--685,
  \href{http://www.arXiv.org/abs/0911.3622}{{\tt 0911.3622}}.

\bibitem{Arai:2010ci}
M.~Arai, K.~Huitu, S.~K. Rai, and K.~Rao, ``{Single production of sleptons with
  polarized tops at the Large Hadron Collider},'' {\em JHEP} {\bf 1008} (2010)
  082, \href{http://www.arXiv.org/abs/1003.4708}{{\tt 1003.4708}}.

\bibitem{Djouadi:2009nb}
A.~Djouadi, G.~Moreau, F.~Richard, and R.~K. Singh, ``{The Forward-backward
  asymmetry of top quark production at the Tevatron in warped extra dimensional
  models},'' {\em Phys.Rev.} {\bf D82} (2010) 071702,
  \href{http://www.arXiv.org/abs/0906.0604}{{\tt 0906.0604}}.

\bibitem{Krohn:2009wm}
D.~Krohn, J.~Shelton, and L.-T. Wang, ``{Measuring the Polarization of Boosted
  Hadronic Tops},'' {\em JHEP} {\bf 07} (2010) 041,
\href{http://www.arXiv.org/abs/0909.3855}{{\tt 0909.3855}}.

\bibitem{Godbole:2010kr}
R.~M. Godbole, K.~Rao, S.~D. Rindani, and R.~K. Singh, ``{On measurement of top
  polarization as a probe of $t \bar t$ production mechanisms at the LHC},''
  {\em JHEP} {\bf 11} (2010) 144,
\href{http://www.arXiv.org/abs/1010.1458}{{\tt 1010.1458}}.

\bibitem{AguilarSaavedra:2010nx}
J.~A. Aguilar-Saavedra and J.~Bernabeu, ``{W polarisation beyond helicity
  fractions in top quark decays},'' {\em Nucl. Phys.} {\bf B840} (2010)
  349--378,
\href{http://www.arXiv.org/abs/1005.5382}{{\tt 1005.5382}}.

\bibitem{Degrande:2010kt}
C.~Degrande, J.-M. Gerard, C.~Grojean, F.~Maltoni, and G.~Servant,
  ``{Non-resonant New Physics in Top Pair Production at Hadron Colliders},''
  {\em JHEP} {\bf 1103} (2011) 125,
  \href{http://www.arXiv.org/abs/1010.6304}{{\tt 1010.6304}}.

\bibitem{Huitu:2010ad}
K.~Huitu, S.~Kumar~Rai, K.~Rao, S.~D. Rindani, and P.~Sharma, ``{Probing top
  charged-Higgs production using top polarization at the Large Hadron
  Collider},'' {\em JHEP} {\bf 04} (2011) 026,
\href{http://www.arXiv.org/abs/1012.0527}{{\tt 1012.0527}}.

\bibitem{Cao:2010nw}
J.~Cao, L.~Wu, and J.~M. Yang, ``{New physics effects on top quark spin
  correlation and polarization at the LHC: a comparative study in different
  models},'' {\em Phys.Rev.} {\bf D83} (2011) 034024,
  \href{http://www.arXiv.org/abs/1011.5564}{{\tt 1011.5564}}.

\bibitem{Jung:2010yn}
D.-W. Jung, P.~Ko, and J.~S. Lee, ``{Longitudinal top polarization as a probe
  of a possible origin of forward-backward asymmetry of the top quark at the
  Tevatron},'' {\em Phys.Lett.} {\bf B701} (2011) 248--254,
  \href{http://www.arXiv.org/abs/1011.5976}{{\tt 1011.5976}}.

\bibitem{Choudhury:2010cd}
D.~Choudhury, R.~M. Godbole, S.~D. Rindani, and P.~Saha, ``{Top polarization,
  forward-backward asymmetry and new physics},'' {\em Phys.Rev.} {\bf D84}
  (2011) 014023, \href{http://www.arXiv.org/abs/1012.4750}{{\tt 1012.4750}}.

\bibitem{Gopalakrishna:2010xm}
S.~Gopalakrishna, T.~Han, I.~Lewis, Z.-g. Si, and Y.-F. Zhou, ``{Chiral
  Couplings of W' and Top Quark Polarization at the LHC},'' {\em Phys.Rev.}
  {\bf D82} (2010) 115020, \href{http://www.arXiv.org/abs/1008.3508}{{\tt
  1008.3508}}.

\bibitem{Godbole:2011hw}
R.~Godbole, C.~Hangst, M.~Muhlleitner, S.~Rindani, and P.~Sharma,
  ``{Model-independent analysis of Higgs spin and CP properties in the process
  $e^+ e^- \to t \bar t \Phi$},'' {\em Eur.Phys.J.} {\bf C71} (2011) 1681,
  \href{http://www.arXiv.org/abs/1103.5404}{{\tt 1103.5404}}.

\bibitem{Krohn:2011tw}
D.~Krohn, T.~Liu, J.~Shelton, and L.-T. Wang, ``{A Polarized View of the Top
  Asymmetry},'' \href{http://www.arXiv.org/abs/1105.3743}{{\tt 1105.3743}}.

\bibitem{Baumgart:2011wk}
M.~Baumgart and B.~Tweedie, ``{Discriminating Top-Antitop Resonances using
  Azimuthal Decay Correlations},'' {\em JHEP} {\bf 1109} (2011) 049,
  \href{http://www.arXiv.org/abs/1104.2043}{{\tt 1104.2043}}.

\bibitem{Rindani:2011pk}
S.~D. Rindani and P.~Sharma, ``{Probing anomalous tbW couplings in single-top
  production using top polarization at the Large Hadron Collider},''
  \href{http://www.arXiv.org/abs/1107.2597}{{\tt 1107.2597}}.

\bibitem{Baglio:2011ap}
J.~Baglio, M.~Beccaria, A.~Djouadi, G.~Macorini, E.~Mirabella, {\em et al.},
  ``{The Left-Right asymmetry of top quarks in associated top-charged Higgs
  bosons at the LHC as a probe of the tan$\beta$ parameter},''
  \href{http://www.arXiv.org/abs/1109.2420}{{\tt 1109.2420}}.

\bibitem{Rindani:2011gt}
S.~D. Rindani and P.~Sharma, ``{CP violation in tbW couplings at the LHC},''
\href{http://www.arXiv.org/abs/1108.4165}{{\tt 1108.4165}}.

\bibitem{Ananthanarayan:2010xs}
B.~Ananthanarayan, M.~Patra, and S.~D. Rindani, ``{Top-spin analysis of new
  scalar and tensor interactions in $e^+ e^-$ collisions with beam
  polarization},'' {\em Phys.Rev.} {\bf D83} (2011) 016010,
  \href{http://www.arXiv.org/abs/1007.5183}{{\tt 1007.5183}}.

\bibitem{Agashe:2006hk}
K.~Agashe, A.~Belyaev, T.~Krupovnickas, G.~Perez, and J.~Virzi, ``{LHC Signals
  from Warped Extra Dimensions},'' {\em Phys.Rev.} {\bf D77} (2008) 015003,
  \href{http://www.arXiv.org/abs/hep-ph/0612015}{{\tt hep-ph/0612015}}.

\bibitem{Falkowski:2011zr}
A.~Falkowski, G.~Perez, and M.~Schmaltz, ``{Spinning the Top},''
  \href{http://www.arXiv.org/abs/1110.3796}{{\tt 1110.3796}}.

\bibitem{Jezabek:1988ja}
M.~Jezabek and J.~H. Kuhn, ``{Lepton Spectra from Heavy Quark Decay},'' {\em
  Nucl.Phys.} {\bf B320} (1989) 20.

\bibitem{Czarnecki:1990pe}
A.~Czarnecki, M.~Jezabek, and J.~H. Kuhn, ``{Lepton spectra from decays of
  polarized top quarks},'' {\em Nucl.Phys.} {\bf B351} (1991) 70--80.

\bibitem{Brandenburg:2002xr}
A.~Brandenburg, Z.~Si, and P.~Uwer, ``{QCD corrected spin analyzing power of
  jets in decays of polarized top quarks},'' {\em Phys.Lett.} {\bf B539} (2002)
  235--241, \href{http://www.arXiv.org/abs/hep-ph/0205023}{{\tt
  hep-ph/0205023}}.

\bibitem{Grzadkowski:1999iq}
B.~Grzadkowski and Z.~Hioki, ``{New hints for testing anomalous top quark
  interactions at future linear colliders},'' {\em Phys.Lett.} {\bf B476}
  (2000) 87--94, \href{http://www.arXiv.org/abs/hep-ph/9911505}{{\tt
  hep-ph/9911505}}.

\bibitem{Grzadkowski:2002gt}
B.~Grzadkowski and Z.~Hioki, ``{Decoupling of anomalous top decay vertices in
  angular distribution of secondary particles},'' {\em Phys.Lett.} {\bf B557}
  (2003) 55--59, \href{http://www.arXiv.org/abs/hep-ph/0208079}{{\tt
  hep-ph/0208079}}.

\bibitem{Grzadkowski:2001tq}
B.~Grzadkowski and Z.~Hioki, ``{Angular distribution of leptons in general $t
  \bar{t}$ production and decay},'' {\em Phys.Lett.} {\bf B529} (2002) 82--86,
  \href{http://www.arXiv.org/abs/hep-ph/0112361}{{\tt hep-ph/0112361}}.

\bibitem{Hioki:2002vg}
Z.~Hioki, ``{A New decoupling theorem in top quark physics},'' in {\em
  {Seogwipo 2002, Linear colliders}}, pp.~333--338.
\newblock 2002.
\newblock \href{http://www.arXiv.org/abs/hep-ph/0210224}{{\tt hep-ph/0210224}}.

\bibitem{Ohkuma:2002iv}
K.~Ohkuma, ``{Effects of top quark anomalous decay couplings at gamma gamma
  colliders},'' {\em Nucl.Phys.Proc.Suppl.} {\bf 111} (2002) 285--287,
  \href{http://www.arXiv.org/abs/hep-ph/0202126}{{\tt hep-ph/0202126}}.

\bibitem{Rindani:2000jg}
S.~D. Rindani, ``{Effect of anomalous t b W vertex on decay lepton
  distributions in e+ e- $\to$ t anti-t and CP violating asymmetries},'' {\em
  Pramana} {\bf 54} (2000) 791--812,
  \href{http://www.arXiv.org/abs/hep-ph/0002006}{{\tt hep-ph/0002006}}.

\bibitem{Godbole:2002qu}
R.~M. Godbole, S.~D. Rindani, and R.~K. Singh, ``{Study of CP property of the
  Higgs at a photon collider using gamma gamma $\to$ t anti-t $\to$ lX},'' {\em
  Phys.Rev.} {\bf D67} (2003) 095009,
  \href{http://www.arXiv.org/abs/hep-ph/0211136}{{\tt hep-ph/0211136}}.

\bibitem{atlasnote}
{\bf ATLAS} Collaboration, ``{ATLAS Public Note no. ATL-PHYS-PUB-2010-008},''
  2010.

\bibitem{Weydert:2009vr}
C.~Weydert {\em et al.}, ``{Charged Higgs boson production in association with
  a top quark in MC@NLO},'' {\em Eur. Phys. J.} {\bf C67} (2010) 617--636,
\href{http://www.arXiv.org/abs/0912.3430}{{\tt 0912.3430}}.

\bibitem{Tait:1999cf}
T.~M.~P. Tait, ``{The $t W^{-}$ mode of single top production},'' {\em Phys.
  Rev.} {\bf D61} (2000) 034001,
\href{http://www.arXiv.org/abs/hep-ph/9909352}{{\tt hep-ph/9909352}}.

\bibitem{Campbell:2005bb}
J.~M. Campbell and F.~Tramontano, ``{Next-to-leading order corrections to W t
  production and decay},'' {\em Nucl. Phys.} {\bf B726} (2005) 109--130,
\href{http://www.arXiv.org/abs/hep-ph/0506289}{{\tt hep-ph/0506289}}.

\bibitem{Zhu:2002uj}
S.~Zhu, ``{Next-to-leading order QCD corrections to b g $\to$ t W- at the CERN
  Large Hadron Collider},'' {\em Phys. Lett.} {\bf B524} (2002)
283--288.

\bibitem{Frixione:2008yi}
S.~Frixione, E.~Laenen, P.~Motylinski, B.~R. Webber, and C.~D. White,
  ``{Single-top hadroproduction in association with a W boson},'' {\em JHEP}
  {\bf 07} (2008) 029,
\href{http://www.arXiv.org/abs/0805.3067}{{\tt 0805.3067}}.

\bibitem{White:2009yt}
C.~D. White, S.~Frixione, E.~Laenen, and F.~Maltoni, ``{Isolating Wt production
  at the LHC},'' {\em JHEP} {\bf 11} (2009) 074,
\href{http://www.arXiv.org/abs/0908.0631}{{\tt 0908.0631}}.

\bibitem{Bernreuther:2008ju}
W.~Bernreuther, ``{Top quark physics at the LHC},'' {\em J.Phys.G} {\bf G35}
  (2008) 083001, \href{http://www.arXiv.org/abs/0805.1333}{{\tt 0805.1333}}.

\bibitem{Alwall:2011uj}
J.~Alwall, M.~Herquet, F.~Maltoni, O.~Mattelaer, and T.~Stelzer, ``{MadGraph 5
  : Going Beyond},'' {\em JHEP} {\bf 1106} (2011) 128,
  \href{http://www.arXiv.org/abs/1106.0522}{{\tt 1106.0522}}.

\bibitem{Alwall:2007st}
J.~Alwall, P.~Demin, S.~de~Visscher, R.~Frederix, M.~Herquet, {\em et al.},
  ``{MadGraph/MadEvent v4: The New Web Generation},'' {\em JHEP} {\bf 0709}
  (2007) 028, \href{http://www.arXiv.org/abs/0706.2334}{{\tt 0706.2334}}.

\bibitem{Frixione:2002ik}
S.~Frixione and B.~R. Webber, ``{Matching NLO QCD computations and parton
  shower simulations},'' {\em JHEP} {\bf 0206} (2002) 029,
  \href{http://www.arXiv.org/abs/hep-ph/0204244}{{\tt hep-ph/0204244}}.

\bibitem{Frixione:2008ym}
S.~Frixione and B.~R. Webber, ``{The MC@NLO 3.4 Event Generator},''
  \href{http://www.arXiv.org/abs/0812.0770}{{\tt 0812.0770}}.

\bibitem{Frixione:2010wd}
S.~Frixione, F.~Stoeckli, P.~Torrielli, B.~R. Webber, and C.~D. White, ``{The
  MC@NLO 4.0 Event Generator},'' \href{http://www.arXiv.org/abs/1010.0819}{{\tt
  1010.0819}}.

\bibitem{Frixione:2003ei}
S.~Frixione, P.~Nason, and B.~R. Webber, ``{Matching NLO QCD and parton showers
  in heavy flavor production},'' {\em JHEP} {\bf 0308} (2003) 007,
  \href{http://www.arXiv.org/abs/hep-ph/0305252}{{\tt hep-ph/0305252}}.

\bibitem{Frixione:2007zp}
S.~Frixione, E.~Laenen, P.~Motylinski, and B.~R. Webber, ``{Angular
  correlations of lepton pairs from vector boson and top quark decays in Monte
  Carlo simulations},'' {\em JHEP} {\bf 04} (2007) 081,
\href{http://www.arXiv.org/abs/hep-ph/0702198}{{\tt hep-ph/0702198}}.

\bibitem{Frixione:2007vw}
S.~Frixione, P.~Nason, and C.~Oleari, ``{Matching NLO QCD computations with
  Parton Shower simulations: the POWHEG method},'' {\em JHEP} {\bf 0711} (2007)
  070, \href{http://www.arXiv.org/abs/0709.2092}{{\tt 0709.2092}}.

\bibitem{Giele:2007di}
W.~T. Giele, D.~A. Kosower, and P.~Z. Skands, ``{A Simple shower and matching
  algorithm},'' {\em Phys.Rev.} {\bf D78} (2008) 014026,
  \href{http://www.arXiv.org/abs/0707.3652}{{\tt 0707.3652}}.

\bibitem{Frixione:2007nw}
S.~Frixione, P.~Nason, and G.~Ridolfi, ``{A Positive-weight
  next-to-leading-order Monte Carlo for heavy flavour hadroproduction},'' {\em
  JHEP} {\bf 0709} (2007) 126, \href{http://www.arXiv.org/abs/0707.3088}{{\tt
  0707.3088}}.

\bibitem{Re:2010bp}
E.~Re, ``{Single-top Wt-channel production matched with parton showers using
  the POWHEG method},'' {\em Eur.Phys.J.} {\bf C71} (2011) 1547,
  \href{http://www.arXiv.org/abs/1009.2450}{{\tt 1009.2450}}.

\bibitem{Weydert:2010cx}
C.~Weydert, ``{Associated production of top quarks and charged Higgs bosons at
  next-to-leading order},'' {\em IL NUOVO CIMENTO 33 C} (2010)
  \href{http://www.arXiv.org/abs/1011.6249}{{\tt 1011.6249}}.

\bibitem{Kauer:2001sp}
N.~Kauer and D.~Zeppenfeld, ``{Finite width effects in top quark production at
  hadron colliders},'' {\em Phys. Rev.} {\bf D65} (2002) 014021,
\href{http://www.arXiv.org/abs/hep-ph/0107181}{{\tt hep-ph/0107181}}.

\bibitem{Kersevan:2006fq}
B.~P. Kersevan and I.~Hinchliffe, ``{A consistent prescription for the
  production involving massive quarks in hadron collisions},'' {\em JHEP} {\bf
  09} (2006) 033,
\href{http://www.arXiv.org/abs/hep-ph/0603068}{{\tt hep-ph/0603068}}.

\bibitem{Denner:2010jp}
A.~Denner, S.~Dittmaier, S.~Kallweit, and S.~Pozzorini, ``{NLO QCD corrections
  to WWbb production at hadron colliders},'' {\em Phys.Rev.Lett.} {\bf 106}
  (2011) 052001, \href{http://www.arXiv.org/abs/1012.3975}{{\tt 1012.3975}}.

\bibitem{Bevilacqua:2010qb}
G.~Bevilacqua, M.~Czakon, A.~van Hameren, C.~G. Papadopoulos, and M.~Worek,
  ``{Complete off-shell effects in top quark pair hadroproduction with leptonic
  decay at next-to-leading order},'' {\em JHEP} {\bf 1102} (2011) 083,
  \href{http://www.arXiv.org/abs/1012.4230}{{\tt 1012.4230}}.

\bibitem{Beenakker:1996ch}
W.~Beenakker, R.~Hopker, M.~Spira, and P.~Zerwas, ``{Squark and gluino
  production at hadron colliders},'' {\em Nucl.Phys.} {\bf B492} (1997)
  51--103, \href{http://www.arXiv.org/abs/hep-ph/9610490}{{\tt
  hep-ph/9610490}}.

\bibitem{Plehn:1998nh}
T.~Plehn, ``{Production of supersymmetric particles at high-energy
  colliders},'' \href{http://www.arXiv.org/abs/hep-ph/9809319}{{\tt
  hep-ph/9809319}}. Ph.D. Thesis.

\bibitem{Plehn:2010gp}
T.~Plehn and C.~Weydert, ``{Charged Higgs production with a top in MC@NLO},''
  in {\em PoS CHARGED2010}, p.~026.
\newblock 2010.
\newblock \href{http://www.arXiv.org/abs/1012.3761}{{\tt 1012.3761}}.

\bibitem{Binoth:2011xi}
T.~Binoth, D.~Goncalves-Netto, D.~Lopez-Val, K.~Mawatari, T.~Plehn, {\em et
  al.}, ``{Automized Squark-Neutralino Production to Next-to-Leading Order},''
  \href{http://www.arXiv.org/abs/1108.1250}{{\tt 1108.1250}}.

\bibitem{Djouadi:2005gi}
A.~Djouadi, ``{The Anatomy of electro-weak symmetry breaking. I: The Higgs
  boson in the standard model},'' {\em Phys. Rept.} {\bf 457} (2008) 1--216,
\href{http://www.arXiv.org/abs/hep-ph/0503172}{{\tt hep-ph/0503172}}.

\bibitem{Djouadi:2005gj}
A.~Djouadi, ``{The Anatomy of electro-weak symmetry breaking. II. The Higgs
  bosons in the minimal supersymmetric model},'' {\em Phys. Rept.} {\bf 459}
  (2008) 1--241,
\href{http://www.arXiv.org/abs/hep-ph/0503173}{{\tt hep-ph/0503173}}.

\bibitem{Bona:2009cj}
{\bf UTfit} Collaboration, M.~Bona {\em et al.}, ``{An Improved Standard Model
  Prediction Of BR(B -> tau nu) And Its Implications For New Physics},'' {\em
  Phys. Lett.} {\bf B687} (2010) 61--69,
\href{http://www.arXiv.org/abs/0908.3470}{{\tt 0908.3470}}.

\bibitem{Martin:2009iq}
A.~D. Martin, W.~J. Stirling, R.~S. Thorne, and G.~Watt, ``{Parton
  distributions for the LHC},'' {\em Eur. Phys. J.} {\bf C63} (2009) 189--285,
\href{http://www.arXiv.org/abs/0901.0002}{{\tt 0901.0002}}.

\bibitem{Martin:2009bu}
A.~D. Martin, W.~J. Stirling, R.~S. Thorne, and G.~Watt, ``{Uncertainties on
  $alpha_S$ in global PDF analyses and implications for predicted hadronic
  cross sections},'' {\em Eur. Phys. J.} {\bf C64} (2009) 653--680,
\href{http://www.arXiv.org/abs/0905.3531}{{\tt 0905.3531}}.

\bibitem{Martin:2010db}
A.~D. Martin, W.~J. Stirling, R.~S. Thorne, and G.~Watt, ``{Heavy-quark mass
  dependence in global PDF analyses and 3- and 4-flavour parton
  distributions},'' {\em Eur. Phys. J.} {\bf C70} (2010) 51--72,
\href{http://www.arXiv.org/abs/1007.2624}{{\tt 1007.2624}}.

\bibitem{Djouadi:1997yw}
A.~Djouadi, J.~Kalinowski, and M.~Spira, ``{HDECAY: A program for Higgs boson
  decays in the standard model and its supersymmetric extension},'' {\em
  Comput. Phys. Commun.} {\bf 108} (1998) 56--74,
\href{http://www.arXiv.org/abs/hep-ph/9704448}{{\tt hep-ph/9704448}}.

\bibitem{Aaltonen:2009jj}
{\bf CDF} Collaboration, T.~Aaltonen {\em et al.}, ``{First Observation of
  Electroweak Single Top Quark Production},'' {\em Phys.Rev.Lett.} {\bf 103}
  (2009) 092002, \href{http://www.arXiv.org/abs/0903.0885}{{\tt 0903.0885}}.

\bibitem{Abazov:2009ii}
{\bf D0} Collaboration, V.~M. Abazov {\em et al.}, ``{Observation of Single
  Top-Quark Production},'' {\em Phys. Rev. Lett.} {\bf 103} (2009) 092001,
\href{http://www.arXiv.org/abs/0903.0850}{{\tt 0903.0850}}.

\bibitem{Abazov:2011rz}
{\bf D0} Collaboration, V.~M. Abazov {\em et al.}, ``{Model-independent
  measurement of $\boldsymbol{t}$-channel single top quark production in
  $\boldsymbol{p\bar{p}}$ collisions at $\boldsymbol{\sqrt{s}=1.96}$ TeV},''
  {\em Phys.Lett.} {\bf B705} (2011) 313--319,
  \href{http://www.arXiv.org/abs/1105.2788}{{\tt 1105.2788}}.

\bibitem{Chatrchyan:2011vp}
{\bf CMS} Collaboration, S.~Chatrchyan {\em et al.}, ``{Measurement of the
  t-channel single top quark production cross section in pp collisions at
  sqrt(s) = 7 TeV},'' {\em Phys. Rev. Lett.} {\bf 107} (2011) 091802,
\href{http://www.arXiv.org/abs/1106.3052}{{\tt 1106.3052}}.

\bibitem{Schwienhorst:2011ng}
R.~Schwienhorst and f.~t.~A. collaboration, ``{Single top-quark production with
  the ATLAS detector in pp collisions at sqrt(s)=7TeV},''
  \href{http://www.arXiv.org/abs/1110.2192}{{\tt 1110.2192}}.

\bibitem{Harris:2002md}
B.~W. Harris, E.~Laenen, L.~Phaf, Z.~Sullivan, and S.~Weinzierl, ``{The fully
  differential single top quark cross section in next-to-leading order QCD},''
  {\em Phys. Rev.} {\bf D66} (2002) 054024,
\href{http://www.arXiv.org/abs/hep-ph/0207055}{{\tt hep-ph/0207055}}.

\bibitem{Campbell:2004ch}
J.~M. Campbell, R.~K. Ellis, and F.~Tramontano, ``{Single top production and
  decay at next-to-leading order},'' {\em Phys. Rev.} {\bf D70} (2004) 094012,
\href{http://www.arXiv.org/abs/hep-ph/0408158}{{\tt hep-ph/0408158}}.

\bibitem{Cao:2004ky}
Q.-H. Cao and C.~P. Yuan, ``{Single top quark production and decay at
  next-to-leading order in hadron collision},'' {\em Phys. Rev.} {\bf D71}
  (2005) 054022,
\href{http://www.arXiv.org/abs/hep-ph/0408180}{{\tt hep-ph/0408180}}.

\bibitem{Cao:2004ap}
Q.-H. Cao, R.~Schwienhorst, and C.~P. Yuan, ``{Next-to-leading order
  corrections to single top quark production and decay at Tevatron. I:
  s-channel process},'' {\em Phys. Rev.} {\bf D71} (2005) 054023,
\href{http://www.arXiv.org/abs/hep-ph/0409040}{{\tt hep-ph/0409040}}.

\bibitem{Cao:2005pq}
Q.-H. Cao, R.~Schwienhorst, J.~A. Benitez, R.~Brock, and C.~P. Yuan,
  ``{Next-to-leading order corrections to single top quark production and decay
  at the Tevatron. II: t-channel process},'' {\em Phys. Rev.} {\bf D72} (2005)
  094027,
\href{http://www.arXiv.org/abs/hep-ph/0504230}{{\tt hep-ph/0504230}}.

\bibitem{Frixione:2005vw}
S.~Frixione, E.~Laenen, P.~Motylinski, and B.~R. Webber, ``{Single-top
  production in MC@NLO},'' {\em JHEP} {\bf 03} (2006) 092,
\href{http://www.arXiv.org/abs/hep-ph/0512250}{{\tt hep-ph/0512250}}.

\bibitem{Alioli:2009je}
S.~Alioli, P.~Nason, C.~Oleari, and E.~Re, ``{NLO single-top production matched
  with shower in POWHEG: s- and t-channel contributions},'' {\em JHEP} {\bf 09}
  (2009) 111,
\href{http://www.arXiv.org/abs/0907.4076}{{\tt 0907.4076}}.

\bibitem{Kidonakis:2011wy}
N.~Kidonakis, ``{Next-to-next-to-leading-order collinear and soft gluon
  corrections for t-channel single top quark production},'' {\em Phys.Rev.}
  {\bf D83} (2011) 091503, \href{http://www.arXiv.org/abs/1103.2792}{{\tt
  1103.2792}}.

\bibitem{Falgari:2011qa}
P.~Falgari, F.~Giannuzzi, P.~Mellor, and A.~Signer, ``{Off-shell effects for
  t-channel and s-channel single-top production at NLO in QCD},'' {\em
  Phys.Rev.} {\bf D83} (2011) 094013,
  \href{http://www.arXiv.org/abs/1102.5267}{{\tt 1102.5267}}.

\bibitem{Falgari:2010sf}
P.~Falgari, P.~Mellor, and A.~Signer, ``{Production-decay interferences at NLO
  in QCD for $t$-channel single-top production},'' {\em Phys.Rev.} {\bf D82}
  (2010) 054028, \href{http://www.arXiv.org/abs/1007.0893}{{\tt 1007.0893}}.

\bibitem{Cao:2007ea}
Q.-H. Cao, J.~Wudka, and C.~P. Yuan, ``{Search for New Physics via Single Top
  Production at the LHC},'' {\em Phys. Lett.} {\bf B658} (2007) 50--56,
\href{http://www.arXiv.org/abs/0704.2809}{{\tt 0704.2809}}.

\bibitem{Sullivan:2005ar}
Z.~Sullivan, ``{Angular correlations in single-top-quark and W j j production
  at next-to-leading order},'' {\em Phys. Rev.} {\bf D72} (2005) 094034,
\href{http://www.arXiv.org/abs/hep-ph/0510224}{{\tt hep-ph/0510224}}.

\bibitem{Sullivan:2004ie}
Z.~Sullivan, ``{Understanding single-top-quark production and jets at hadron
  colliders},'' {\em Phys. Rev.} {\bf D70} (2004) 114012,
\href{http://www.arXiv.org/abs/hep-ph/0408049}{{\tt hep-ph/0408049}}.

\bibitem{Motylinski:2009kt}
P.~Motylinski, ``{Angular correlations in t-channel single top production at
  the LHC},'' {\em Phys. Rev.} {\bf D80} (2009) 074015,
\href{http://www.arXiv.org/abs/0905.4754}{{\tt 0905.4754}}.

\bibitem{Catani:1993hr}
S.~Catani, Y.~L. Dokshitzer, M.~Seymour, and B.~Webber, ``{Longitudinally
  invariant $K_t$ clustering algorithms for hadron hadron collisions},'' {\em
  Nucl.Phys.} {\bf B406} (1993) 187--224.

\end{thebibliography}\endgroup
\end{document}